\documentclass[aps,a4paper,showkeys,nofootinbib,longbibliography,notitlepage,twocolumn]{revtex4-1}
\usepackage{filecontents}

\usepackage{natbib}
\usepackage{amsmath,amssymb,bm,amsthm}
\usepackage{subfigure}
\usepackage{graphicx}% Include figure files
\usepackage{dcolumn}% Align table columns on decimal point
\usepackage{bm}% bold math
\usepackage[mathlines]{lineno}% Enable numbering of text and display math
\usepackage{color}
\newcounter{one}
\usepackage{url}

\usepackage{textcase}

\usepackage{colortbl}
\usepackage{tabularx}
\usepackage{verbatim}
\usepackage{multirow}

\usepackage[T1]{fontenc} 
\usepackage{lmodern}
\usepackage{bbm}
\usepackage[utf8]{inputenc}
\usepackage{amsfonts}
\usepackage{array}

\usepackage{mathtools}
\def\multiset#1#2{\ensuremath{\left(\kern-.3em\left(\genfrac{}{}{0pt}{}{#1}{#2}\right)\kern-.3em\right)}}

\usepackage{bbm}
\usepackage{enumitem}
\usepackage{umoline}
\usepackage[usenames,svgnames]{xcolor}
\usepackage{natbib}
\usepackage[hyperindex,breaklinks]{hyperref}
\hypersetup{
     colorlinks=true,       		% false: boxed links; true: colored links
     linkcolor=Navy,          	% color of internal links
     citecolor=Navy,            % color of links to bibliography
     filecolor=Navy,      		% color of file links
     urlcolor=Navy,           	% color of external links
    runcolor=cyan,
 }
\usepackage{graphicx}
\usepackage{amsfonts}
\usepackage{amssymb}
\usepackage{amsmath}
\usepackage{bbm}
\usepackage{enumitem}

\newcommand{\bra}[1]{\langle #1 |}
\newcommand{\ket}[1]{| #1 \rangle}

\newcommand{\tr}[0]{ {\rm tr}}

\newcommand{\half}[1]{{ \rm h}}
\newcommand{\Oorderof}{\mathcal{O}}
\newcommand{\orderof}[1]{\Oorderof(#1)} 

\newcommand{\for}[0]{\quad \textrm{for} \quad}

\newcommand{\dist}{{\rm dist}}

\newcommand{\Prod}{\mathcal{P}}
\newcommand{\co}{{\rm c}}

\newcommand{\diam}{{\rm diam}}

\newcommand{\tc}{{\rm t}}

\newcommand{\Gs}{0}

\newcommand{\poly}{{\rm poly}}

\newcommand{\Or}{\quad {\rm or} \quad}
\newcommand{\with}{\quad {\rm with} \quad}

\newcommand{\ad}{{\rm ad}}

\def\beq{\begin{equation}}
\def\eeq{\end{equation}}
\def\nbeq{\begin{equation*}}
\def\neeq{\end{equation*}}
\def\<{\langle}
\def\>{\rangle}

\def\tr{{\rm tr}}

\theoremstyle{definition}
\newtheorem{theorem}{Theorem}
\newtheorem{lemma}{Lemma}
\newtheorem{corol}{Corollary}
\newtheorem{assump}{Assumption} 

\newtheorem{prop}[theorem]{Proposition} 
\newtheorem{claim}{Claim} 

\usepackage{minitoc}

\usepackage[toc,page,titletoc]{appendix}
\usepackage[capitalise,nameinlink]{cleveref}

\crefname{supp}{Supplement}{Supplements}
\usepackage{chngcntr}

\begin{document}
\title{Area law of non-critical ground states in 1D long-range interacting systems}

\author{Tomotaka Kuwahara$^{1,2}$}

%\altaffiliation{Present address: Mathematical Science Team, RIKEN Center for Advanced Intelligence Project (AIP),1-4-1 Nihonbashi, Chuo-ku, Tokyo 103-0027, Japan}
\email{tomotaka.kuwahara@riken.jp}
\affiliation{$^{1}$
Mathematical Science Team, RIKEN Center for Advanced Intelligence Project (AIP),1-4-1 Nihonbashi, Chuo-ku, Tokyo 103-0027, Japan
}
%\affiliation{$^{2}$Department of Mathematics, Faculty of Science and Technology, Keio University, 3-14-1 Hiyoshi, Kouhoku-ku, Yokohama 223-8522, Japan}
\affiliation{$^{2}$Interdisciplinary Theoretical \& Mathematical Sciences Program (iTHEMS) RIKEN 2-1, Hirosawa, Wako, Saitama 351-0198, Japan}

\author{Keiji Saito$^{3}$}
%\affiliation{
%Department of Physics, Graduate School of Science,
%University of Tokyo, Kashiwa 277-8574, Japan
%}
%\author{Keiji Saito}
%\email{saitoh@rk.phys.keio.ac.jp}
\affiliation{$^{3}$Department of Physics, Keio University, Yokohama 223-8522, Japan}

%\begin{abstract}
% Introducition:
% --------------
% 
% summarizing the background, rationale, main results (introduced by
% "Here we show" or some equivalent phrase) and implications of the
% study.
%

\begin{abstract}
The area law for entanglement provides one of the most important connections between information theory and quantum many-body physics.
 It is not only related to the universality of quantum phases, but also to efficient numerical simulations in the ground state.
Various numerical observations have led to a strong belief that the area law is true for every non-critical phase in short-range interacting systems. 
However, the area law for long-range interacting systems is still elusive, as the long-range interaction results in correlation patterns similar to those in critical phases. 
Here, we show that for generic non-critical one-dimensional ground states with locally bounded Hamiltonians, the area law robustly holds without any corrections, even under long-range interactions. 
Our result guarantees an efficient description of ground states by the matrix-product state in experimentally relevant long-range systems, which justifies the density-matrix renormalization algorithm. 
\end{abstract}

\maketitle

\section{Introduction}

The quantum entanglement plays a crucial role in characterizing the low-temperature physics of quantum many-body systems in terms of quantum information science.
It is often measured by the quantum entanglement entropy between two subsystems and its scaling is deeply related to the universality of the ground state~\cite{PhysRevLett.90.227902,Calabrese_2004}.
When the interactions in quantum many-body systems are local, the quantum correlation is typically expected to be short-range. 
This intuition leads to the conjecture that the entanglement entropy naturally scales as the boundary area of the subregion.
This \textit{area-law conjecture} is numerically verified in various quantum many-body systems and is expected to be true in all gapped ground states (i.e., in non-critical phases)~\cite{RevModPhys.82.277}.

\begin{figure}[tt]
\centering
\includegraphics[clip, scale=0.35]{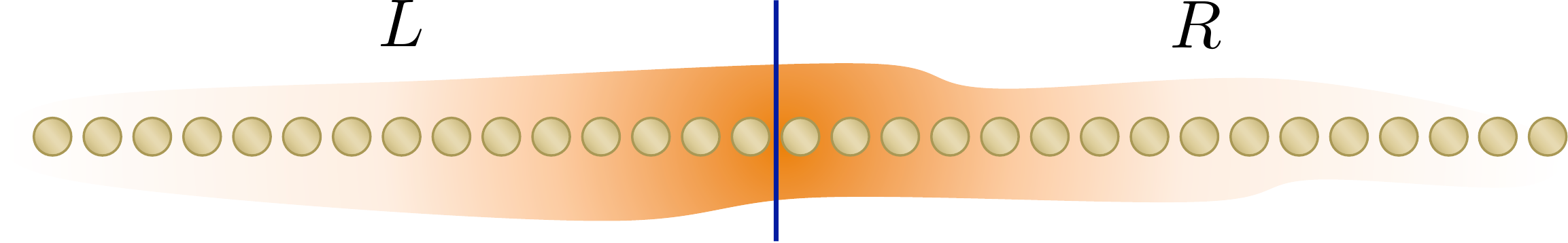}
\caption{(Area law in 1D system) When we decompose the total system into two subsystems $L$ and $R$, the boundary area between the two subsystems is described by points. 
The area law simply argues that the entanglement entropy is bounded from above by a constant value, as in Eq.~\eqref{entropy_1D_area_law}.
We investigate the robustness of the area law under long-range interactions, which induce non-local quantum correlations.  
}
\label{fig:area_law}
\end{figure}

In one-dimensional (1D) systems, for an arbitrary decomposition of the total system, the area law for a ground state is simply stated as follows (Fig.~\ref{fig:area_law}):
\begin{align}
S(\rho_L) &\le {\rm const.} ,\quad  \rho_L = \tr_{R} (\ket{\Gs}\bra{\Gs}), \label{entropy_1D_area_law}
\end{align}
where the ground state is denoted as $\ket{\Gs}$ and $S(\rho_L)$ is the von Neumann entropy, namely $S(\rho_L)=\tr(-\rho_L \log \rho_L)$.
%The area-law conjecture states that in every gapped ground states (i.e., in non-critical phases), the entanglement entropy scales at most as the boundary area of subregion.
Over the past dozen years or so, the area-law conjecture has attracted much attention, as it characterizes the universal structure of many-body physics in simple and beautiful ways~\cite{RevModPhys.82.277}. However, providing detailed proof of the area law is still an extremely challenging problem. 
So far, the proof of this law is limited to gapped 1D systems~\cite{
Hastings_2007,
PhysRevB.85.195145,
arad2013area,
Arad2017}, 
1D quantum states with finite correlation lengths~\cite{
brandao2013area,
PhysRevX.8.031009}, 
gapped harmonic lattice systems~\cite{PhysRevA.66.042327,
PhysRevLett.94.060503}, tree-graph systems~\cite{abrahamsen2019polynomial}, and high-dimensional systems with specific assumptions~\cite{PhysRevA.80.052104,PhysRevB.92.115134,PhysRevB.76.035114,PhysRevB.92.115134,PhysRevLett.113.197204,10.1145/3357713.3384292} (see Ref.~\cite{
RevModPhys.82.277} 
for a comprehensive review).
The area law is the backbone of the density-matrix renormalization algorithm~\cite{SCHOLLWOCK201196}, as it implicitly assumes the area law structure of the ground states.  
The results pertaining to the 1D area law~\cite{
Hastings_2007,
arad2013area} 
rigorously justify the efficient description of the ground states using the matrix-product state (MPS), which facilitates the calculation of the ground states by the classical polynomial-time algorithm~\cite{
landau2015polynomial,
Arad2017}.
Finally, in the characterization of ground states, complete classification of 1D quantum phases has been achieved under the MPS ansatz~\cite{
PhysRevB.84.235128}.

Recent experimental advances have enabled the fine-tuning of the interactions between particles~\cite{
%yan2013observation,
richerme2014non,
jurcevic2014quasiparticle,
Islam583,
zhang2017observation}.
These advances push the long-range interacting systems from the theoretical playground to the field relevant to practical applications.
One of the examples of controllable 1D long-range interacting spin systems is the following long-range transverse Ising model: 
\begin{align}
H= \sum_{i<j} \frac{J_{i,j} }{r_{i,j}^\alpha} \sigma_i^x  \sigma_j^x + B \sum_{i}  \sigma_i^z   ,
\label{exp_long_range_trans_Ising}
\end{align}
with $\{\sigma^x,\sigma^y,\sigma^z\}$ as the Pauli matrices,
where $r_{i,j}$ is the distance between the two sites $i$ and $j$, and the exponent is tunable from $\alpha=0$ to $\alpha=3$~\cite{
jurcevic2014quasiparticle,
zhang2017observation} (also $\alpha=6$ by van-der-Waals interactions~\cite{PhysRevX.7.041063,zeiher2016many}).
In theoretical studies, new types of quantum phases induced by long-range interactions have been reported in the transverse Ising model~\cite{
PhysRevLett.109.267203,PhysRevB.94.075156}, 
the Kitaev chain~\cite{
PhysRevLett.113.156402,
LEPORI201635}, 
the XXZ-model~\cite{
PhysRevLett.119.023001}, 
the Heisenberg model~\cite{
PhysRevLett.104.137204,
PhysRevB.93.205115}, as well as other models. 
Typically, non-trivial quantum phases are induced by long-range interactions with power exponents smaller than three ($\alpha\le 3$). 
For $\alpha>3$, the universality class is the same as that of short-range interacting systems~\cite{
PhysRevLett.29.917,
PhysRevB.64.184106}
(i.e., $\alpha=\infty$).
This means that the regime of $\alpha\le 3$ is essentially important to the discussion of the area law in long-range interacting systems.

We can now turn to the question of whether the area law of the entanglement entropy~\eqref{entropy_1D_area_law} is still satisfied in the presence of long-range interactions. 
Typically, long-range interacting systems show a power-law decay of the correlations even in non-critical ground states~\cite{
PhysRevLett.109.267203,
PhysRevLett.113.156402}; 
 this property is similar to critical ground states in short-range interacting systems. 
To date, it has been a challenge, both numerically and theoretically, to identify the regime of $\alpha$ to justify the area law.
Although several numerical studies suggest that the area law holds for short-range regimes (i.e., $\alpha>3$), 
the possibility of a sub-logarithmic violation to the standard area law~\eqref{entropy_1D_area_law} has also been indicated for $\alpha\le 3$~\cite{
PhysRevLett.109.267203}.
On the other hand, most theoretical analyses regarding the area law rely on the strict locality of the interactions and cannot be directly applied to the power-law decay of interactions even for sufficiently large $\alpha$ values.

One of the natural routes to prove the area law under long-range interactions is to connect the entanglement entropy to the power-law decay of the bi-partite correlation 
by extending the area-law proof from exponential clustering~\cite{
brandao2013area,
PhysRevX.8.031009}.
% that is ensured in gapped ground states from the well-known clustering theorem~\cite{ref:Hastings2006-ExpDec}.
However, such a connection cannot be generalized because of the existence of strange quantum states~\cite{
hastings2016random} 
that have arbitrarily large entanglement entropy values while maintaining a correlation length of order $\mathcal{O}[\log (n)]$ (i.e., corresponding to $\alpha = \infty$).
The other route relies on assuming the existence of the quasi-adiabatic path~\cite{
PhysRevB.72.045141} 
to a trivial ground state satisfying the area law. 
Using the small-incremental-entangling theorem, this assumption ensures the area law in generic gapped short-range interacting systems~\cite{
PhysRevLett.111.170501}.
However, regarding 1D long-range interacting systems, the area law has been proved 
only for short-range regimes $\alpha>4$ even under this strong assumption~\cite{
PhysRevLett.119.050501}.

Based on the above discussion, we report a general theorem on the area law in 1D long-range interacting systems in this work.
It applies to generic 1D gapped systems with $\alpha>2$ and 
ensures a constant-bounded entanglement entropy even in long-range regimes ($\alpha\le 3$) in which non-trivial quantum phases appear owing to their long-range nature.
We provide an outline of the proof in Method section.

\section{Results}

\subsection{Main statement on the area law}

We consider a 1D system with $n$ sites, each of which has a $d$-dimensional Hilbert space. 
We focus on the Hamiltonian $H$ with power-law decaying interactions:
\begin{align}
H=\sum_{i<j}  h_{i,j} +  \sum_{i=1}^n h_i  \label{Ham:general_power}
\end{align} 
with $\|h_{i,j}\|\le J/r_{i,j}^\alpha$ and $\|h_i\| \le B$ for $\forall i,j$, where $\{h_{i,j}\}_{i<j}$ are the bi-partite interaction operators, $\{h_i\}_{i=1}^n$ are the local potentials, $J$ and $B$ are  constants of $\orderof{1}$.
One typical example is given by the long-range Ising model, shown in Eq.~\eqref{exp_long_range_trans_Ising}, where $d=2$, $h_{i,j} =  J_{i,j} \sigma_i^x  \sigma_j^x /r_{i,j}^\alpha $ and $h_i=B\sigma_i^z$. 
As long as the local energy is finitely bounded, our result can also be extended to fermionic and bosonic systems (e.g., hard-core bosons).
For simplicity, we here restrict ourselves to two-body interactions, but our results are generalized to generic $k$-body interactions with $k=\orderof{1}$.  
We consider the entanglement entropy of the ground state $\ket{\Gs}$ in terms of the spectral gap $\Delta$ just above the ground-state energy.
We assume that the ground state is not degenerate.

 \begin{figure}[tt]
\centering
\includegraphics[clip, scale=0.35]{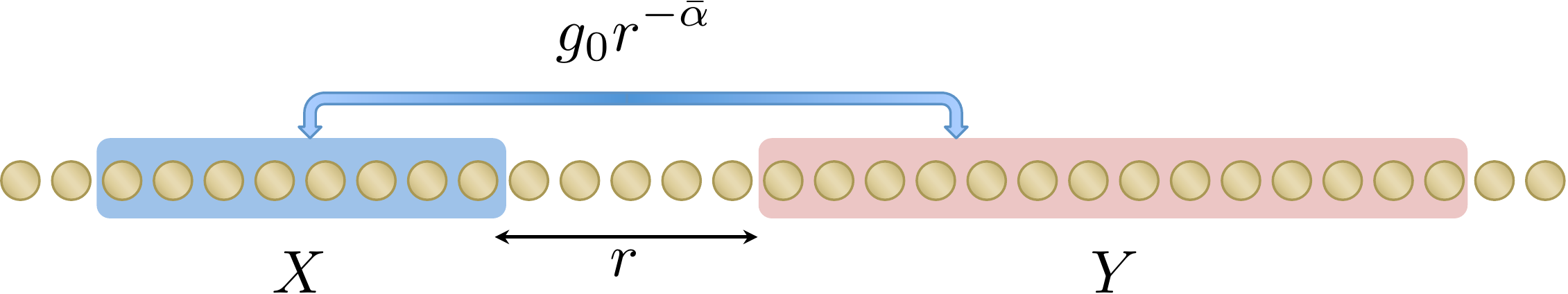}
\caption{(Condition for the area law in long-range interactions). For arbitrary subsystems $X$ and $Y$ separated by $r$ from each other, we assume the total interaction strength between $X$ and $Y$ decay as $r^{-\bar{\alpha}}$, as shown in \eqref{cond_area_LR}. This condition implies that $\alpha >2$ in Eq.~\eqref{Ham:general_power} if we consider the most general class of long-range interacting systems. On the other hand, if we restrict ourselves to a special class of fermionic systems with long-range hopping~\eqref{eq:ham_1D_free}, the condition is relaxed to $\alpha >3/2$. }
\label{fig:block_interaction}
\end{figure}

We now discuss our main theorem. 
We define the interaction between two concatenated subsystems $X$ and $Y$ as follows (Fig.~\ref{fig:block_interaction}):
\begin{align}
V_{X,Y} = \sum_{i \in X}\sum_{j \in Y}  h_{i,j}. \label{V_X_Y_def}
\end{align} 
It simply selects all the interaction terms $\{h_{i,j}\}_{i<j}$ between two sites in $X$ and $Y$.
Here, we assume the existence of a constant $g_0\ge 1$ such that
\begin{align}
\| V_{X,Y} \| \le g_0 r^{-\bar{\alpha}} \quad (\bar{\alpha}>0) \label{cond_area_LR}
\end{align} 
for arbitrary choices of $X$ and $Y$, separated by a distance $r$.
Under the condition~\eqref{cond_area_LR}, the entanglement entropy $S(\rho_L)$ is bounded from above by
\begin{align}
S(\rho_L) &\le c \log^2 (d) \mathcal{G} \left(\frac{\log (d)}{\Delta} \right) 
\label{long_range_area_law}
\end{align}
for arbitrary choices of $L$ and $R$, where $\mathcal{G}(x):= x^{1+2/\bar{\alpha}}\log^{3+3/\bar{\alpha}}(x)$ and $c$ is a constant that depends on $\alpha$, $J$, $B$, $\bar{\alpha}$ and $g_0$.
When the local dimension $d$ and the spectral gap $\Delta$ are independent of the system size $n$, the above inequality results in a constant upper bound for the entanglement entropy. 
Our area-law result can also be applied to quasi-1D systems (e.g., ladder systems) by appropriately choosing the local dimension $d$.

\subsection{Why does the area law holds for $\pmb{\alpha>2}$ ?}

We here show a physical intuition behind our long area law~\eqref{long_range_area_law}.
Naively, the area law might be derived from the power-law decay of the bi-partite correlations~\cite{ref:Hastings2006-ExpDec}.
However, this behavior of the correlation functions is also observed in critical ground states, where the area law is usually known to be violated~\cite{PhysRevLett.90.227902,Calabrese_2004}. 
Moreover, as has been mentioned, the entanglement entropy can obey the volume law for a quantum state with super-polynomially decaying correlations~\cite{
hastings2016random}. 
At first glance, these points are contradictory to our results.
In order to resolve this, we need to focus on the fact that the gap condition imposes much stronger restrictions on the entanglement structure of the ground states than the decay of bi-partite correlations (see Refs.~\cite{Kuwahara_2016_gs,Kuwahara_2017} for example).
Our proof approach fully utilizes the gap condition. 
This point is reflected to the approximation of the ground state using a polynomial of the Hamiltonian, where the approximation error increases as the spectral gap shrinks (see Claim~\ref{Chebyshev_AGSP} in Method section).

We also mention why the condition $\alpha>2$ is a natural condition for the long-range area law. 
If the exponent $\alpha$ is small enough such that the condition~\eqref{cond_area_LR} breaks down, 
the norm of the boundary interaction along a cut (i.e., $V_{X,Y}$ with $X=L$ and $Y=R$) diverges in the thermodynamic limit ($n\to \infty)$.
Then, the system energy possesses a high-dimensional character, and hence its 1D character should be lost.

In order to study this point in more detail, let us consider the area law for thermal equilibrium states, namely $\rho=e^{-\beta H}/\tr(e^{-\beta H})$. 
A natural extension of the ground state's area law is to consider the mutual information $I_{\rho}(L:R):=S(\rho_L)+S(\rho_R) - S(\rho)$.
Note that the mutual information is equal to the entanglement entropy in the limit of $\beta\to\infty$.
At arbitrary temperatures, Ref.~\cite{
PhysRevLett.100.070502} has provided the upper bound of $I_{\rho}(L:R)\le 2\beta \|V_{L,R}\|$ (see also~\cite{2007.11174}), which becomes a constant upper bound (i.e., the area law) if $\|V_{L,R}\|=\orderof{1}$. 
On the other hand, the area law may collapse for $\alpha\le 2$, where the norm of $V_{L,R}$ can diverge to infinity in the thermodynamic limit.
It is natural to expect that the condition for the area law in the thermal state should be looser than that in the ground state.  
This intuition indicates that the condition of $\alpha>2$ should be, at least, a necessary condition for the area law of the entanglement entropy in the ground state. 
We have actually proved that $\alpha>2$ is the sufficient condition. 
We thus believe that our condition of $\alpha>2$ is already optimal (see also the Discussion section below).

\subsection{Several remarks on the area law}

There are several remarks pertaining to the above area law results.
First, in the short-range limit (i.e., $\bar{\alpha} \to \infty$), our area-law bound reduces to the following upper bound:
$$
S(\rho_L) \le c \frac{\log^3 (d)}{\Delta}\log^3\left(\frac{\log (d)}{\Delta}\right) \for \bar{\alpha} \to \infty.
$$
This upper bound reproduces the state-of-the-art bound in short-range interacting systems~\cite{
arad2013area,
Arad2017}. 
This implies that our result provides a natural generalization from the short-range area law to the long-range area law.

Second, the assumption~\eqref{cond_area_LR} is always satisfied for $\alpha>2$ because of $\bar{\alpha} \ge \alpha -2$ (see Method section).
This condition covers important classes of long-range interactions such as  van der Waals interactions ($\alpha=6$) and dipole-dipole interactions ($\alpha=3$).
The condition $\alpha>2$ is the most general sufficient condition for the inequality~\eqref{cond_area_LR} to be satisfied.
Hence, when considering special classes of Hamiltonians, this condition can be relaxed. 
As one such example, we consider fermionic %(or boson) 
systems with long-range hopping as follows:
\begin{align}
H= \sum_{i<j} \frac{1}{r_{i,j}^\alpha} ( A_{i,j} a_i a_j^\dagger +B_{i,j}a_i a_j + {\rm h.c.} )   +V,
\label{eq:ham_1D_free}
\end{align}
where $\{a_i^\dagger, a_i\}_{i=1}^n$ are the creation and the annihilation operators for the fermion, 
%(or boson), 
and $V$ is composed of arbitrary finite-range interaction terms such as $a_ia_i^\dagger a_ja_j^\dagger $ with $r_{i,j}=\orderof{1}$.
In the above cases, we can prove that for $\alpha > 3/2$, the condition~\eqref{cond_area_LR} is satisfied (see Lemma~2 in Supplementary Note~1).
For $V=0$, this model is integrable and exactly solvable. For example, the Kitaev chain with long-range hopping corresponds to this class.
Interestingly, in the long-range Kitaev chain, 
the point $\alpha_c=3/2$ is linked to a phase transition resulting from conformal-symmetry breaking~\cite{
PhysRevLett.113.156402}.

Finally, we mention the relevance to experimental observations regarding the long-range area law.
Recent advances in experimental setups have achieved direct observation of the second-order R\'enyi entropy~\cite{islam2015measuring}.
The second-order R\'enyi entropy for a subsystem $L$ (as in Fig.~\ref{fig:area_law}) is defined as $S_2(\rho_L):= -\log [\tr (\rho_L^2)]$, 
 and $S_2(\rho_L)$ provides a lower bound for the entanglement entropy $S(\rho_L)$ in Eq.~\eqref{entropy_1D_area_law}.
Hence, we can obtain the same area-law bound as \eqref{long_range_area_law} for $S_2(\rho_L)$.
Recently, the measurement of R\'enyi entropy was reported~\cite{Brydges260} in long-range $XY$ models with tunable power exponents $0< \alpha <3$.
We expect that our area-law bound would support the outcome of experimental observations regarding entanglement entropy of ground states.

\subsection{Matrix product state approximation}

Based on our analysis, we can also determine the efficiency of approximation of ground states $\ket{\Gs}$ in terms of the matrix-product representation. 
We approximate the exact ground state $\ket{\Gs}$ using the following quantum state $\ket{\psi_D}$:
\begin{align}
\ket{\psi_D} = \sum_{s_1,s_2,\ldots,s_n =1}^d \tr(A_{1}^{[s_1]}A_{2}^{[s_2]} \cdots A_{n}^{[s_n]}) \ket{s_1,s_2,\ldots,s_n} \notag, 
\end{align}
where each of the matrices $\{A_i^{[s_i]}\}_{i,s_i}$ is described by the $D\times D$ matrix.
We refer to the matrix size $D$ as the bond dimension.
This MPS has entanglement entropy less than $\log D$ for an arbitrary cut of the system. 
Although arbitrary quantum states can be described by the MPS, generic quantum states require exponentially large bond dimensions, namely $D=\exp[\orderof{n}]$~\cite{SCHOLLWOCK201196}.
If a quantum state is well approximated by the MPS with small bond dimensions, we can efficiently calculate the expectation values of local observables (e.g., energy).

The MPS is the basic ansatz for various types of variational methods (e.g. the density-matrix renormalization group~\cite{
SCHOLLWOCK201196}) and it is crucial to determine whether ground states can be well approximated by the MPS with a small bond dimension.
On the MPS representation of the ground state $\ket{\Gs}$, we prove the following statement: If the condition~\eqref{cond_area_LR} is satisfied and the spectral gap is nonvanishing, there exists an MPS $\ket{\psi_D}$ with bond dimensions $D=\exp[c' \bar{\alpha}^{-1}\log^{5/2}(1/\delta)]$ [$c'$: constant, $\bar{\alpha}=\orderof{1}$] such that
\begin{align}
\| \tr_{X^\co} (\ket{\psi_D}\bra{\psi_D} ) - \tr_{X^\co} (\ket{\Gs}\bra{\Gs} )\|_1 \le \delta  |X| \label{MPS_approx}
\end{align}
for an arbitrary concatenated subregion $X$, where $\|\cdot\|_1$ is the trace norm and $|X|$ denotes the cardinality of $X$. 
We show the proof in the Method section.

From the approximation~\eqref{MPS_approx}, to achieve an approximation error of $\delta=1/\poly(n)$, we need quasi-polynomial bond dimensions, namely $D=\exp[\orderof{\log^{5/2}(n)}]$. 
Our result justifies the MPS ansatz with small bond dimensions, obtained at a moderate computational cost.
This in turn explains the empirical success of the density-matrix-renormalization-group algorithm in long-range interacting systems~\cite{
PhysRevLett.109.267203, PhysRevLett.113.156402, PhysRevB.93.205115}.
On the other hand, our estimation is still slightly weaker than polynomial-size bond dimensions $D=\exp[\orderof{\log(n)}]$.
This is in contrast to the short-range interacting cases, where only sub-linear bond dimensions $D=\exp[\orderof{\log^{3/4} (n)}]$ are required
to represent the gapped ground states using the MPS~\cite{
arad2013area}.

\section{Discussion}

We discuss several future research directions and open questions.
First, could we find an explicit example that violates the entanglement area law for $\alpha\le 2$ or for $\alpha\le 3/2$ in free fermionic systems?
So far, rigorous violations of the area law have been observed for $\alpha=1$ in gapped free fermionic systems~\cite{PhysRevLett.97.150404}.
Moreover, at $\alpha \approx 1$, all existing area-law violations are at most logarithmic, namely $S(\rho_L) \lesssim \log (|L|)$. 
The existence of a natural long-range interacting gapped system where the entanglement entropy obeys the sub-volume law as $S(\rho_L) \lesssim |L|^{\gamma}$ ($0<\gamma<1$) is an intriguing issue. 
Conversely, it is also challenging to generalize our area-law bound to the sub-volume-law bound for $\alpha\le 2$. 
This regime is more relevant to high-dimensional systems, and any entropic bound better than the volume law would be helpful in tackling the high-dimensional area-law conjecture.

Second, can we develop an efficiency-guaranteed algorithm to calculate the ground state under the gap condition?
In the inequality~\eqref{MPS_approx}, we have proved the existence of an efficient MPS description of the ground state, but how to find such a description is not clear. 
In short-range interacting systems, this problem has been extensively investigated in popular works by Vidick et al.~\cite{
landau2015polynomial,
Arad2017}.
We expect that their formalism would be generalized to the present cases 
and leads to a quasi-polynomial-time algorithm for calculating ground states within a polynomial error $1/\poly(n)$.
Furthermore, we still have scope to improve the quasi-polynomial bond dimension of $\exp[\orderof{\log^{5/2}(n)}]$ to approximate the ground states.
Whether this bound can be relaxed to a polynomial form of $\exp[\orderof{\log (n)}]=\poly(n)$ is a question that will be addressed in the future.

{\footnotesize
\section{Method}

\subsection{Derivation of $\pmb{\bar{\alpha}\ge \alpha-2}$}

We here show the proof of $\bar{\alpha}\ge \alpha-2$ for the Hamiltonian~\eqref{Ham:general_power}. 
More general cases including fermionic systems are given in Supplementary Note 1.  
For the proof, we estimate the upper bound of
$
\| V_{X,Y} \| \le  \sum_{i \in X} \sum_{j \in Y} \| h_{i,j}\| \le J \sum_{i \in X} \sum_{j \in Y}  r_{i,j}^{-\alpha}, 
$
where we use the power-law decay of the interaction as $\|h_{i,j}\| \le J/r_{i,j}^{\alpha}$. 
Let us define $\dist(X,Y)=r$. Then, we obtain 
$$
J\sum_{i \in X} \sum_{j \in Y} r_{i,j}^{-\alpha} \le J\sum_{x=0}^\infty \sum_{y=0}^\infty (r+x+y)^{-\alpha}, 
$$
where we use the fact that $X$ and $Y$ are concatenated subsets.
For arbitrary integer $r_0\in \mathbb{N}$, we have 
$
\sum_{x=0}^\infty (x+r_0)^{-\alpha}
\le  r_0^{-\alpha}+ \int_{r_0}^\infty x^{-\alpha} dx \le \frac{\alpha}{\alpha-1} r_0^{-\alpha+1} ,
$
and hence 
$$
J\sum_{x=0}^\infty \sum_{y=0}^\infty (r+x+y)^{-\alpha} \le \frac{\alpha J}{\alpha-2} r^{-\alpha+2}. 
$$
We thus prove that $\| V_{X,Y} \|$ decays at least faster than $r^{-\alpha+2}$.

\begin{figure}[tt]
\centering
{\includegraphics[clip, scale=0.4]{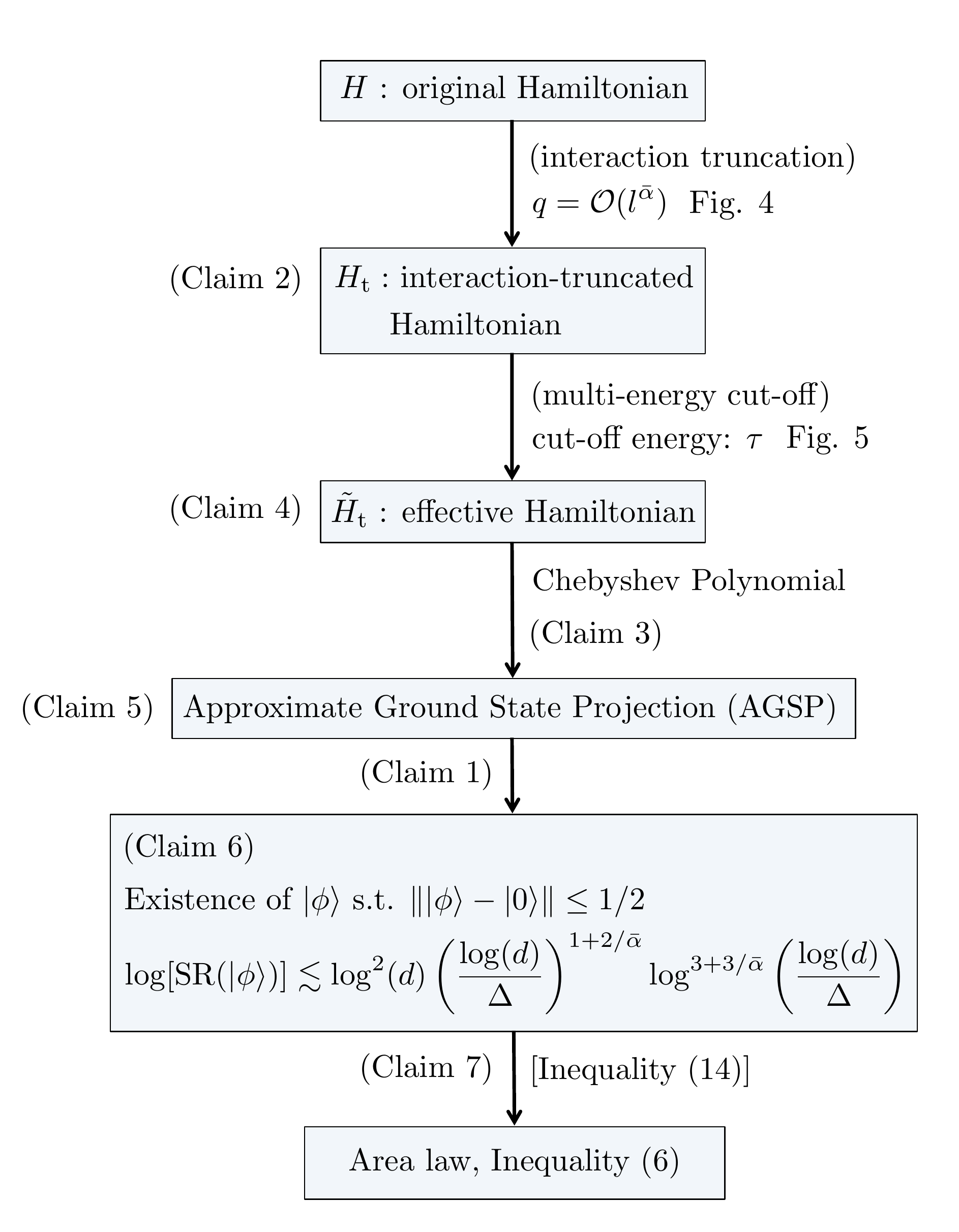}}
\caption{Flow chart of the area law proof.  The proof consists of several key claims. The details of the proof for these claims are given in Supplementary Notes 2-4. 
}
\label{fig:Flow chart of the area law proof}
\end{figure}

\subsection{Proof sketch of the main result}
We here show the sketch of the proof for the area-law inequality~\eqref{long_range_area_law}.
The full proof is quite intricate, and we show the details in Supplementary Notes~2, 3 and 4. 
In Fig.~\ref{fig:Flow chart of the area law proof}, we have summarized a flow of the discussions in this section. 

For the proof, we take the Approximate-Ground-State-Projection (AGSP) approach~\cite{PhysRevB.85.195145,arad2013area}. 
The AGSP operator $K$ is roughly given by the operator that satisfies $K\ket{\Gs}  \simeq \ket{\Gs}$ and $\| K (1 - \ket{\Gs}\bra{\Gs}) \| \simeq 0$. 
The ground state $\ket{\Gs}$ does not change by the AGSP $K$, while any excited state approximately vanishes by $K$.
In more formal definitions, the AGSP is defined by three parameters $\delta_K$, $\epsilon_K$ and $D_K$.
Let $\ket{\Gs_K}$ be a quantum state that does not change by $K$, namely $K\ket{\Gs_K}=\ket{\Gs_K}$. 
Then, the three parameters are defined by the following three inequalities:
$$
\|\ket{\Gs}  -\ket{\Gs_K} \|  \le \delta_K    , \quad  \| K (1- \ket{\Gs_K}\bra{\Gs_K})  \|\le   \epsilon_K ,\quad  {\rm SR}(K) \le D_K,
$$
where ${\rm SR}(K)$ is the Schmidt rank of $K$ with respect to the given partition $\Lambda=L\sqcup R$. 
The essential point of this approach is that a good AGSP ensures the existence of a quantum state that has small Schmidt rank and large overlap with the ground state.
It is mathematically formulated by the following statement:
\begin{claim}[Proposition~2 in Supplementary Note~2] \label{claim:Supplementary Proposition2}
Let $K$ be an AGSP operator for $\ket{\Gs}$ with the parameters ($\delta_K, \epsilon_K, D_K$). 
If we have $\epsilon_K^2 D_K \le (1/2)$, 
there exists a quantum state  $\ket{\psi}$ with ${\rm SR}(\ket{\psi})\le D_{K}$ such that
 \begin{align}
\left \| \ket{\psi} - \ket{\Gs} \right\|\le \epsilon_{K}\sqrt{2 D_{K}   } + \delta_{K}.
\label{ineq:AGSP_bootstrap_norm_distance}
\end{align}
where ${\rm SR}(\ket{\psi})$ is the Schmidt rank of $\ket{\psi}$ with respect to the given partition. 
\end{claim}
\noindent 
From this statement, the primary problem reduces to one of finding a good AGSP to satisfy the condition $\epsilon_K^2 D_K \le (1/2)$. 

\begin{figure}[tt]
\centering
{\includegraphics[clip, scale=0.33]{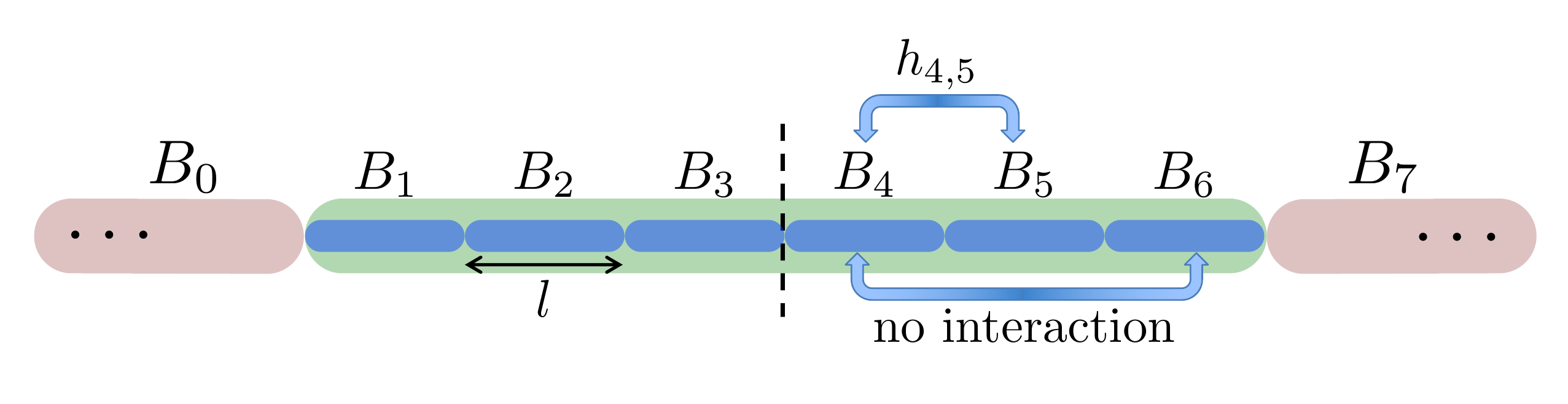}
}
\caption{Interaction-truncated Hamiltonian $H_\tc$. We truncate the long-range interactions only around the boundary. In the figure above, the interactions between non-adjacent blocks (i.e., $\{B_s\}_{s=0}^{7}$) are truncated. By this truncation, the properties of the Hamiltonian $H_\tc$ are proved to be almost the same as the original one $H$, as long as $ql^{-\bar{\alpha}} \lesssim 1$ (see Claim~\ref{claim_truncated_Ham}). 
}
\label{fig:Hamiltonian_define_1}
\end{figure}

In the construction of the AGSP operator with the desired properties, we usually utilize a polynomial of the Hamiltonian. 
The obstacle here is that the long-range interactions induce an infinitely large Schmidt rank in the thermodynamic limit; that is, the Hamiltonian $H$ has the Schmidt rank of ${\rm poly}(n)$.
In order to avoid this, we truncate the long-range interactions of the Hamiltonian.
If we truncate all the long-range interactions, the norm difference between the original Hamiltonian and the truncated one is on the order of $\orderof{n}$, and hence the spectral gap condition cannot be preserved. 
The first central idea in the proof is to truncate the long-range interaction only around the boundary (see Fig.~\ref{fig:Hamiltonian_define_1}).
In more detail, we first decompose the total system into $(q+2)$ blocks with $q$ an even integer.
The edge blocks $B_0$ and $B_{q+1}$ have arbitrary sizes, but the bulk blocks $B_1,B_2,\ldots,B_q$ have the size $l$ (i.e., $|B_s|=l$).
Then, we truncate all the interactions between non-adjacent blocks, which yields the Hamiltonian $H_\tc$ as
\begin{align}
H_\tc =\sum_{s=0}^{q+1} h_{s} + \sum_{s=0}^q h_{s,s+1},
\end{align}
where $h_s$ is the internal interaction in the block $B_s$, and $h_{s,s+1}$ is the interaction between two blocks $B_s$ and $B_{s+1}$.
By using the notation~\eqref{V_X_Y_def}, we have $h_{s,s+1}= V_{B_s,B_{s+1}}$. 
In the Hamiltonian $H_\tc$, long-range interactions only around the boundary are truncated, and hence 
the norm difference between the original Hamiltonian and the truncated Hamiltonian can be sufficiently small for large $l$:
\begin{claim}[Lemmas~3 and 4 in Supplementary Note~2] \label{claim_truncated_Ham}
The norm distance between $H$ and $H_\tc$ is bounded from above by
$$
\|H-H_\tc\|\le  g_0 q l^{-\bar{\alpha}}  . 
$$
Also, the spectral gap $\Delta_\tc$ of $H_\tc$ and the norm difference between $\ket{\Gs}$ and $\ket{\Gs_\tc}$ are upper-bounded by
$$
\Delta_\tc \ge \Delta - 2g_0 q l^{-\bar{\alpha}} ,\quad \| \ket{\Gs}-\ket{\Gs_\tc} \| \le \frac{\|H-H_\tc\|}{\Delta - 4\|H-H_\tc\|},
$$
where $\ket{\Gs_\tc}$ is the ground state of $H_\tc$.  
\end{claim}
\noindent 
From this statement, if $q l^{-\bar{\alpha}} \lesssim 1$, the truncated Hamiltonian $H_\tc$ possesses almost the same properties as the original one.

The second technical obstacle is the norm of the Hamiltonian. 
The gap condition provides us an efficient construction of the AGSP operator, which is expressed by the following statement: 
\begin{claim}[Lemma 11 in Supplementary Note~2] \label{Chebyshev_AGSP}
By using the Chebyshev polynomial, we can find a $m$-degree polynomial $K(m,H_\tc)$ such that 
\begin{align}
\| K(m,H_\tc) (1-\ket{\Gs_\tc}\bra{\Gs_\tc}) \| \le 2\exp\left( -\frac{2m}{\sqrt{\|H_\tc\|/\Delta_{\tc}}}\right) , \label{Chebyshev_Based_AGSP}
\end{align}
where the explicit form of the inequality is given in Supplemental Lemma 11.
\end{claim}
\noindent
We notice that the gap condition plays a crucial role in this claim.
Here, the norm of $\|H_\tc\|$ is as large as $\orderof{n}$, which necessitates the polynomial degree of $m=\orderof{\sqrt{n}}$. 
Polynomials with such a large degree cannot be utilized to prove the condition for the AGSP in Claim~\ref{claim:Supplementary Proposition2}.  
To overcome this difficulty, we aim to construct an effective Hamiltonian with a small norm that retains the similar low-energy properties to the original Hamiltonian.
For this purpose, in each of the blocks, we cut-off the energy spectrum up to some truncation energy (see Fig.~\ref{fig:Hamiltonian_define_2}).
Then, the block-block interactions (i.e., $h_{s,s+1}$) do not change, and the internal Hamiltonian $h_s$ is transformed to $\tilde{h}_s$.
By this energy cut-off, the total norm of the effective Hamiltonian $\tilde{H}_\tc$ is roughly given by $q\tau$.
The question is whether this effective Hamiltonian possesses the ground state property similar to $H$.
By extending the original result in Ref.~\cite{Arad_2016}, which considers a cut-off in a Hamiltonian of one region, we prove the statement as follows:
\begin{claim}[Theorem~5 in Supplementary Note~2] \label{thm:multi_ver_Arad_kuwahara_Landau}
Let us choose $\tau$ such that
$$
\tau \gtrsim \log(q)  .
$$
Then, the spectral gap $\tilde{\Delta}_\tc$ of the effective Hamiltonian is preserved as 
$
\tilde{\Delta}_\tc \ge \orderof{ \Delta_\tc} .
$
Moreover, the norm distance between the original ground state $\ket{\Gs_\tc}$ and the effective one $\ket{\tilde{\Gs}_\tc}$ is exponentially small with respect to the cut-off energy $\tau$:
$$
\| \ket{\tilde{\Gs}_\tc}-\ket{\Gs_\tc} \| \le  e^{-\orderof{\tau}} . 
$$
\end{claim}
\noindent
As long as $\tau$ is larger than $\orderof{log(q)}$, the spectral gap is preserved, and the norm of the effective Hamiltonian is as large as $q\log(q)$, namely $\|\tilde{H}_\tc\| \lesssim q\log(q)$.
In the standard construction of the effective Hamiltonian~\cite{arad2013area,Arad_2016}, we perform the energy cut-off only in the edge blocks (i.e., $B_0$ and $B_{q+1}$). 
However, this simple procedure allows us to prove the long-range area law only in the short-range power-exponent regimes (i.e., $\alpha>3$).
The multi-energy cut-off is crucial to prove the area law even in the long-range power-exponent regimes (i.e., $2< \alpha\le 3$).

\begin{figure}[tt]
\centering
{\includegraphics[clip, scale=0.33]{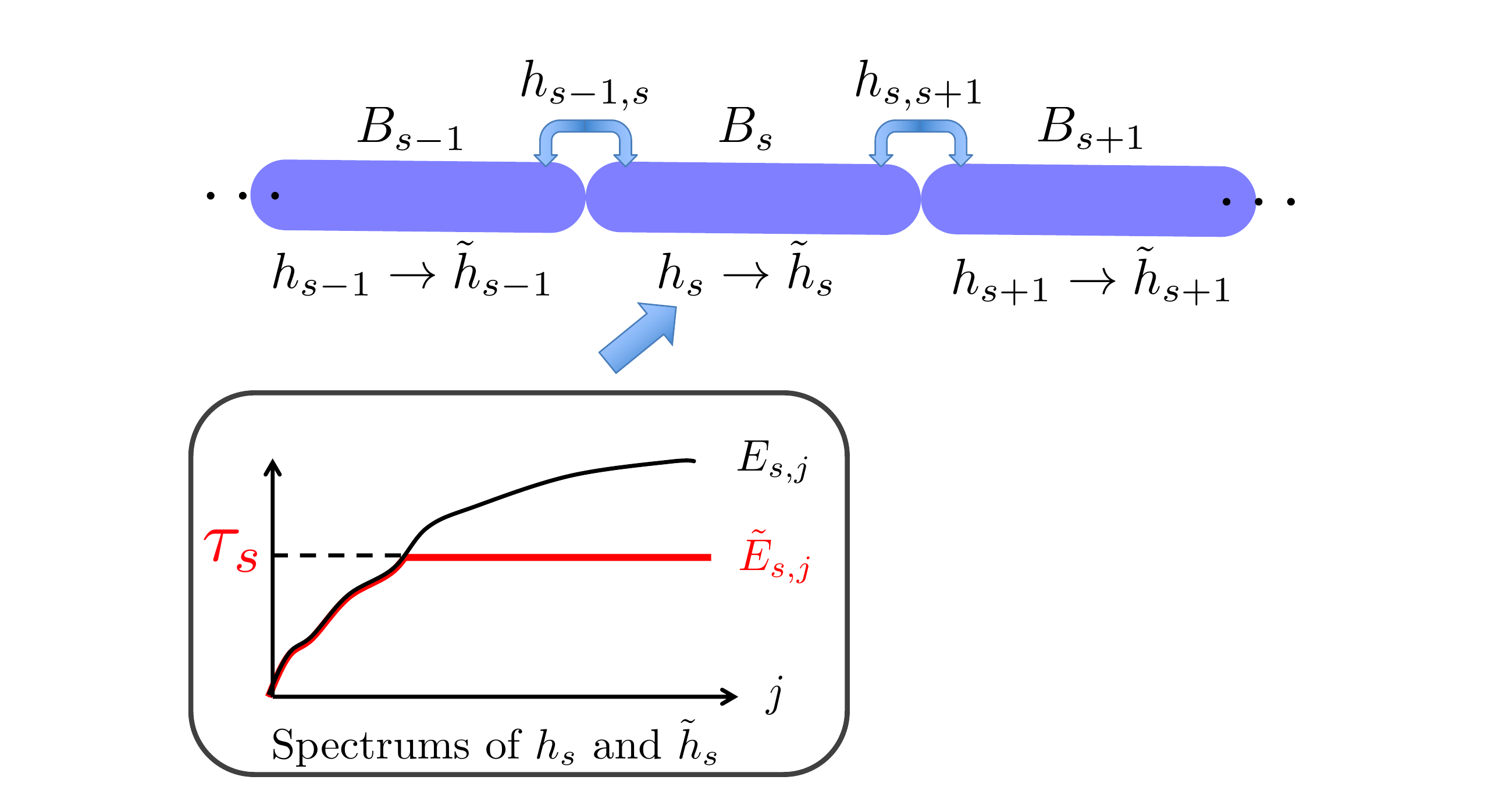}
}
\caption{Effective Hamiltonian $\tilde{H}_\tc$ by multi-energy cut-off. In each of the internal Hamiltonians $\{h_s\}_{s=0}^{q+1}$, we perform the energy cutoff up to the energy $\tau_s=E_{s,0} + \tau$. Here, $\{E_{s,j},\ket{E_{s,j}}\}$ are the energy eigenvalues and the corresponding eigenstates of $h_s$, respectively. The internal Hamiltonians $h_s$ and $\tilde{h}_s$ have the same eigenstates $\{\ket{E_{s,j}}\}$ and the same eigenvalues (i.e., $E_{s,j}=\tilde{E}_{s,j}$), as long as $E_{s,j} \le \tau_s$, above which the eigenvalues differ from each other.   
}
\label{fig:Hamiltonian_define_2}
\end{figure}

By using the polynomial $K(m,x)$ in~\eqref{Chebyshev_Based_AGSP} with $x=\tilde{H}_\tc$, we can obtain the AGSP operator $K_\tc$ for the ground state $\ket{\Gs_\tc}$ of $H_\tc$.
Before showing the AGSP parameter for $K_\tc$, we discuss the Schmidt rank of the polynomial of the Hamiltonian. 
Now, the effective Hamiltonian $\tilde{H}_\tc$ is given by the form of $\sum_{s=0}^{q+1} \tilde{h}_s + \sum_{s=0}^q h_{s,s+1}$. 
By extending the Schmidt rank estimation in Ref.~\cite{PhysRevB.85.195145,arad2013area}, we can derive the following statement:
\begin{claim}[Proposition 4 in Supplementary Note~2] \label{Schmidt rank of the power of the effective Hamiltonian}
The Schmidt rank of the power of the effective Hamiltonian ${\rm SR} (H_\tc^m)$ is bounded from above by
$$
{\rm SR} (\tilde{H}_\tc^m) \le e^{\orderof{ql} + \orderof{m/q} \log (ql) } .
$$
\end{claim}
\noindent 
This inequality gives the upper bound of the Schmidt rank for $K(m,\tilde{H}_\tc)$. 

We have obtained all the ingredients to estimate the parameters $\delta_{K_\tc}$, $\epsilon_{K_\tc}$ and $D_{K_\tc}$ for the AGSP $K_\tc= K(m,\tilde{H}_\tc)$. 
They are given by Claim~\ref{thm:multi_ver_Arad_kuwahara_Landau},  the inequality~\eqref{Chebyshev_Based_AGSP} and Claim~\ref{Schmidt rank of the power of the effective Hamiltonian} as follows:
\begin{align}
\label{AGSP_parameters_3}
&\delta_{K_\tc} = e^{-\orderof{\tau}} ,\quad \epsilon_{K_\tc}= e^{-\orderof{m}/\sqrt{q\log(q)}} ,\notag \\
&{\rm and} \quad D_{K_\tc} = e^{\orderof{ql} + \orderof{m/q} \log (ql) } ,
\end{align}
where we omit the $\bar{\alpha}$-dependence of the parameters.
Let us apply Claim~\ref{claim:Supplementary Proposition2} to the AGSP $K_\tc$ and the ground state $\ket{\Gs_\tc}$. 
Under the condition of $q l^{-\bar{\alpha}} \lesssim 1$, we can find $q$, $m$ and $l$ such that $\epsilon_{K_\tc}^2 D_{K_\tc} \le (1/2)$, where $\{q,m,l\}$ have quantities of $\orderof{1}$.
This leads to the following statement:
\begin{claim}[Proposition~6 in Supplementary Note~3.] \label{prop1:truncate_gs_overlap}
There exists a quantum state $\ket{\phi}$ such that 
$
\|\ket{\Gs} -\ket{\phi} \| \le 1/2
$
with
\begin{align}
\label{upp_Schmidt_phi_ineq}
\log \left[{\rm SR}(\ket{\phi}) \right]\le  c^\ast  \log^2 (d) \left(\frac{\log (d)}{\Delta}\right)^{1 +2/\bar{\alpha}}\log^{3+3/\bar{\alpha}}\left(\frac{\log (d)}{\Delta} \right),
\end{align}
where $ c^\ast$ is a constant which depends only on $g_0$ and $\bar{\alpha}$, which is finite in the limit of $\bar{\alpha}\to \infty$.
\end{claim}

Finally, we construct a set of the AGSP operators $\{K_p\}_{p=1}^\infty$ for the ground state $\ket{\Gs}$, where the AGSP parameters are denoted by $\delta_p$, $\epsilon_p$ and $D_p$.
The errors $\epsilon_{p}$ and $\delta_{p}$ decrease with the index $p$, namely $\epsilon_1\ge \epsilon_2\ge\cdots$ and $\delta_1\ge \delta_2\ge\cdots$. 
In the limit of $p\to \infty$, the AGSP $K_p$ approaches the exact ground-state projection $K_{\infty}=\ket{\Gs}\bra{\Gs}$, namely $\lim_{p\to\infty}\delta_p=0$ and $\lim_{p\to\infty}\epsilon_p=0$.  
These AGSP operators allow the derivation of an upper bound of the entanglement entropy as well as the approximation of the ground state by quantum states with small Schmidt ranks:
\begin{claim}[Proposition~3 in Supplementary Note~2] \label{prop:entropy_and_AGSP}
Let $\ket{\phi}$ be an arbitrary quantum state with 
$\| \ket{\Gs}-\ket{\phi} \|\le 1/2$. 
Also, let $\{K_p\}_{p=1}^{\infty}$ be AGSP operators defined as above. 
Then, we prove for each of $\{K_p\}_{p=1}^{\infty}$
$$
 \left \| \frac{ K_p e^{-i\theta_p}\ket{\phi} }{\| K_p \ket{\phi}\|} - \ket{\Gs} \right\|  \le \gamma_p := \frac{\epsilon_{p}}{ 1/2- \delta_p} + \delta_{p} ,
$$
where the phase $\theta_p \in \mathbb{R}$ is appropriately chosen. 
Moreover, under the condition $\gamma_p\le 1$ for all $p$, the entanglement entropy $S(\ket{0})$ is bounded from above by
$$
S(\ket{\Gs}) \le \log \left[{\rm SR}(\ket{\phi}) \right] -  \sum_{p=0}^\infty \gamma_p^2  \log  \frac{\gamma_p^2 }{3D_{p+1}} , 
$$
where we set $\gamma_0:=1$.
\end{claim}
\noindent
In Proposition~7 of Supplementary Note~3, we show a construction of the AGSP set $\{K_p\}_{p=1}^\infty$ such that 
$\gamma_p^2 = 1/p^2$ and 
\begin{align}
\label{D_p_gamma_p_1/p^2}
\log (3D_p) \le c_1 \bar{\alpha}^{-1} \frac{\log^{5/2} (3p/\Delta)}{\sqrt{\Delta}} + c_2 \frac{\log^{3/2} (3p/\Delta) \log (d) }{\sqrt{\Delta}}  ,
\end{align}
where $c_1$ and $c_2$ are constants that depend on $g_0$. 
We have obtained the quantum state $\ket{\phi}$ with the Schmidt rank as in \eqref{upp_Schmidt_phi_ineq}, and hence from Claim~\ref{prop:entropy_and_AGSP}, 
the above AGSP operators give the upper bound of the entanglement entropy in \eqref{long_range_area_law}. 
This completes the proof of the area law in long-range interacting systems. $\square$

\subsection{MPS approximation of the ground state}

We here prove the inequality~\eqref{MPS_approx}.
For simplicity, let us consider $X$ to be the total system (i.e., $X=\Lambda$). 
Generalization to $X \subset \Lambda$ is straightforward. 
Our proof relies on the following statement:
\begin{claim}[Lemma~1 in Ref.~\cite{VC06mps}] \label{claim_Verstraete_Cirac}
Let $\ket{\psi}$ be an arbitrary quantum state.
We define the Schmidt decomposition between the subsets $\{1,2,\ldots, i\}$ and $\{i+1,i+2,\ldots,n\}$, as follows:
\begin{align}
\label{psi_def_schmidt}
\ket{\psi} = \sum_{m=1}^\infty \mu_m^{(i)}  \ket{\psi_{\le i,m}} \otimes \ket{\psi_{> i,m}},
\end{align}
where $\{\mu_m^{(i)}\}_{m=1}^\infty$ are the Schmidt coefficients in the descending order.
Then, there exists an MPS approximation $\ket{\psi_D}$ with the bond dimension $D$ that approximates the quantum state $\ket{\psi}$ as 
$$
\| \ket{\psi}- \ket{\psi_D} \|^2 \le 2 \sum_{i=1}^{n-1} \delta_i, \quad \delta_i:= \sum_{m>D} |\mu_m^{(i)}|^2.
$$
\end{claim}
\noindent 
From this claim, if we can obtain the truncation error of the Schmidt rank, we can also derive the approximation error by the MPS.

In the following, we give the truncation error by using Claim~\ref{prop:entropy_and_AGSP}.
Let us consider a fixed decomposition as $\Lambda=L \sqcup R$. 
Then, Claim~\ref{prop:entropy_and_AGSP} ensures the existence of the approximation of the ground state $\ket{\Gs}$ with the approximation error $\gamma_p$, which is achieved by the quantum state $\ket{\psi_p}:= K_p e^{i\theta_p} \ket{\phi}/\|K_p\ket{\phi}\|$ with its Schmidt rank of 
$$
\log [{\rm SR} (\ket{\psi_p}) ] \le \log(D_p) + \log [{\rm SR} (\ket{\phi}) ], 
$$
where  $\ket{\phi}$ has the Schmidt rank of \eqref{upp_Schmidt_phi_ineq} at most. 
We have already proved that for $\gamma_p=1/p^2$, the quantity $D_p$ is upper-bounded by \eqref{D_p_gamma_p_1/p^2}. 
Thus, for $p\ge (1/\delta)^{1/4}$ (or $\gamma_p\le \delta^{1/2}$), the Schmidt rank $\log [{\rm SR} (\ket{\psi_p}) ] $ satisfies the following inequality:
\begin{align}
\log [{\rm SR} (\ket{\psi_p}) ] \lesssim   (\bar{\alpha}^{-1} \log (1/\delta) +1)  \log^{3/2} (1/\delta)
\label{Schmidt_rank_psi_p}
\end{align}
for $1/\Delta=\orderof{1}$ and sufficiently small $\delta \ll 1$, where we use the fact that $\log [{\rm SR} (\ket{\phi}) ]$ is a constant of $\orderof{1}$.    

In order to connect the inequality~\eqref{Schmidt_rank_psi_p} to the truncation error of the Schmidt decomposition, we use the following statement:
\begin{claim}[Eckart-Young theorem~\cite{Eckart1936}] \label{claim:Eckart-Young theorem}
Let us consider a normalized state $\ket{\psi}$ as in Eq.~\eqref{psi_def_schmidt}.
Then, for an arbitrary quantum state $\ket{\psi'}$, we have the inequality of  
$
\sum_{m>{\rm SR} (\ket{\psi'})} |\mu_m^{(i)}|^2 \le \| \ket{\phi} -\ket{\psi'}\|^2 , 
$
where the Schmidt rank ${\rm SR} (\ket{\psi'})$ is defined for the decomposition of $\{1,2,\ldots, i\}$ and $\{i+1,i+2,\ldots,n\}$.
\end{claim}
\noindent 
%When the state $\ket{\psi}$ is well-approximated by a quantum state with a small Schmidt rank $\ket{\psi'}$, the Schmidt-rank truncation up to $m\approx {\rm SR} (\ket{\psi'}$ also gives 
In the above claim, we choose $\ket{\Gs}$ and $\ket{\psi_p}$ as $\ket{\psi}$ and $\ket{\psi'}$, respectively, and obtain the inequality of
\begin{align}
\sum_{m>{\rm SR} (\ket{\psi_p})} |\mu_m^{(i)}|^2 \le \gamma_p^2 , \label{thm:The Eckart-Young theorem}
\end{align}
where we use $\|\ket{\psi_p}-\ket{\Gs}\|\le \gamma_p$.
By applying the inequalities~\eqref{Schmidt_rank_psi_p} and \eqref{thm:The Eckart-Young theorem} to Claim~\ref{claim_Verstraete_Cirac}, we can achieve 
$$
\| \ket{\Gs}- \ket{\psi_D} \|^2 \le 2n \delta
$$
if $\log(D)$ is as large as $\bar{\alpha}^{-1} \log^{5/2} (1/\delta)$ [$\bar{\alpha}=\orderof{1}$]. 
This completes the proof. $\square$

}
\begin{acknowledgments}

The work of TK was supported by the RIKEN Center for AIP and JSPS KAKENHI Grant No. 18K13475.
TK gives thanks to God for his wisdom.
KS was supported by JSPS Grants-in-Aid for Scientific Research (JP16H02211 and  JP19H05603).

\end{acknowledgments}

\def\bibsection{\section*{References}} 

\bibliography{Area_LR_main}

\clearpage
\newpage

\renewcommand{\figurename}{Supplementary Figure}

\renewcommand{\tablename}{Supplementary Table}

\renewcommand{\thesection}{Supplementary Note~\arabic{section}}

\addtocounter{section}{-4}

\counterwithout{equation}{section}
\addtocounter{equation}{-17}

\renewcommand{\theequation}{S.\arabic{equation}}

\begin{widetext}

\begin{center}
{\large \bf Supplementary Information for  \protect \\ 
  ``Area law of non-critical ground states in 1D long-range interacting systems'' }\\
\vspace*{0.3cm}
Tomotaka Kuwahara$^{1,2}$ and Keiji Saito$^{3}$ \\
\vspace*{0.1cm}
$^{1}${\small \it Mathematical Science Team, RIKEN Center for Advanced Intelligence Project (AIP),1-4-1 Nihonbashi, Chuo-ku, Tokyo 103-0027, Japan \protect \\
$^{2}$Interdisciplinary Theoretical \& Mathematical Sciences Program (iTHEMS) RIKEN 2-1, Hirosawa, Wako, Saitama 351-0198, Japan} \\
$^{3}${\small \it Department of Physics, Keio University, Yokohama 223-8522, Japan} 
\end{center}

\setcounter{tocdepth}{1}

\tableofcontents

\section{Outline of area law proof}

\subsection{Set up and assumption}
We here restate the setup of the system.
We consider a one-dimensional quantum system with $n$ sites, where each of the sites has $d$-dimensional Hilbert space. 
We denote the total set of the sites by $\Lambda$, namely $\Lambda=\{1,2,\ldots,n\}$.

In Supplementary Table~\ref{tab:fund_para}, we give a list of parameters which are used throughout the proof.
In~\ref{Sec:List of notations and definitions}, we give a list of definitions and notations which we use several times in the proof.

\subsubsection{Definition of the Hamiltonian} \label{Sec:Definition of the Hamiltonian}
%論文中では簡単のために2体相互作用を仮定しておき、suppleで一般の場合を扱う。

We define the system Hamiltonian $H$ as
\begin{align}
H= \sum_{|Z|\le k} h_Z  
\label{eq:ham_1D}
\end{align}
with $|Z|$ the cardinality of $Z$, where each of $\{h_Z\}_{|Z| \le k}$ denotes an interaction between the sites in $Z \subset \Lambda$.
For example, in the case of $k=2$, the Hamiltonian is given in the form of
\begin{align}
H= \sum_{Z: |Z|=2} h_Z  + \sum_{Z: |Z|=1} h_Z = \sum_{i<j} h_{i,j} + \sum_{i=1}^n h_i  . 
\label{eq:ham_1D_k=2}
\end{align}
This is the Hamiltonian that we considered in the main paper.

The Hamiltonian~\eqref{eq:ham_1D} describes a generic $k$-body-interacting system.
We assume the power-law decaying interaction as
\begin{align}
\max_{i\in \Lambda}\sum_{Z:Z\ni i, \diam (Z)= r} \|h_Z\| \le \frac{J}{r^{\alpha}}  \quad (\alpha>1),    \label{eq:ham_Power_law}
\end{align}
and
\begin{align}
\max_{i\in \Lambda} \|h_{\{i\}}\| \le B,\label{eq:ham_1_local_term}
\end{align}
where $\diam (Z) = \max_{i,j \in Z} (|i-j|)$ and $\|\cdots\|$ is the operator norm.
In the proof, we often use the notation of $\sum_{Z:{\rm condition}}$ which means the summation over all $Z$ satisfying the condition.
Thus, $\sum_{Z:Z\ni i, \diam (Z)= r}$ means the summation which picks up all the subsets $Z\subset \Lambda$ such that  $Z\ni i$ and $\diam (Z)= r$.
From Ineqs.~\eqref{eq:ham_Power_law} and \eqref{eq:ham_1_local_term}, we immediately obtain
\begin{align}
\max_{i\in \Lambda}\sum_{Z:Z\ni i} \|h_Z\| =B+ \max_{i\in \Lambda} \sum_{r=1}^\infty \sum_{Z:Z\ni i, \diam (Z)= r} \|h_Z\| \le B+\frac{\alpha J}{\alpha-1}=:g,       \label{g_exensiveness}
\end{align}
where in the last inequality we use
\begin{align}
\sum_{r=1}^\infty \sum_{Z:Z\ni i, \diam (Z)= r} \|h_Z\| \le J \sum_{r=1}^\infty r^{-\alpha} \le J+ J\int_1^\infty x^{-\alpha}dx=\frac{\alpha J}{\alpha-1} .
\end{align}
We assume a non-degenerate ground state $\ket{\Gs}$ with a spectral gap $\Delta$.
We notice that the spectral gap is always smaller than $2g$ (see below for the derivation): 
\begin{align}
0< \Delta \le 2g. \label{gap_2g_bound}
\end{align}
Throughout the paper, by appropriately choosing the energy unit,  we set $g=1$, or equivalently $ B+\alpha J/(\alpha-1)=1$.

{~}\\

\textit{Proof of the inequality~\eqref{gap_2g_bound}.} \label{sub_sub_sec_gap}
We here would like to prove 
\begin{align}
\Delta \le 2 \max_{i\in \Lambda}\sum_{Z:Z\ni i} \|h_Z\|=2g ,
\end{align}  
where we use the definition of $g$ in \eqref{g_exensiveness}.
It has been already given in Ref.~\cite{Nachtergaele2007}, but we show the proof here.
Let us decompose the Hamiltonian as 
$H=H_{\Lambda_i} + V_i$, where $H_{\Lambda_i}$ acts only on the sites $\Lambda_i:= \Lambda \setminus\{i\}$ and $V_i:=H-H_{\Lambda_i}$.
We note that $\|V_i\|\le g$ from the inequality~\eqref{g_exensiveness}.
we consider a quantum state 
$\ket{\phi}=\ket{\phi_i} \otimes \ket{\Gs_{\Lambda_i}}$
with 
$\ket{\Gs_{\Lambda_i}}$
the ground state of $H_{\Lambda_i}$. 
By choosing 
$\ket{\phi_i}$
such that 
$\langle \Gs | \phi \rangle=0$, 
we have
\begin{align}
&\bra{\phi} H \ket{\phi}- \bra{\Gs} H \ket{\Gs} \ge \Delta .
\label{phi_H_Gs_lower}
\end{align}  
On the other hand, we have 
\begin{align}
\bra{\phi} H \ket{\phi} \le \bra{\Gs_{\Lambda_i}} H_{\Lambda_i} \ket{\Gs_{\Lambda_i}} + \|V_i\| \le \bra{\Gs_{\Lambda_i}}H_{\Lambda_i} \ket{\Gs_{\Lambda_i}} + g
\end{align}  
and 
\begin{align}
\bra{\Gs} H \ket{\Gs} \ge \bra{\Gs_{\Lambda_i}} H_{\Lambda_i} \ket{\Gs_{\Lambda_i}} - \|V_i\| \ge\bra{\Gs_{\Lambda_i}} H_{\Lambda_i} \ket{\Gs_{\Lambda_i}} - g,
\end{align}  
which yields 
\begin{align}
&\bra{\phi} H \ket{\phi}- \bra{\Gs} H \ket{\Gs} \le 2g .
\label{phi_H_Gs_upper}
\end{align}  
By combining the inequalities~\eqref{phi_H_Gs_lower} and \eqref{phi_H_Gs_upper}, we obtain the inequality~\eqref{gap_2g_bound}. $\square$

\subsubsection{Main assumption}

In order to give the condition under which the area law is obtained, we define the interaction operator $V_{X,Y}(\Lambda_0)$ between two subsystems $X\subset \Lambda$ and $Y\subset \Lambda$ as follows  (see Supplementary Figure~\ref{fig:block_block_interaction}):
\begin{align}
V_{X,Y}(\Lambda_0) :=\sum_{\substack{Z: Z\subset \Lambda_0 \\ Z\cap X\neq \emptyset , Z\cap Y\neq \emptyset} }h_Z  .
\label{Def:V_X_Y}
\end{align} 
Here, $V_{X,Y}(\Lambda_0)$ is defined for an arbitrary subset $\Lambda_0\subset \Lambda $ such that $X\sqcup Y \subset \Lambda_0 $.
Note that the operator $V_{X,Y}(\Lambda_0)$ composed of the interaction terms $h_Z$ between $X$ and $Y$ which are supported on $\Lambda_0\subset \Lambda$.
In the case of $k=2$ as in Eq.~\eqref{eq:ham_1D_k=2}, $V_{X,Y}(\Lambda_0)$ does not depend on the choice of $\Lambda_0$ and is simply given by
\begin{align}
V_{X,Y}(\Lambda_0) = V_{X,Y}(X\sqcup Y) = \sum_{i \in X}\sum_{j \in Y}  h_{i,j}, \label{2-local_Hamiltonian_V_X_Y}
\end{align} 
where each of the terms $h_Z$ is given by $h_{i,j}$. 
We utilized the form~\eqref{2-local_Hamiltonian_V_X_Y} in Eq.~(4) of the main paper.
On the other hand, in the case of $k\ge 3$, $V_{X,Y}(\Lambda_0)$ usually depends on choice of the subsystem $\Lambda_0$.

Throughout the paper, we assume the following algebraic decay of $\|V_{X,Y}(\Lambda_0)\|$.
\begin{assump} \label{assump:main_assumption}
Let $X$ and $Y$ be arbitrary concatenated subsystems with $\dist(X,Y)=r$.
Then, for arbitrary choice of $\Lambda_0 \subset \Lambda$, there exists a constant $g_0$ ($\ge 1$) such that
\begin{align}
\| V_{X,Y} (\Lambda_0)\| \le g_0 r^{-\bar{\alpha}}  \label{cond_area_LR_sup}
\end{align} 
with $\bar{\alpha}>0$.  
In~\ref{discuss_assumption_decay}, we will discuss how the parameters $\{g_0,\bar{\alpha}\}$ are given in terms of $\{J,\alpha\}$ in Eq.~\eqref{eq:ham_Power_law}.
\end{assump}
\noindent
The assumption is utilized in deriving the inequalities~\eqref{truncated_Hamiltonian_block_interaction} and \eqref{ineq_H_minus_H_t}.  
The former inequality~\eqref{truncated_Hamiltonian_block_interaction} implies a finite upper bound of the boundary interaction along a cut (see Supplementary Figure~\ref{fig:Area_law_Ham_truncate}  in~\ref{Outline of the proof:Preliminaries}).
The latter inequality~\eqref{ineq_H_minus_H_t} is an essential tool to upper-bound the error in truncating the long-range interaction (see also Supplementary Figure~\ref{fig:Area_law_Ham_truncate}).

In order to discuss the entanglement entropy, we spatially decompose the total space into two subsystems $L$ and $R$ (see Supplementary Figure~\ref{fig:Area_law_Ham_truncate} for example), respectively.
We denote the reduced density matrix of the ground state in $L$ by $\rho_L$:
\begin{align}
\rho_L = \tr_{R}( \ket{\Gs} \bra{\Gs} ),
\end{align}
where $\tr_{R}(\cdots)$ denotes the partial trace operation with respect to the subsystem $R$.
We define the entanglement entropy of this decomposition $\Lambda=L\sqcup R$ as 
\begin{align}
S(L) :=  -\tr (\rho_L \log\rho_L) =  -\tr (\rho_R \log\rho_R) .
\end{align}
Our purpose is to bound the entropy $S(L)$ from above by a function of $d$, $\Delta$ and $\{k, g_0, \bar{\alpha}\}$ (see also Table~\ref{tab:fund_para}).

\begin{table}[b]%The best place to locate the table environment is directly after its first reference in text
  \caption{Fundamental parameters in our statement}
  \label{tab:fund_para} 
\begin{ruledtabular}
\begin{tabular}{lr}
\textrm{\textbf{Parameters}}&\textrm{\textbf{Definition}} 
\\
\colrule
    $d$ 
& Dimension of the Hilbert space of one site\\
    $\Delta$ 
& Spectral gap between the ground state and the first excited state \\
    $k$ 
& Maximum number of sites involved in interactions (see Eq.~\eqref{eq:ham_1D})    \\
    $g_0$ 
& Defined in Assumption~\ref{assump:main_assumption} (see Ineq.~\eqref{cond_area_LR_sup}) \\
    $\bar{\alpha}$
&Defined in Assumption~\ref{assump:main_assumption} (see Ineq.~\eqref{cond_area_LR_sup}) 
\end{tabular}
\end{ruledtabular}
\end{table}

   \begin{figure}
  \begin{center}
    \includegraphics[scale=0.5]{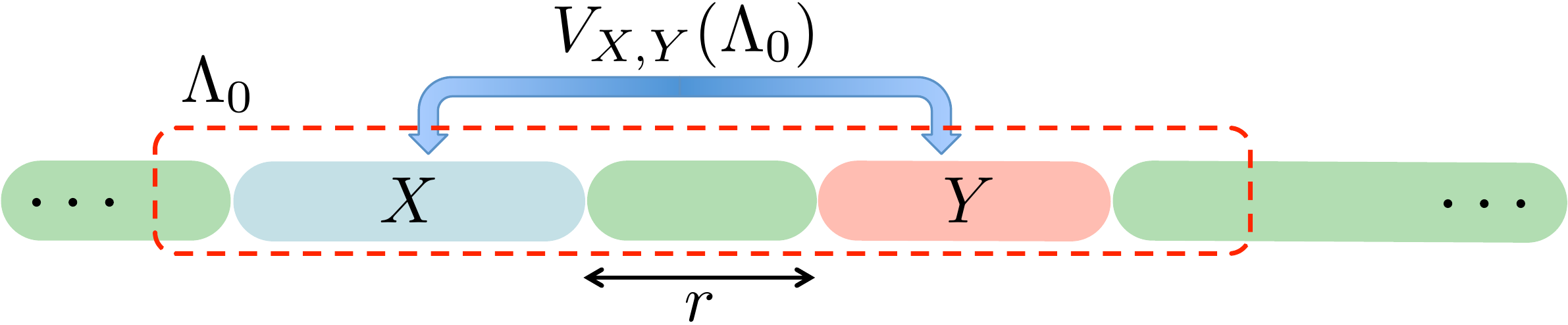}
  \end{center}
  \caption{Subsystem-subsystem interaction. 
  In order to define $V_{X,Y}(\Lambda_0)$, we pick up all the terms $h_Z$ in $\Lambda_0\subset \Lambda$ which connects $X$ and $Y$. 
  Our condition for the area law is that the norm $\|V_{X,Y}(\Lambda_0)\|$ decays algebraically with respect to the distance between $X$ and $Y$.
  }\label{fig:block_block_interaction}
\end{figure}

\subsubsection{Schmidt rank} \label{Subsec:Schmidt rank}
 
We  consider an operator $O$ and define the Schmidt rank ${\rm SR} (O,X)$ for $X\subseteq \Lambda$ as the minimum integer such that
 \begin{align}
O = \sum_{m=1}^{{\rm SR} (O,X)} O_{X,m} \otimes O_{X^\co,m},
\end{align}
where $O_{X,m}$ and $O_{X^\co,m}$ are supported on the subsystems $X$ and $X^\co$ (the complementary set of $X$), respectively.  
We also define the Schmidt rank ${\rm SR} (\ket{\psi},X)$ of a state $\ket{\psi}$ as follows:
 \begin{align}
\ket{\psi} = \sum_{m=1}^{{\rm SR} (\ket{\psi},X)}  \mu_m \ket{\psi_{X,m}} \otimes \ket{\psi_{X^\co,m}}, 
\label{Schmidt_rank_state}
\end{align}
which is the Schmidt decomposition.
Especially in considering ${\rm SR} (O,L)$ (or ${\rm SR} (\ket{\psi},L)$) for the target decomposition $\Lambda=L\sqcup R$, we simply denote ${\rm SR} (O)$ (or ${\rm SR} (\ket{\psi})$)  omitting the subsystem dependence.

\subsection{Main results}

\begin{theorem}[Area law for 1D long-range interacting systems] \label{main_theorem_area_law}
For an arbitrary bipartition of the system $\Lambda=L\sqcup R$. 
The entanglement entropy $S(\ket{\Gs})$ is bounded from above by
\begin{align}
S(\ket{\Gs}) &\le c_0 \log^2 (d) \left(\frac{\log (d)}{\Delta}\right)^{1 +2/\bar{\alpha}}\log^{3+3/\bar{\alpha}}\left(\frac{\log (d)}{\Delta} \right),
\end{align}
where $c_0$ is a constant which depends only on $k$, $g_0$, $\bar{\alpha}$, which has a finite value in the limit of $\bar{\alpha}\to \infty$\footnote{In the main text, for the sake of readability, we do not introduce the quantity~\eqref{g_exensiveness}, and hence
we explain that the coefficient $c$ in Ineq. (6) depends on $\{\alpha,J,B,k,g_0,\bar{\alpha}\}$. However, we here set $g=1$ without loss of generality taking an appropriate unit, and hence the coefficient $c_0$ depends only on $\{k,g_0,\bar{\alpha}\}$. We emphasize that there are no inconsistencies between these.}.
Also, there exists a quantum state $\ket{\psi}$ such that 
 \begin{align}
\|\ket{\Gs} -\ket{\psi} \| \le \delta 
\end{align}
with the Schmidt rank of 
\begin{align}
\log \left[{\rm SR}(\ket{\psi}) \right] =\orderof{\bar{\alpha}^{-1} \log^{5/2} (1/\delta) } + \orderof{\log^{3/2} (1/\delta) }
\end{align}
for sufficiently small $\delta$, where ${\rm SR}(\ket{\psi})$ was defined in \eqref{Schmidt_rank_state}.
\end{theorem}

This theorem implies that in the limit of $\bar{\alpha} \to \infty$, the entanglement entropy is given by
\begin{align}
S(\ket{\Gs}) &\lesssim c_0 \frac{\log^3 (d)}{\Delta}  \for \bar{\alpha} \to \infty
\end{align}
up to a logarithmic correction.
This reproduces the results by Arad-Kitaev-Landau-Vazirani for short-range interacting systems~\cite{arad2013area}. 

By applying Lemma~1 in Ref.~\cite{VC06mps} to the above theorem, we immediately obtain the efficiency of the MPS representation of the ground state $\ket{\Gs}$ (see Method section in the main text for the proof).
\begin{corol}
Let us assume $\bar{\alpha}=\orderof{1}$. 
Then, under the same set up of Theorem~\ref{main_theorem_area_law}, 
there exists a matrix product state $\ket{\psi_D}$ with its bond dimension $D=\exp[c' \bar{\alpha}^{-1} \log^{5/2}(1/\delta)]$ ($c'$: constant) such that 
\begin{align}
\| \tr_{X^\co} (\ket{\psi_D}\bra{\psi_D} ) - \tr_{X^\co} (\ket{\Gs}\bra{\Gs} )\|_1 \le \delta  |X| \label{MPS_approx_sup}
\end{align}
for an arbitrary concatenated subregion $X$, where $\|\cdot\|_1$ is the trace norm and $|X|$ denotes the cardinality of $X$. 
We here denote the complementary subset of $X$ by $X^\co := \Lambda\setminus X$.
\end{corol}
From the corollary, in order to approximate the ground state $\ket{\Gs}$ by using the matrix product states with $\delta=1/\poly(n)$, we need the bond dimension $D_{\rm MPS}$ of order 
\begin{align}
D_{\rm MPS} = \exp[c\bar{\alpha}^{-1} \log^{5/2} (n)] = n^{c \bar{\alpha}^{-1}\log^{3/2} (n)}.
\end{align}
Hence, the simulation of the gapped ground states requires quasi-polynomial computational time. 
This contrasts to the short-range interacting cases, where the sufficient bond-dimension for $\delta=1/\poly(n)$ is sub-linear~\cite{arad2013area} as 
\begin{align}
D'_{\rm MPS} = \exp[c' \log^{3/4} (n)]  . 
\end{align}

\subsection{Specific values of $\bar{\alpha}$ and $g_0$}  \label{discuss_assumption_decay}

We first derive the upper bound for $\|V_{X,Y}(\Lambda_0)\|$ only from the inequality~\eqref{eq:ham_Power_law}. 
For this purpose, we prove the following lemma (see~\ref{proof:lem:block_block_interaction}):
\begin{lemma} \label{lem:block_block_interaction}
Let $X\subset \Lambda$ be an arbitrary concatenated subsystem and $Y\subset \Lambda$ be the subsystem such that $\dist(X,Y)=r$ (see Supplementary Figure~\ref{fig:block_block_interaction}).
Then, the norm of $V_{X,Y}(\Lambda_0)$ is bounded from above by
 \begin{align}
\|V_{X,Y}(\Lambda_0)\| \le \frac{J \alpha}{\alpha-2}  r^{-\alpha+2}.  \label{definition_of_H_X_l0}
\end{align}
Hence, we have  
 \begin{align}
g_0=\frac{\alpha J}{\alpha-2}, \quad \bar{\alpha} = \alpha - 2 ,
\end{align}
in the inequality~\eqref{cond_area_LR_sup}. 
\end{lemma}
From the lemma, the assumption~\ref{assump:main_assumption} is always satisfied for $\alpha>2$.
This lower bound of $\alpha$ is the most general one and applied to arbitrary quantum many-body systems.
On the other hand, the condition $\alpha>2$ can be relaxed if we consider a specific class of Hamiltonians. 
For example, we here consider a fermion system with  long-range hopping as follows:
\begin{align}
H= \sum_{i<j} \frac{1}{r_{i,j}^\alpha} ( A_{i,j} a_i a_j^\dagger +B_{i,j}a_i a_j + {\rm h.c.} )   +V \with |A_{i,j}| ,\ |B_{i,j}|  \le \tilde{J} ,
\label{eq:ham_1D_free_sup}
\end{align}
where $r_{i,j}:=\dist(i,j)$ and $\{a_i^\dagger, a_i\}_{i=1}^n$ are the creation and the annihilation operators for fermion, 
%(or boson), 
and $V$ is arbitrary short-range interacting terms such as $a_ia_i^\dagger a_ja_j^\dagger $ with $r_{i,j}\le \orderof{1}$.
In this case, we can prove the following lemma:
\begin{lemma} \label{lem:block_block_interaction_ferm}
Let $X\subseteq \Lambda$ be an arbitrary concatenated subsystem and $Y\subseteq \Lambda$ be the subsystem such that $\dist(X,Y)=r$.
We assume that the distance $r$ is larger than the short-range interaction length which is given by $V$.
Then, the norm of $V_{X,Y}(\Lambda_0)$ is bounded from above by
 \begin{align}
\|V_{X,Y}(\Lambda_0)\|  \le 4\tilde{J}  \sqrt{\frac{2\alpha}{2\alpha-1}} \frac{2\alpha-1}{2\alpha-3} r^{-\alpha+3/2},  \label{upper_V_X_Y_ferm}
\end{align}
Hence, we have  
 \begin{align}
g_0=4 \tilde{J}  \sqrt{\frac{2\alpha}{2\alpha-1}} \frac{2\alpha-1}{2\alpha-3}  ,\quad \bar{\alpha} =\alpha-3/2 ,
\end{align}
in the inequality~\eqref{cond_area_LR_sup}.
\end{lemma}

From the lemma,  the assumption~\ref{assump:main_assumption} is satisfied for $\alpha > 3/2$ (instead of $\alpha>2$).
In this way, depending on the situation, the condition for the power exponent $\alpha$ can be loosen. 
Other cases which gives a better condition than $\alpha>2$ include quantum many-body systems with random long-range interactions~\cite{PhysRevLett.120.200601}.

\subsubsection{Proof of Lemma~\ref{lem:block_block_interaction}} \label{proof:lem:block_block_interaction}
For the proof, we estimate the upper bound of
\begin{align}
\overline{V}_{X,Y} := \sum_{Z:Z\cap X\neq \emptyset, Z\cap Y \neq \emptyset} \| h_Z\| , \label{upper_bound_overline_H_X_l0}
\end{align}
which clearly gives an upper bound of $\|V_{X,Y}(\Lambda_0)\|$ for arbitrary choices of $\Lambda_0\subset \Lambda$.
From the inequality~\eqref{eq:ham_Power_law}, we first obtain for an arbitrary integer $\tilde{r}$
\begin{align}
\sum_{Z:Z\ni i, \diam (Z)\ge \tilde{r}} \|h_Z\| \le \sum_{x=\tilde{r}}^\infty J x^{-\alpha}
\le J \tilde{r}^{-\alpha}+J \int_{\tilde{r}}^\infty x^{-\alpha} dx \le \frac{\alpha J}{\alpha-1} \tilde{r}^{-\alpha+1} ,
\label{eq:ham_Power_law_2_lem}
\end{align}
where we use $\tilde{r}^{-\alpha} \le \tilde{r}^{-\alpha+1}$ in the last inequality.

Second, in order to estimate the upper bound of $\overline{V}_{X,Y}$, we define $Y=\{i_0+1,i_0+2,\ldots, i_0+n_Y \}$ with $n_Y=|Y|$.
Without loss of generality, we assume that the subsystem $Y$ locates on the right side of $X$.  
Then, because of $\dist(i_0+j,X) = r-1+j$ we have 
\begin{align}
\sum_{Z:Z\cap X\neq \emptyset, Z\cap Y \neq \emptyset} \| h_Z\| & \le 
\sum_{j=1}^{n_Y} \sum_{\substack{Z:Z\ni i_0+j\\ \diam (Z)\ge r+j-1}}  \|h_Z\| 
\le \sum_{j=1}^{n_Y} \frac{\alpha J}{\alpha-1} \frac{1}{(r+j-1)^{\alpha-1}}   \notag \\
&\le  \frac{\alpha J}{\alpha-1} \left(r^{-\alpha+1} + \int_{r}^\infty x^{-\alpha+1} dx \right)
\le  \frac{\alpha J}{\alpha-2}  r^{-\alpha+2} , 
 \label{Interaction_between_X_Y_L_1}
%\le \int_l^\infty x^{-\alpha} dx =  \frac{l^{-\alpha+1}}{\alpha-1},
 \end{align}
where we use the inequality~\eqref{eq:ham_Power_law_2_lem} in the second inequality.  This completes the proof. $\square$

\subsubsection{Proof of Lemma~\ref{lem:block_block_interaction_ferm}} \label{proof:lem:block_block_interaction_ferm}
Without loss of generality, we assume that the subsystem $Y$ locates on the right side of $X$.  
Then, we notice that $V_{X,Y}(\Lambda_0)$ is given by
\begin{align}
V_{X,Y}(\Lambda_0) &=V_{X,Y}(X\sqcup Y)=  \sum_{i \in X}\sum_{j \in Y}  \frac{1}{r_{i,j}^\alpha} ( A_{i,j} a_i a_j^\dagger +B_{i,j}a_i a_j + {\rm h.c.} )   \notag \\
&=\sum_{i \in X}  a_i  \sum_{j \in Y}  \frac{A_{i,j} a_j^\dagger + B_{i,j} a_j}{r_{i,j}^\alpha}  + {\rm h.c.}  ,
\end{align} 
which gives the upper bound of $\|V_{X,Y}(\Lambda_0) \|$ as 
\begin{align}
\|V_{X,Y}(\Lambda_0) \| \le 2\sum_{i \in X} \left\|  \sum_{j \in Y}  \frac{A_{i,j} a_j^\dagger }{r_{i,j}^\alpha} \right\|  + 2\sum_{i \in X} \left\|  \sum_{j \in Y}  \frac{ B_{i,j} a_j}{r_{i,j}^\alpha} \right\| . \label{upper_bound_V_X_Y}
\end{align} 
By using the condition $\|A_{i,j}\| \le \tilde{J}$ in~\eqref{eq:ham_1D_free_sup}, the first term is bounded from above as 
\begin{align}
\left\|  \sum_{j \in Y}  \frac{A_{i,j} a_j^\dagger }{r_{i,j}^\alpha} \right\|  \le  \left(\sum_{j \in Y}  \frac{|A_{i,j}|^2}{r_{i,j}^{2\alpha}}  \right)^{1/2} 
\le \tilde{J}  \left(\sum_{x=1}^\infty (r_i+x-1)^{-2\alpha}  \right)^{1/2} \le \tilde{J} \sqrt{\frac{2\alpha}{2\alpha-1}} r_i^{-\alpha+1/2},
\end{align} 
where we define $r_i= \dist(i,Y)$ and utilize the inequality
\begin{align}
\sum_{x=1}^\infty (r_i+x-1)^{-2\alpha} \le  r_i^{-2\alpha} +  \int_{r_i}^\infty   x^{-2\alpha} dx =  r_i^{-2\alpha}  + \frac{r_i^{-2\alpha+1}}{2\alpha-1} \le \frac{2\alpha}{2\alpha-1} r_i^{-2\alpha+1}.
\end{align} 
The summation with respect to $i\in X$ reduces to the summation from $r_i=r$ to $r_i=r+|X|-1$.
Hence, we obtain 
\begin{align}
\sum_{i \in X} \left\|  \sum_{j \in Y}  \frac{A_{i,j} a_j^\dagger }{r_{i,j}^\alpha} \right\| \le \sum_{r_i=r}^\infty \tilde{J}  \sqrt{\frac{2\alpha}{2\alpha-1}} r_i^{-\alpha+1/2}  
\le \tilde{J}  \sqrt{\frac{2\alpha}{2\alpha-1}} \frac{2\alpha-1}{2\alpha-3} r^{-\alpha+3/2}.
\end{align} 
We can derive the same inequality for the summation of $B_{i,j} a_j/r_{i,j}^\alpha$ with respect to $i\in X$ and $j\in Y$. 
By applying the above inequality to \eqref{upper_bound_V_X_Y}, we prove the inequality~\eqref{upper_V_X_Y_ferm}. $\square$

\subsection{Outline of the proof}

\subsubsection{Preliminaries} \label{Outline of the proof:Preliminaries}

%%%%%%%%%%%%%%%%%%%%%%%%%%%%%%%%%%%%%%%%%%%%%%%%%%%%%%%%%%%%%%%%%%%%%%%%%%%%%%%%%%%%%%%%%%%%%%%%%%
%%%%%%%%%%%%%%%%%%%%%%%%%%%%%%%%%%%%%%%%%%%%%%%%%%%%%%%%%%%%%%%%%%%%%%%%%%%%%%%%%%%%%%%%%%%%%%%%%%
{\bf[Approximate ground state projection (AGSP)]}

{~}\\
We here introduce the projection operator onto the ground state.
It is usually difficult to construct the exact ground-state projection operator, and hence we consider an approximate one as
 \begin{align}
K\ket{\Gs}  \simeq \ket{\Gs}    \quad {\rm and} \quad   \| K (1 - \ket{\Gs}\bra{\Gs}) \| \simeq 0,
\label{AGSP:formal_def}
\end{align}
where $(1 - \ket{\Gs}\bra{\Gs}) $ is equivalent to the projection operator onto the space of the excited eigenstates.
We assume that $K$ is a Hermitian operator (i.e., $K=K^\dagger$).
In the following, we characterize the approximate ground state projection (AGSP) operators by three parameters $\{\delta_K, \epsilon_K, D_K \}$.
Let $\ket{\Gs_K}$ be a quantum state that is invariant by $K$ such that
 \begin{align}
 K \ket{\Gs_K} = \ket{\Gs_K} \label{def_of_Gs_K}.
\end{align}
Then, the parameters are defined by the following inequalities:
 \begin{align}
\|\ket{\Gs}  -\ket{\Gs_K} \|  \le \delta_K    , \quad  \| K (1- \ket{\Gs_K}\bra{\Gs_K})  \|\le   \epsilon_K, \quad {\rm and} \quad  {\rm SR}(K) \le D_K . \label{Def_AGSP_error_chap2}
\end{align}
The second inequality implies for arbitrary $\ket{\psi_\bot}$ which is orthogonal to $\ket{\Gs_K}$ (i.e., $\langle \psi_\bot \ket{\Gs_K}=0$)  
 \begin{align}
\| K \ket{\psi_\bot} \|= \| K (1- \ket{\Gs_K}\bra{\Gs_K}) \ket{\psi_\bot} \| \le   \epsilon_K. \label{Def_AGSP_error_second}
\end{align}
Recall that in~\ref{Subsec:Schmidt rank} we denote ${\rm SR} (O,L)$ by ${\rm SR} (O)$ for the simplicity.

Note that the state $\ket{\Gs_K} $ is an approximate ground state if $\delta_K \simeq0$.
When $\delta_K=\epsilon_K=0$, the operator $K$ is the exact ground state projection, namely $K=\ket{\Gs}\bra{\Gs}$. 
In the standard definition of the AGSP~\cite{arad2013area,PhysRevB.85.195145,Kuwahara_2017}, we do not need to consider the parameter $\delta_K$ explicitly.
However, in the present case of the long-range interacting systems, the error of $\|\ket{\Gs}  -\ket{\Gs_K} \|$ is too large to ignore and we have to correctly take the effect of $\delta_K$ into account.

%%%%%%%%%%%%%%%%%%%%%%%%%%%%%%%%%%%%%%%%%%%%%%%%%%%%%%%%%%%%%%%%%%%%%%%%%%%%%%%%%%%%%%%%%%%%%%%%%%
{~}

{~}

{~}\\
%%%%%%%%%%%%%%%%%%%%%%%%%%%%%%%%%%%%%%%%%%%%%%%%%%%%%%%%%%%%%%%%%%%%%%%%%%%%%%%%%%%%%%%%%%%%%%%%%%
%%%%%%%%%%%%%%%%%%%%%%%%%%%%%%%%%%%%%%%%%%%%%%%%%%%%%%%%%%%%%%%%%%%%%%%%%%%%%%%%%%%%%%%%%%%%%%%%%%
{\bf[Interaction-truncated Hamiltonian]}

   \begin{figure}
\centering
   \includegraphics[scale=0.5]{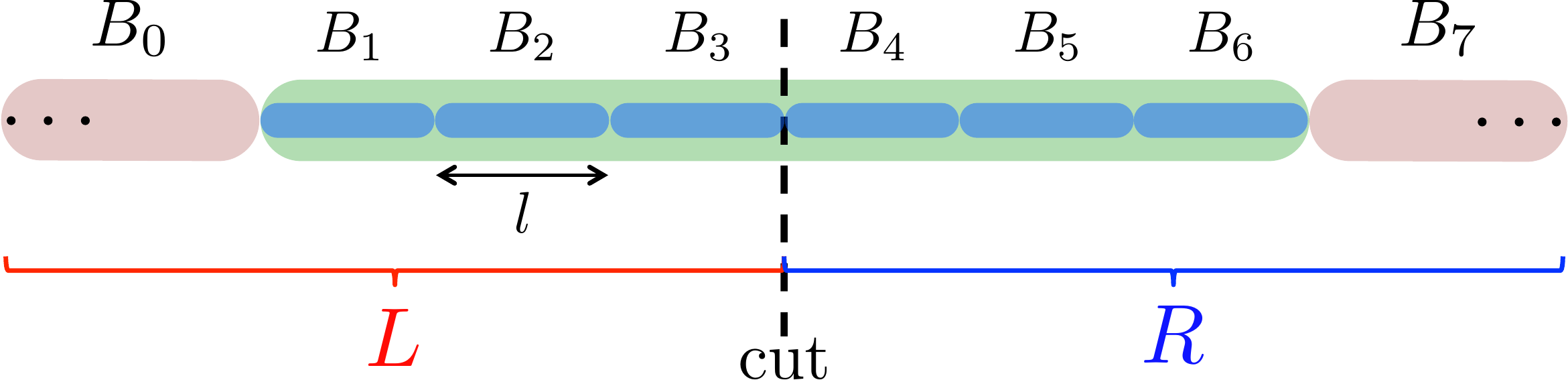}
  \caption{Interaction truncation in the Hamiltonian. Across the dashed line indicated at the center of the system, we decompose the system into $(q+2)$-blocks ($q=6$ in the above picture).
  Each of the blocks $\{B_s\}_{s=1}^q$ has a length $l$, and the edge blocks $B_0$ and $B_{q+1}$ extend to the left and right ends of the system, respectively.
  Then, we truncate all the interactions between separated blocks. Because we only truncate the long-range interactions around the cut, the truncated Hamiltonian $H_\tc$ in Eq.~\eqref{def:truncated_Hamiltonian} is still close to the original Hamiltonian $H$ as shown in Lemma~\ref{thm:locality_exp_effectiveHam}. 
     }\label{fig:Area_law_Ham_truncate}
\end{figure}

{~}\\
We first decompose the total system into $B_0$, $\{B_s\}_{s=1}^q$ and $B_{q+1}$ with $\bigcup_{s=0}^{q+1} B_s=\Lambda$, where $q$ is an even integer ($q \ge 2$) and we 
choose $B_s$ ($1\le s \le q$) such that $|B_s|=l$.  
Note that we express the subsets $L$ and $R$ in terms of these blocks:
\begin{align}
L=\bigcup_{s=0}^{q/2} B_s,\quad R=\bigcup_{s=q/2+1}^{q+1} B_s. \label{Block_definition_LR}
\end{align}
We now truncate all the interactions between the non-adjacent blocks. 
After the truncation, only the interactions between the adjacent blocks exist, namely
\begin{align}
H_\tc =\sum_{s=0}^{q+1} h_{s} + \sum_{s=0}^q h_{s,s+1} , \label{def:truncated_Hamiltonian}
\end{align}
where $h_{s,s+1}:=V_{B_s,B_{s+1}}(B_s\sqcup B_{s+1})$ by choosing $X=B_s$, $Y=B_{s+1}$ and $\Lambda_0=B_s\sqcup B_{s+1}$ in the definition of $V_{X,Y}(\Lambda_0)$ in Eq.~\eqref{Def:V_X_Y}, and $h_s$ collects all the terms supported only on $B_s$.
We notice that the assumption~\ref{assump:main_assumption} with $r=1$ gives 
\begin{align}
\| h_{s,s+1}\| \le g_0  . \label{truncated_Hamiltonian_block_interaction}
\end{align}
In the following, we describe $H_\tc$ as
\begin{align}
& H_\tc = \sum_{s=0}^{q} H_s ,  \with H_0=h_{s=0}+ h_{s=1} + h_{s=0,s=1},\quad H_s=h_{s+1} + h_{s,s+1} \ (s\ge 1)  . \label{truncated_Hamiltonia_simple_form}
\end{align}
We denote by $\ket{\Gs_\tc}$ the ground state of the truncated Hamiltonian~$H_\tc$.
Throughout the paper, we take the origin of the energy so that $E_{\tc,0}=0$, where $E_{\tc,0}$ is the ground-state energy of $H_\tc$.
Note that we set the origin of the energy through not the original Hamiltonian $H$, but the truncated Hamiltonian $H_\tc$.
%; note that we originally take it as $E_{\tc,0}=0$. 

%%%%%%%%%%%%%%%%%%%%%%%%%%%%%%%%%%%%%%%%%%%%%%%%%%%%%%%%%%%%%%%%%%%%%%%%%%%%%%%%%%%%%%%%%%%%%%%%%%
{~}

{~}

{~}\\
%%%%%%%%%%%%%%%%%%%%%%%%%%%%%%%%%%%%%%%%%%%%%%%%%%%%%%%%%%%%%%%%%%%%%%%%%%%%%%%%%%%%%%%%%%%%%%%%%%
%%%%%%%%%%%%%%%%%%%%%%%%%%%%%%%%%%%%%%%%%%%%%%%%%%%%%%%%%%%%%%%%%%%%%%%%%%%%%%%%%%%%%%%%%%%%%%%%%%
{\bf[Effective Hamiltonian by multi-energy cut-off]}

   \begin{figure}
  \begin{center}
    \includegraphics[scale=0.4]{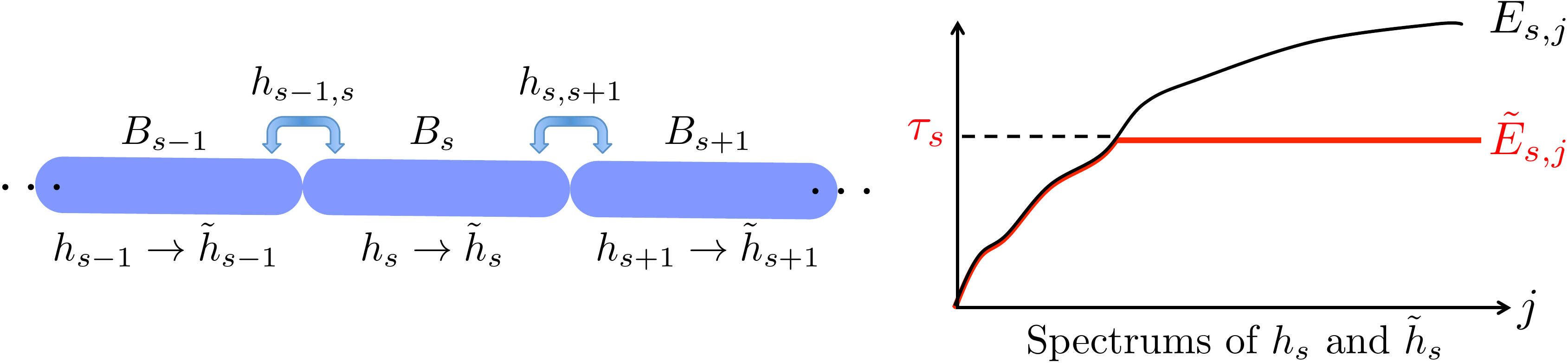}
  \end{center}
  \caption{Schematic picture of our effective Hamiltonian. In the effective Hamiltonian~$\tilde{H}_\tc$, we modify the energy spectrum so that the energy higher than 
  $\tau_s$ is constant in each of the Hamiltonians $\{h_s\}_{s=0}^{q+1}$, whereas the other part of the Hamiltonian (i.e., $\{h_{s,s+1}\}_{s=0}^q$) is the same as the original Hamiltonian. 
  As long as we focus on the low-energy spectrum, the effective Hamiltonian looks almost the same as the original Hamiltonian. 
It will be shown in Theorem~\ref{Effective Hamiltonian_multi_truncation} that the accuracy exponentially approaches with the cut-off energy $\tau$. 
   }\label{fig:effective_Ham}
\end{figure}

{~}\\
In the construction of the AGSP operator~\eqref{AGSP:formal_def}, we need an effective Hamiltonian $\tilde{H}_\tc$ which has a small norm but possesses almost the same low-energy properties as the original Hamiltonian $H_\tc$.
For the construction of such an effective Hamiltonian, we apply the energy cut-off in Ref.~\cite{Arad_2016} to the Hamiltonian $H_\tc$ in Eq.~\eqref{def:truncated_Hamiltonian}.
For each of the block Hamiltonian $\{h_{s}\}_{s=0}^{q+1}$, we apply the following energy cut-off (Supplementary Figure~\ref{fig:effective_Ham}):
 \begin{align}
\tilde{h}_{s} %&:= \int_{-\infty}^\tau h_{s} \delta (h_{s}-x) dx + \int_{\tau}^\infty \tau \delta (h_{s}-x) dx  \notag \\
 &=\sum_{E_{s,j} < \tau_s} E_{s,j} \ket{E_{s,j}}\bra{E_{s,j}} + \sum_{E_{s,j} \ge \tau_s} \tau_s\ket{E_{s,j}}\bra{E_{s,j}}  
\label{truncation_effective_Hamiltonian_h_s}
\end{align}
with
 \begin{align}
\tau_s = E_{s,0} + \tau, \label{tau_s_tau}
\end{align}
where $\{ E_{s,j}, \ket{E_{s,j}}\}_j$ are the eigenvalues and the eigenstates of $h_{s}$, respectively.
Then, the effective Hamiltonian $\tilde{H}_\tc$ is given by
 \begin{align}
\tilde{H}_\tc =\sum_{s=0}^{q+1} \tilde{h}_{s} + \sum_{s=0}^q h_{s,s+1} . \label{truncation_effective_Hamiltonian_H_tc}
\end{align}
Note that we do not make any changes for the interaction terms $\{h_{s,s+1}\}_{s=0}^q$.
We notice that Propositions~\ref{thm:Schmit_rank_Ham_power} and \ref{thm:Schmit_rank_Ham_power_0} on the Schmidt rank 
are applicable to the effective Hamiltonian $\tilde{H}_\tc$.
We denote by $\ket{\tilde{\Gs}_\tc}$ the ground state of the effective Hamiltonian~$\tilde{H}_\tc$.

\subsubsection{Brief outline}

   \begin{figure}
  \begin{center}
    \includegraphics[scale=0.6]{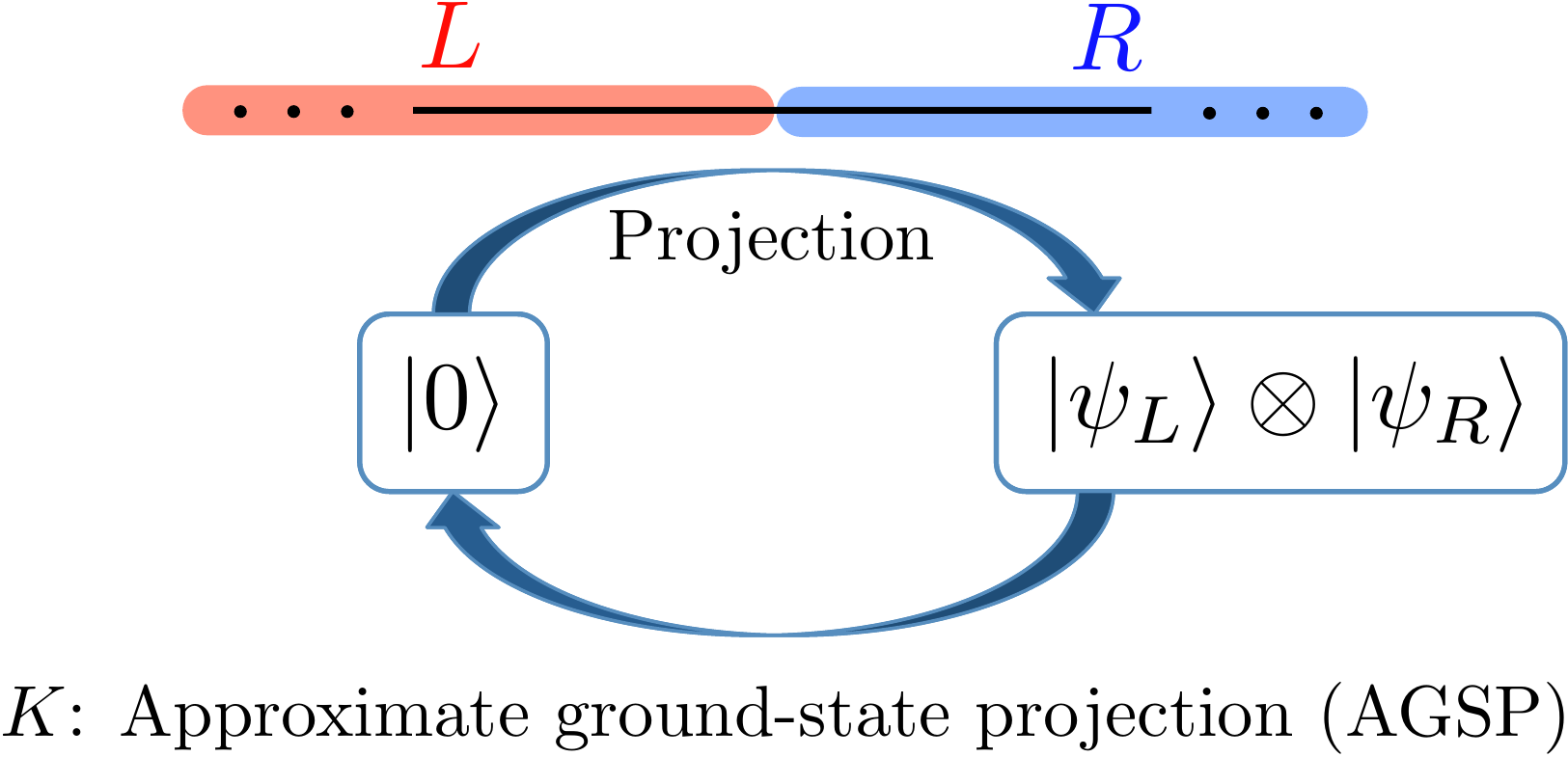}
  \end{center}
  \caption{Schematic picture of the proof of the one-dimensional area law. We decompose the total system $\Lambda$ into the two subsystems $L$ and $R$.
In the proof of the area law, we first perform a projection onto a product state which has the maximum overlap with the ground state $\ket{\Gs}$.
We then recover the original state from this product state by the use of the approximate ground state projector $K$ (AGSP) as in Eq.~\eqref{AGSP:formal_def}.
The entanglement entropy can be bounded from above by the entanglement generation of AGSP (the Schmidt rank of $K$) because the product state contains no entanglement. 
  }\label{fig:Area_law_outline}
\end{figure}

   \begin{figure}
  \begin{center}
    \includegraphics[scale=0.48]{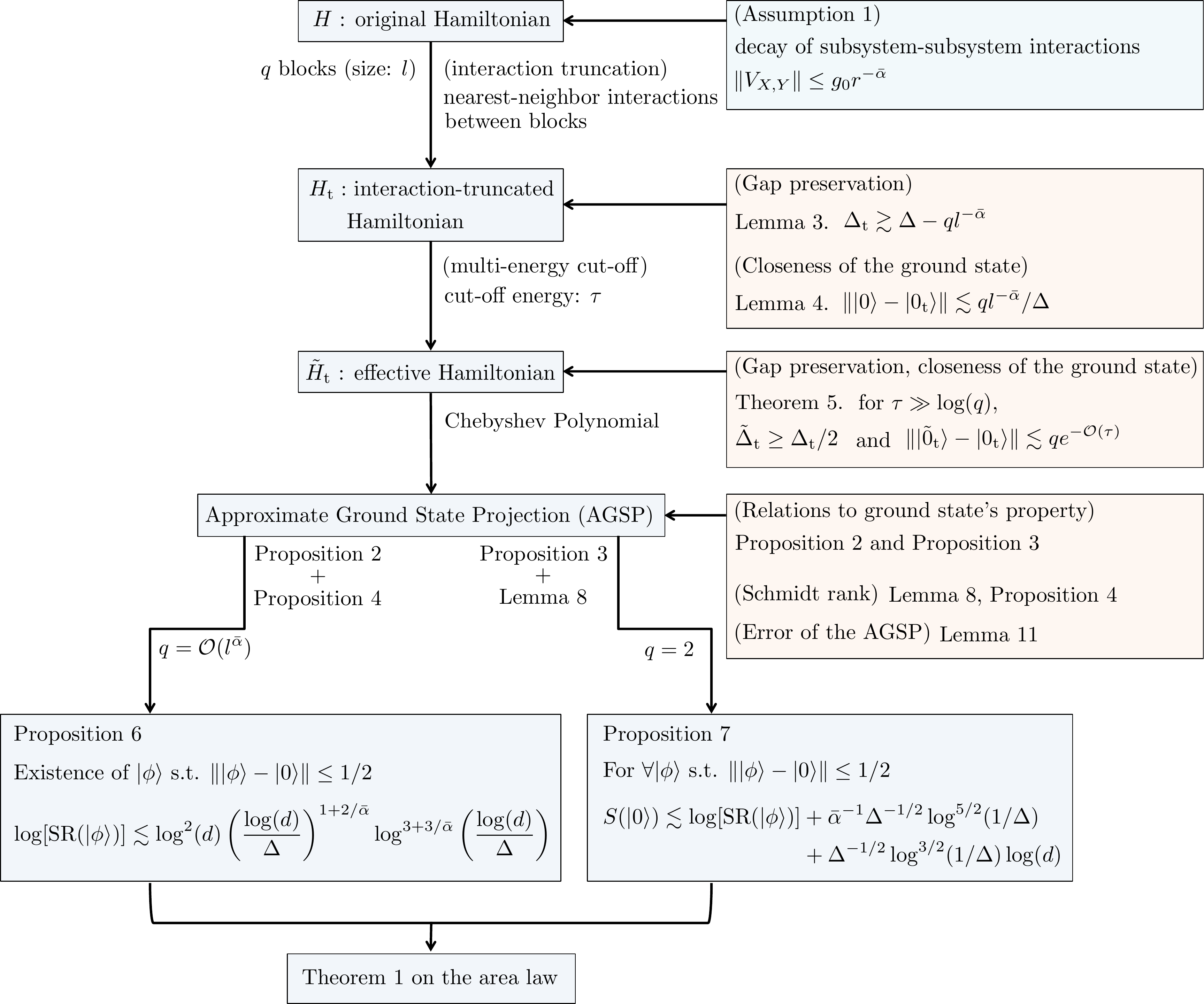}
  \end{center}
  \caption{Flow chart of the proof. 
  }\label{fig:Outline_proof}
\end{figure}

We here show the high-level overview of the area-law proof in one-dimensional long-range interacting systems (see~Supplementary Figure~\ref{fig:Area_law_outline}).
We have also shown it in Method section in the main text. 

We first completely break the entanglement entropy of the ground state by performing a projection operator onto a product state with respect to the partition $L\sqcup R$. Note that the entanglement entropy is equal to zero for product states.
We second consider a reverse operator $K$ from the product state to the ground state $\ket{\Gs}$.
The operator $K$ is now taken as an approximate ground state projector (AGSP) as in Eq.~\eqref{AGSP:formal_def}.

We need to consider the following problems: how small is the overlap between the ground state and the product state? 
On this problem, we can utilize the bootstrapping lemma~\cite{arad2013area,Hastings_2007} (i.e., Lemma~\ref{lem:Bootstrapping_lemma_} in our manuscript); that is,
under a good choice of the AGSP operator, an overlap between the ground state and a product state is lower-bounded by using the AGSP parameters $\epsilon_K$ and $D_K$. 
Roughly speaking, we need to find an AGSP operator which satisfies $\epsilon_K^2  D_K \le 1/2$.

%。
The primary problem is how to construct the AGSP operators with appropriate properties to apply the basic strategy. 
For the purpose, we first perform the truncation of long-range interactions in order to suppress the Schmidt rank $D_K$.
If we simply truncate all the long-range interactions in the entire region, the truncated Hamiltonian $H_\tc$ and the original Hamiltonian $H$ is extensively different, namely $\|H-H_\tc\|\approx \orderof{n}$. 
This may completely change the ground state's property.
To avoid it, we truncate the long-range interaction only around the cut between $L$ and $R$ (see Supplementary Figure~\ref{fig:Area_law_Ham_truncate}).
This truncation ensures the small norm distance between the original Hamiltonian and the truncated Hamiltonian, which preserves the gap condition of $H_\tc$ (Lemma~\ref{thm:locality_exp_effectiveHam}) and ensures the closeness between both of the ground states (Lemma~\ref{overlap_Lemma_truncate_original}).
This imposes the following condition for the block size $l$ and the block number $q$:
 \begin{align}
ql^{-\bar{\alpha}} \lesssim 1. \label{cond_block_size_numb}
\end{align}

In the construction of the AGSP operator from the Chebyshev polynomial, the projection error $\epsilon_K$ strongly depends on the norm of the Hamiltonian (see the inequality~\eqref{error_k_m_poly_AGSP}).
Hence, in the second step, we perform the multi-energy cut-off in each of the blocks as in Supplementary Figure~\ref{fig:effective_Ham} to define an effective Hamiltonian $\tilde{H}_\tc$.
Roughly speaking, the norm of the effective Hamiltonian in Eq.~\eqref{truncation_effective_Hamiltonian_H_tc} is given by $\orderof{q\tau}$.
The gap preservation and the closeness of the ground state is still ensured as long as the cut-off energy satisfies $\tau \gg \log (q)$ (Theorem~\ref{Effective Hamiltonian_multi_truncation}).
In the standard construction of the effective Hamiltonian~\cite{arad2013area,Arad_2016}, we suppose to perform the energy cut-off only in the edge blocks (i.e., $B_0$ and $B_{q+1}$). However, this simple procedure allows us to prove the long-range area law only in the short-range power-exponent regimes (i.e., $\alpha>3$).
The multi-energy cut-off is crucial to prove the area law even in the long-range power-exponent regimes (i.e., $\alpha\le 3$).

We then need to derive basic properties of the AGSP so that they meet our present setup and purposes.
In Proposition~\ref{prop:Overlap between the ground state and  low-entangled state}, we lower-bound the overlap between the ground state and low-entangled state.
We then derive the upper bound of the entanglement entropy by using a sequence of the AGSP operators (Proposition~\ref{prop:entropy_and_AGSP}).
As for the connection between the AGSP parameters $\{D_K,\epsilon_K,\delta_K\}$ and polynomials of the effective Hamiltonian $\tilde{H}_\tc$, we derive 
Lemma~\ref{thm:Schmit_rank_Ham_power_0} and Proposition~\ref{thm:Schmit_rank_Ham_power} for the Schmidt rank, and give Lemma~\ref{ChebyShev_box} to 
upper-bound the projection error $\epsilon_K$ in terms of the norm of the effective Hamiltonian $\|\tilde{H}_\tc\|$. 
Then, by using the $m$th order Chebyshev polynomial, the Schmidt rank $D_K$ and the error of the AGSP $\epsilon_K$ is roughly given by
 \begin{align}
D_K\sim e^{m \log (q)/q + q^{1+1/\bar{\alpha}}},\quad \epsilon_K \sim e^{-m/\sqrt{q \log q}}
\end{align}
under the condition~\eqref{cond_block_size_numb} (see the inequality~\eqref{epsilon_K_D_K_formal_form}), 
where we use Proposition~\ref{thm:Schmit_rank_Ham_power} in estimating the Schmidt rank.
These estimations for $D_K$ and $\epsilon_K$ ensure the existence of $\{q,m\}$ which satisfies $\epsilon_K^2  D_K \le 1/2$ for the condition of the bootstrapping lemma 
(Proposition~\ref{prop:Overlap between the ground state and  low-entangled state}).
We then prove the existence of a quantum state which is close to the original ground state with an error smaller than $1/2$ and has a small Schmidt rank (Proposition~\ref{prop1:truncate_gs_overlap}).

Finally, based on Proposition~\ref{prop:entropy_and_AGSP} and Lemma~\ref{thm:Schmit_rank_Ham_power_0}, we construct the complete ground state $\ket{\Gs}$ to upper-bound the entanglement entropy (Proposition~\ref{prop0:overlap_AGSP_entropy_bound}).
This yields our main Theorem~\ref{main_theorem_area_law}.
In Supplementary Figure~\ref{fig:Outline_proof}, we give the flow chart of the whole proof.

\section{Details of technical lemmas, propositions and sub-theorems}

Before giving the proof of Theorem~\ref{main_theorem_area_law}, we show technical lemmas, propositions and sub-theorems, which are the key ingredients to prove the theorem.
Several lemmas can be trivially derived from the previous analyses in Refs~\cite{arad2013area,Hastings_2007,PhysRevB.85.195145,Kuwahara_2017,Arad_2016} by extending their setups to the present setup.
We show the details of almost all the lemmas, propositions, and sub-theorems so that all the readers can follow the proofs.

In the following analyses, we assume the open-boundary condition.
In the periodic-boundary condition, we can also define the truncated Hamiltonian in the same way (see Supplementary Figure~\ref{fig:Area_law_Ham_truncate_periodic}) by regarding 
the system as a one-dimensional ladder.

\subsection{Gap condition for the truncated Hamiltonian $H_\tc$ }

First of all, we analyze the ground state of the truncated Hamiltonian $H_\tc$.
For the purpose, we need to clarify the gap condition for $H_\tc$. 
It is ensured by the following lemma which gives the norm difference between $H$ and $H_\tc$:
\begin{lemma} \label{thm:locality_exp_effectiveHam}
The norm distance between $H$ and $H_\tc$ is bounded from above by
 \begin{align}
\|\delta H_\tc\|\le  g_0 q l^{-\bar{\alpha}}  , \label{lemma:truncate__ineq}
\end{align}
where we define $\delta H_\tc := H-H_\tc$.
Also, the spectral gap $\Delta_\tc$ of $H_\tc$  is bounded from below by
 \begin{align}
\Delta_\tc \ge \Delta - 2g_0 q l^{-\bar{\alpha}}  . \label{lemma:truncate_ineq_gap}
\end{align}
\end{lemma}

   \begin{figure}
\centering
    \includegraphics[scale=0.5]{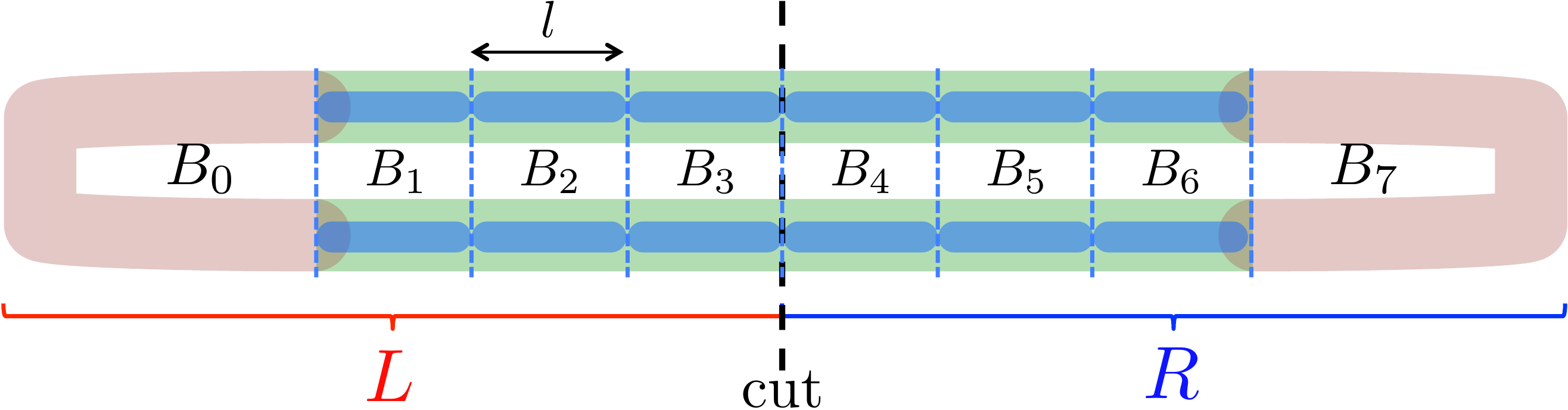}
  \caption{Interaction truncation in the Hamiltonian in the periodic-boundary condition. 
  In this case, we can apply the same discussion as in the case of the open-boundary condition by regarding the system as a one-dimensional ladder.
     }\label{fig:Area_law_Ham_truncate_periodic}
\end{figure}

\textit{Proof of Lemma~\ref{thm:locality_exp_effectiveHam}}.
We define $X_s := \bigcup_{j\ge s+2} B_{j}$ and $\Lambda_s=\bigcup_{j\ge s} B_{j}$ for $s=0,1,2,\ldots, q-1$.
Then, from the definition of the truncated Hamiltonian~\eqref{def:truncated_Hamiltonian}, we obtain 
 \begin{align}
\|H-H_\tc\| \le \sum_{s=0}^{q-1} \| V_{B_s ,X_s} (\Lambda_s)\|  . \label{ineq_H_minus_H_t}
\end{align}
By using the assumption~\ref{assump:main_assumption}, the inequality~\eqref{cond_area_LR_sup} gives 
 \begin{align}
\| V_{B_s ,X_s} (\Lambda_s)\|   \le  g_0 l^{-\bar{\alpha}}  . \label{ineq_H_assump_H_t}
\end{align}
From the inequalities~\eqref{ineq_H_minus_H_t} and \eqref{ineq_H_assump_H_t}, we obtain the inequality~\eqref{lemma:truncate__ineq}.

In order to derive the inequality~\eqref{lemma:truncate_ineq_gap}, we utilize the Weyl's inequality:
\begin{align}
|E_j - E_{\tc,j}| \le  \|\delta H_\tc\| \le g_0 q l^{-\bar{\alpha}},
 \end{align}
where $\{E_j\}_{j\ge 0}$ and $\{E_{\tc,j}\}_{j\ge 0}$ are eigenvalues of $H$ and $H_\tc$  in ascending order (i.e., $E_j\le E_{j'}$ for $j\le j'$), respectively.
We thus obtain the inequality~\eqref{lemma:truncate_ineq_gap} as follows:
 \begin{align}
\Delta_\tc = E_{\tc,1}- E_{\tc,0} \ge E_{1}- E_{0} - | E_{1} - E_{\tc,1}|- | E_{0} - E_{\tc,0}| \ge \Delta - 2 \|\delta H_\tc\| .\label{ineq_Delta_t_gap_delta_H_t}
\end{align}
This completes the proof. $\square$

\subsection{Perturbation of the ground state}

\begin{lemma} \label{overlap_Lemma_truncate_original}
Under the assumption of $4\|\delta H_\tc\| < \Delta $, the original ground state $\ket{\Gs}$ have an overlap with that of the truncated Hamiltonian $\ket{\Gs_\tc}$ as follows:
 \begin{align}
\| \ket{\Gs}-\ket{\Gs_\tc} \| \le \frac{ \|\delta H_\tc\|}{\Delta - 4\|\delta H_\tc\|} . \label{overlap_Gs_t_Gs}
\end{align}
Also, for an arbitrary quantum state $\ket{\phi}$, the norm distance between $\ket{\Gs}$ and $\ket{\phi}$ is bounded from above by
\begin{align}
\| \ket{\Gs}-\ket{\phi} \|\le  \| \ket{\Gs_\tc}-\ket{\phi} \| +\frac{ \|\delta H_\tc\|}{\Delta -4 \|\delta H_\tc\|}.
\label{overlap_Gs_phi_norm}
\end{align}
\end{lemma}

\textit{Proof of Lemma~\ref{overlap_Lemma_truncate_original}.}
The inequality~\eqref{overlap_Gs_phi_norm} is simply derived from the triangle inequality, and hence
we need to prove the inequality~\eqref{overlap_Gs_t_Gs}.
We first expand $\ket{\Gs}$ as follows:
\begin{align}
\ket{\Gs} =\zeta_1 \ket{\Gs_\tc} + \zeta_2 \ket{\psi_{\tc,\bot}} ,
\end{align}
where $\langle \Gs_\tc  \ket{\psi_{\tc,\bot}} =0$ and we choose the phase term of $\ket{\Gs_\tc}$ so that $\bra{ \Gs } \Gs_\tc  \rangle $ 
has a positive real value, namely $|\zeta_1|= | \bra{ \Gs} \Gs_\tc  \rangle |= \bra{ \Gs} \Gs_\tc  \rangle =\zeta_1 $. 
Then, the coefficients $\{\zeta_1,\zeta_2\}$ is determined by the eigen-problem of the following matrix:
\begin{align}
\begin{pmatrix}
\bra{\Gs_\tc} H \ket{\Gs_\tc} & \bra{\Gs_\tc }  H     \ket{\psi_{\tc,\bot}} \\
 \bra{\psi_{\tc,\bot}}   H  \ket{\Gs_\tc} &  \bra{\psi_{\tc,\bot}} H \ket{\psi_{\tc,\bot}} 
\end{pmatrix}
 =: 
 \begin{pmatrix}
f_0 & f \\
f^\ast &  f_\bot
\end{pmatrix}.
\end{align}
Then, the ground-state energy of $H$ is formally given by
\begin{align}
E_{0} = \frac{f_0+f_\bot - \sqrt{(f_0-f_\bot )^2 +4 |f|^2 } }{2}, \label{effective_ground_state_formal_0}
\end{align}
and the corresponding coefficients $\{\zeta_1,\zeta_2\}$ are 
\begin{align}
\{\zeta_1,\zeta_2\} \propto \left \{f_\bot -f_0 + \sqrt{(f_\bot -f_0)^2 +4 |f|^2 } , -2 f^\ast  \right \}.
\end{align}
Then, if $f_\bot > f_0$, we have 
\begin{align}
\frac{|\zeta_2|}{\zeta_1} = \frac{2|f|/(f_\bot -f_0)}{1 + \sqrt{1 +4 |f|^2/(f_\bot -f_0)^2}} \le \frac{|f|}{f_\bot -f_0}   ,  \label{mu_2_mu_1_frac_0}
\end{align}
where we will prove the assumption $f_\bot > f_0$ afterward.

From the equation $\zeta_1^2 +|\zeta_2|^2=1$, we obtain
\begin{align}
\zeta_1 = \frac{1}{\sqrt{1+|\zeta_2/\zeta_1|^2}} \ge 1- \frac{1}{2 }\left| \frac{\zeta_2}{\zeta_1} \right|^2. \label{lower_bound_mu_1_mu2/mu1_0}
\end{align}
On the other hand, we have
\begin{align}
\| \ket{\Gs}-\ket{\Gs_\tc} \|^2 = (\zeta_1-1)^2 + |\zeta_2|^2 = 2 - 2 \zeta_1, \label{norm_distance_Gs_Gs_tilde_0}
\end{align}
where we use the fact that $\zeta_1 \in \mathbb{R}^+$.
By combining the inequalities~\eqref{mu_2_mu_1_frac_0}, \eqref{lower_bound_mu_1_mu2/mu1_0} and \eqref{norm_distance_Gs_Gs_tilde_0}, we obtain
\begin{align}
\| \ket{\Gs}-\ket{\Gs_\tc} \|^2 \le \left| \frac{\zeta_2}{\zeta_1} \right|^2 \le \left( \frac{|f|}{f_\bot -f_0}\right)^2,
\end{align}
which reduces to
\begin{align}
\| \ket{\Gs}-\ket{\Gs_\tc} \| \le \frac{|f|}{f_\bot -f_0}. \label{norm_distance_Gs_Gs_tc}
\end{align}

The remaining task is to obtain the upper bound or the lower bound of $|f|$, $f_\bot$ and $f_0$.
First, we have
\begin{align}
|f|=| \bra{\Gs_\tc }  H     \ket{\psi_{\tc,\bot}} |  \le \|H\ket{\Gs_\tc}\| =  \|\delta H_\tc \ket{\Gs_\tc}\|  \le \|\delta H_\tc \|, 
\label{norm_distance_Gs_Gs_tc_f}
\end{align}
where we use $H_\tc\ket{\Gs_\tc}=0$.
Also, we have
\begin{align}
f_\bot =  \bra{\psi_{\tc,\bot}} H \ket{\psi_{\tc,\bot}}=  \bra{\psi_{\tc,\bot}} H_\tc \ket{\psi_{\tc,\bot}} +\bra{\psi_{\tc,\bot}} \delta H_\tc \ket{\psi_{\tc,\bot}}    \ge \Delta_\tc  - \|\delta H_\tc\| \ge \Delta - 3\|\delta H_\tc\| ,
\label{norm_distance_Gs_Gs_tc_bot}
\end{align}
where we use $\bra{\psi_{\tc,\bot}} H_t \ket{\psi_{\tc,\bot}} \ge \Delta_\tc$ and Ineq.~\eqref{ineq_Delta_t_gap_delta_H_t} in the first and second inequalities, respectively.
Finally, we have
\begin{align}
f_0 = \bra{\Gs_\tc} H \ket{\Gs_\tc}= \bra{\Gs_\tc} \delta H_\tc \ket{\Gs_\tc}  \le \|\delta H_\tc \|.
\label{norm_distance_Gs_Gs_tc_0}
\end{align}
By combining the above three inequalities~\eqref{norm_distance_Gs_Gs_tc_f}, \eqref{norm_distance_Gs_Gs_tc_bot} and  \eqref{norm_distance_Gs_Gs_tc_0} with \eqref{norm_distance_Gs_Gs_tc}, we obtain the main inequality~\eqref{overlap_Gs_t_Gs}.

Finally, under the assumption of $\Delta >4\|\delta H_\tc\|$, the inequalities \eqref{norm_distance_Gs_Gs_tc_bot} and  \eqref{norm_distance_Gs_Gs_tc_0} ensures $f_\bot > f_0$ which we have utilized 
in \eqref{mu_2_mu_1_frac_0}:
\begin{align}
f_\bot - f_0 \ge  (\Delta - 3\|\delta H_\tc\|) -  \|\delta H_\tc\| =  \Delta - 4\|\delta H_\tc\|  >0.
\end{align}
This completes the proof. $\square$

%We thus derive the inequality~\eqref{overlap_Gs_t_Gs}.

\subsection{Convenient lemmas on the Schmidt rank}

We here show several convenient lemmas on the Schmidt rank.
\begin{lemma} \label{thm:Schmit_rank_basic}
{~} \\
i) For arbitrary quantum state $\ket{\psi}$ and operator $O$, the Schmidt rank of $O\ket{\psi}$ is bounded from above by
 \begin{align}
{\rm SR} (O\ket{\psi},X) \le {\rm SR} (O,X) {\rm SR} (\ket{\psi},X) .
\end{align}
ii) For arbitrary two operators $O_1$ and $O_2$, the Schmidt rank ${\rm SR} (O_1O_2,X)$ and ${\rm SR} (O_1+O_2,X)$ are bounded from above by
 \begin{align}
{\rm SR} (O_1O_2,X) \le {\rm SR} (O_1,X) {\rm SR} (O_2,X) \quad {\rm and} \quad {\rm SR} (O_1+O_2,X)\le  {\rm SR} (O_1,X) + {\rm SR} (O_2,X),
\end{align}
respectively.\\
iii) For an arbitrary decomposition of $X=Y\sqcup Z$, the Schmidt rank ${\rm SR} (O,Y)$ is bounded from above by
 \begin{align}
{\rm SR} (O,Y) \le d^{2|Z|}{\rm SR} (O,X)  .
\label{Schmidt_rank_X_Y_Zupp}
\end{align}
iv) For arbitrary operator $O_Y$ which is supported on the subset $Y$, the Schmidt rank of ${\rm SR} (O_Y,X)$ is bounded from above by
\begin{align}
{\rm SR} (O_Y,X) \le d^{|Y|} \label{maximum_Schmidt_rank_local_operator}
\end{align}
for $\forall X \subseteq \Lambda$.\\
v) If an operator $O_Y$ is supported on $Y\subseteq X$ (or $Y\subseteq X^\co$), the Schmidt rank ${\rm SR} (O_Y,X)$ is equal to $1$:
 \begin{align}
{\rm SR} (O_Y,X) =1 \for Y \subseteq X \Or Y\subseteq X^\co.
\label{minimum_Schmidt_rank_local_operator}
\end{align}
\end{lemma}
Lemma~\ref{thm:Schmit_rank_basic} is immediately derived from the definition.

On the Schmidt rank of the Hamiltonian~\eqref{eq:ham_1D}, we prove the following lemma.
\begin{lemma} \label{thm:Schmit_rank_Ham}
Let us define $L_s := \bigcup_{j=0}^s B_j $. 
Then, an interaction term $H_s$ of $H_\tc$  in Eq.~\eqref{truncated_Hamiltonia_simple_form} satisfies
\begin{align}
{\rm SR} (H_s,L_s) \le (2d l)^{k}.
\end{align}
\end{lemma}

\textit{Proof of Lemma~\ref{thm:Schmit_rank_Ham}.}
We first recall the definitions of $\{H_s\}_{s=0}^q$: $H_0=h_{s=0}+ h_{s=1} + h_{s=0,s=1}$ and $H_s=h_{s+1} + h_{s,s+1}$ ($s\ge 1$).
From Eq.~\eqref{minimum_Schmidt_rank_local_operator}, we immediately obtain
\begin{align}
{\rm SR} (h_s,L_s) =1 \label{Schmidt_rank_H_bj}
\end{align}
for $0\le s \le q$.
Also, the inequality~\eqref{maximum_Schmidt_rank_local_operator} implies 
\begin{align}
{\rm SR} (h_Z,L_s) \le d^{|Z|} \le d^{k}
\end{align}
for arbitrary interaction terms $\{h_Z\}_{Z\subset \Lambda, |Z|\le k}$.
The block-block interaction $h_{s,s+1}$ contains at most 
\begin{align}
\sum_{j=1}^k\left[ \binom{2l}{j} - 2\binom{l}{j}\right]
\end{align}
interaction terms $h_Z$ with $Z\subseteq B_s \sqcup B_{s+1}$ and $|Z|\le k$.
Therefore, we obtain
\begin{align}
{\rm SR} (h_{s,s+1} ,L_s) \le   d^{k}\sum_{j=1}^k\left[ \binom{2l}{j} - 2\binom{l}{k}\right] \le (2dl)^k -2  .\label{Schmidt_rank_V_bjb_j1}
\end{align}
By combining the inequalities~\eqref{Schmidt_rank_H_bj} and \eqref{Schmidt_rank_V_bjb_j1}, we obtain
\begin{align}
{\rm SR} (H_s ,L_s) \le  {\rm SR} (h_{s,s+1} ,L_s)  +2 \le (2d l)^{k}.
\end{align}
This completes the proof. $\square$

\subsection{The Eckart-Young theorem}

We here show the Eckart-Young theorem~\cite{Eckart1936} without the proof. 
Let us consider a normalized state $\ket{\psi}$ and give its Schmidt decomposition as
 \begin{align}
\ket{\psi} = \sum_{m=1}^{D_\psi} \mu_m  \ket{\psi_{1,m}} \otimes \ket{\psi_{2,m}},
\end{align}
where $\mu_1\ge\mu_2 \ge \mu_3 \cdots \ge  \mu_{D_\psi}$, and $\{\ket{\psi_{1,m}}\}_{m=1}^{D_\psi}$ and $\{\ket{\psi_{2,m}}\}_{m=1}^{D_\psi}$ are orthonormal states, respectively.
We then consider another normalized state $\ket{\phi}$ with its Schmidt rank $D_\phi$ and define the overlap with the state $\ket{\psi}$ as 
 \begin{align}
\| \ket{\phi} -\ket{\psi}\|.
\end{align}
The Eckart-Young theorem gives the following inequality:
 \begin{align}
\sum_{m>D_\phi} \mu_m^2 \le \| \ket{\phi} -\ket{\psi}\|^2 . \label{thm:The Eckart-Young theorem}
\end{align}

 \subsection{Overlap between the ground state and  low-entangled state}
 We relate the AGSP operator to the overlap between the ground state $\ket{\Gs_\tc}$ and the low-entangled state.
 Note that we here make the AGSP operator not for $\ket{\Gs}$ but for $\ket{\Gs_\tc}$.
 On this point, we can prove the following proposition:
 \begin{prop} \label{prop:Overlap between the ground state and  low-entangled state}
Let $K_\tc$ be an AGSP operator for $\ket{\Gs_\tc}$ with the parameters ($\delta_{K_\tc}, \epsilon_{K_\tc}, D_{K_\tc}$). 
If the following inequality holds 
 \begin{align}
\epsilon_{K_\tc}^2 D_{K_\tc} \le  \frac{1}{2}, \label{cond:Boot_strapping_lemma}
\end{align}
there exists a quantum state  $\ket{\psi}$ with ${\rm SR}(\ket{\psi})\le D_{K_\tc}$ such that
 \begin{align}
\left \| \ket{\psi} - \ket{\Gs_\tc} \right\|\le \epsilon_{K_\tc}\sqrt{2 D_{K_\tc}   } + \delta_{K_\tc}.
\label{ineq:AGSP_bootstrap_norm_distance}
\end{align}
\end{prop}

\textit{Proof of Proposition~\ref{prop:Overlap between the ground state and  low-entangled state}.} 
Let  $\ket{\Gs_{K_\tc}}$ be a quantum state  such that  $K_\tc \ket{\Gs_{K_\tc}}=\ket{\Gs_{K_\tc}}$ as in Eq.~\eqref{Def_AGSP_error_chap2}, namely
 \begin{align}
\|\ket{\Gs_\tc}  -\ket{\Gs_{K_\tc}}\|  \le \delta_{K_\tc}    , \quad  \| K (1- \ket{\Gs_{K_\tc}}\bra{\Gs_{K_\tc}})  \|\le   \epsilon_{K_\tc}, \quad {\rm and} \quad  {\rm SR}(K_\tc) \le D_{K_\tc} . \label{Def_AGSP_error_chap2_prop_2}
\end{align}
We then expand the state $\ket{\Gs_{K_\tc}}$ by the use of the Schmidt decomposition with respect to the partition $\Lambda=L\sqcup R$:
 \begin{align}
\ket{\Gs_{K_\tc}} = \sum_{m\ge 1} \mu_{K_\tc,m}  \ket{\Prod_{K_\tc,m}}, \label{Schmidt_decomp_appro_gs}
\end{align}
where the Schmidt coefficients $\{\mu_{K_\tc,m}\}$ are positive real numbers and defined in non-ascending order as $\mu_{K_\tc,1} \ge \mu_{K_\tc,2} \ge \mu_{K_\tc,3} \cdots$.
Each of $\{\ket{\Prod_{K_\tc,m}}\}$ is a product state with respect to the partition $\Lambda=L\sqcup R$ (see Supplementary Figure~\ref{fig:Area_law_Ham_truncate}).
Hence,  for an arbitrary product state $\ket{\Prod}$, the overlap with $\ket{\Gs_{K_\tc}}$ is smaller than $\bra{\Gs_{K_\tc}}\Prod_{K_\tc,1}\rangle=\mu_{K_\tc,1}$: 
 \begin{align}
|\langle \Prod \ket{\Gs_{K_\tc}}|\le  \mu_{K_\tc,1} .\label{Schmidt_decomp_arbtirary_overlap_gs}
\end{align}

Let us choose the target state $\ket{\psi}$ in the inequality~\eqref{ineq:AGSP_bootstrap_norm_distance} as
 \begin{align}
\ket{\psi} = \frac{K_\tc \ket{\Prod_{K_\tc,1}} }{\| K_\tc \ket{\Prod_{K_\tc,1}}\|}.
 \label{Def:gamma_K_tc_Gs_t}
\end{align}
Then, our task is to upper-bound the following quantity $\Gamma_{K_\tc}$:
 \begin{align}
\Gamma_{K_\tc}:= \left \|\ket{\Gs_\tc} - \frac{K_\tc \ket{\Prod_{K_\tc,1}} }{\| K_\tc \ket{\Prod_{K_\tc,1}}\|} \right\|  .
 \label{Def:gamma_K_tc_Gs_t}
\end{align}
The value of $\Gamma_{K_\tc}$ is bounded from above as follows.
First, from the triangle inequality, we  obtain
 \begin{align}
\Gamma_{K_\tc} = \biggl \| \ket{\Gs_\tc}  - \frac{K_\tc \ket{\Prod_{K_\tc,1}}}{\| K_\tc \ket{\Prod_{K_\tc,1}}\|} \biggr \|
 &\le \left \| \ket{\Gs_{K_\tc}} - \frac{K_\tc \ket{\Prod_{K_\tc,1}}}{\| K_\tc \ket{\Prod_{K_\tc,1}} \|} \right\| + \| \ket{\Gs_\tc} -\ket{\Gs_{K_\tc}}\|\notag \\ 
 & \le  \left \| \ket{\Gs_{K_\tc}} - \frac{K_\tc \ket{\Prod_{K_\tc,1}}}{\| K_\tc \ket{\Prod_{K_\tc,1}} \|} \right\|  + \delta_{K_\tc}, \label{upper_bound_gamma_s_chap2_ineq}
\end{align}
where the last inequality is given by the inequality~\eqref{Def_AGSP_error_chap2_prop_2}.

Second, we prove the inequality of
 \begin{align}
\left \| \ket{\Gs_{K_\tc}} - \frac{K_\tc \ket{\Prod_{K_\tc,1}}}{\| K_\tc \ket{\Prod_{K_\tc,1}} \|} \right\|  \le \frac{\epsilon_{K_\tc} }{\mu_{K_\tc,1}} ,\label{upper_bound_gamma_s_chap2_ineq_223}
\end{align}
which reduces the inequality~\eqref{upper_bound_gamma_s_chap2_ineq} to 
 \begin{align}
\Gamma_{K_\tc} \le \frac{\epsilon_{K_\tc} }{\mu_{K_\tc,1}} + \delta_{K_\tc}. \label{upper_bound_gamma_K_tc_ineq}
\end{align}\
In order to derive the inequality~\eqref{upper_bound_gamma_s_chap2_ineq_223}, we express the product state $\ket{\Prod_{K_\tc,1}}$ in the definition~\eqref{Schmidt_decomp_appro_gs} as 
\begin{align}
\ket{\Prod_{K_\tc,1}} = \mu_{K_\tc,1} \ket{\Gs_{K_\tc}} + \sqrt{1-\mu_{K_\tc,1}^2} \ket{\psi_{K_\tc,\bot}},
\end{align}  
where $\ket{\psi_{K_\tc,\bot}}$ is a state orthogonal to $\ket{\Gs_{K_\tc}}$.
Note that $\mu_{K_\tc,1} \in \mathbb{R}^+$ from the definition of the Schmidt decomposition.  
From $K_\tc \ket{\Gs_{K_\tc}}= \ket{\Gs_{K_\tc}}$, we have
\begin{align}
&K_\tc^2 \ket{\Prod_{K_\tc,1}} = \mu_{K_\tc,1} \ket{\Gs_{K_\tc}} + \sqrt{1-\mu_{K_\tc,1}^2} K_\tc^2 \ket{\psi_{K_\tc,\bot}}, \notag \\
& \|K_\tc \ket{\Prod_{K_\tc,1}}\|^2 =\bra{\Prod_{K_\tc,1}}K_\tc^2 \ket{\Prod_{K_\tc,1}} =  \mu_{K_\tc,1}^2  + (1-\mu_{K_\tc,1}^2)\bra{\psi_{K_\tc,\bot}} K_\tc^2 \ket{\psi_{K_\tc,\bot}},  \label{K_p_2Prod_p_1}
\end{align}  
where we use $\bra{\Gs_{K_\tc}} K_\tc^2 \ket{\psi_{K_\tc,\bot}}=\langle \Gs_{K_\tc}  \ket{\psi_{K_\tc,\bot}}=0$ in the second equation.

From the above equation, we obtain
 \begin{align}
\left \| \ket{\Gs_{K_\tc}} - \frac{K_\tc \ket{\Prod_{K_\tc,1}}}{\| K_\tc \ket{\Prod_{K_\tc,1}} \|} \right\|^2=& \left \| \ket{\Gs_{K_\tc}} - \frac{\mu_{K_\tc,1} \ket{\Gs_{K_\tc}} + \sqrt{1-\mu_{K_\tc,1}^2} K_\tc \ket{\psi_{K_\tc,\bot} }}{\| K_\tc \ket{\Prod_{K_\tc,1}}\| } \right \|^2\notag \\
=& \frac{(\| K_\tc \ket{\Prod_{K_\tc,1}}\|-\mu_{K_\tc,1})^2+(1-\mu_{K_\tc,1}^2) \bra{\psi_{K_\tc,\bot}}K_\tc^2\ket{\psi_{K_\tc,\bot}} }{\| K_\tc \ket{\Prod_{K_\tc,1}}\|^2} \notag \\
=& \frac{2\| K_\tc \ket{\Prod_{K_\tc,1}}\|^2 - 2 \mu_{K_\tc,1}\| K_\tc \ket{\Prod_{K_\tc,1}}\| }{\| K_\tc \ket{\Prod_{K_\tc,1}}\|^2}   \notag \\
=&2- 2 \biggl(1+\frac{1-\mu_{K_\tc,1}^2}{\mu_{K_\tc,1}^2}  \bra{\psi_{K_\tc,\bot}}K_\tc^2\ket{\psi_{K_\tc,\bot}}  \biggr)^{-1/2} ,\label{K_p_2Prod_p_1_2}
\end{align}
where we use Eq.~\eqref{K_p_2Prod_p_1} in derivations of the third and the fourth equations.
Then, from the inequality~\eqref{Def_AGSP_error_second}, we have  $\bra{\psi_{K_\tc,\bot}}K_\tc^2\ket{\psi_{K_\tc,\bot}} \le \epsilon_{K_\tc}^2 $, and hence  
 \begin{align}
\left\|  \ket{\Gs_{K_\tc}} - \frac{K_\tc \ket{\Prod_{K_\tc,1}}}{\| K_\tc \ket{\Prod_{K_\tc,1}}\|} \right\|^2
&=2- 2 \biggl(1+\frac{1-\mu_{K_\tc,1}^2}{\mu_{K_\tc,1}^2}  \bra{\psi_{K_\tc,\bot}}K_\tc^2\ket{\psi_{K_\tc,\bot}}  \biggr)^{-1/2}  \notag \\
&\le \frac{1-\mu_{K_\tc,1}^2}{\mu_{K_\tc,1}^2}  \bra{\psi_{K_\tc,\bot}}K_\tc^2\ket{\psi_{K_\tc,\bot}} \le \frac{ \epsilon_{K_\tc}^2}{\mu_{K_\tc,1}^2}  ,\label{K_p_2Prod_p_1_2_3}
\end{align}
where we use $(1+x)^{-1/2} \ge 1-x/2$ for $x\ge 0$ in the first inequality.
We thus obtain the inequality~\eqref{upper_bound_gamma_s_chap2_ineq_223}, and hence the inequality~\eqref{upper_bound_gamma_K_tc_ineq} is also proven.

To finish the proof, we need to derive a relationship between the coefficient $\mu_{K_\tc,1}$ and the AGSP parameters $\{\delta_{K_\tc}, \epsilon_{K_\tc}, D_{K_\tc} \}$. 
It allows us to obtain the upper bound of $\Gamma_{K_\tc}$ in Eq.~\eqref{Def:gamma_K_tc_Gs_t} only by the AGSP parameters.
For the purpose, we here utilize the following statement called \textit{the bootstrapping lemma};
\begin{lemma}[Bootstrapping lemma~\cite{arad2013area}] \label{lem:Bootstrapping_lemma_}
If the AGSP operator $K_\tc$ satisfies $\epsilon^2_{K_\tc} D_{K_\tc} \le 1/2$, $\mu_{K_\tc,1}$ is bounded from below  by
\begin{align}
\mu_{K_\tc, 1} \ge  \frac{1}{\sqrt{2D_{K_\tc}}},  \label{Bootstrapping_lemma_chap2}
%\ge \sqrt{\epsilon_{p}}.
\end{align}  
where $\mu_{K_\tc,1}$ was defined in Eq.~\eqref{Schmidt_decomp_appro_gs}.
\end{lemma}
By combining the inequalities~\eqref{Bootstrapping_lemma_chap2} and \eqref{upper_bound_gamma_K_tc_ineq}, we have
 \begin{align}
\Gamma_{K_\tc} \le \frac{\epsilon_{K_\tc} }{\mu_{K_\tc,1}} + \delta_{K_\tc}\le  \epsilon_{K_\tc} \sqrt{2D_{K_\tc}   } + \delta_{K_\tc}. \label{upper_bound_gamma_bar_s}
\end{align}
This completes the proof of Proposition~\ref{prop:Overlap between the ground state and  low-entangled state}.

\subsubsection{Proof of Lemma~\ref{lem:Bootstrapping_lemma_}}
We first denote the Schmidt decomposition of $K_\tc \ket{\Prod_{K_\tc,1}}$ by
\begin{align}
K_\tc \ket{\Prod_{K_\tc,1}} &= \sum_{j=1}^{D_{K_\tc}}  \mu'_{K_\tc,j}  \ket{\Prod'_j}.
\end{align}  
Note that $K_\tc \ket{\Prod_{K_\tc,1}}$ is not normalized.
We obtain
\begin{align}
\bra{\Gs_{K_\tc}} K_\tc \ket{\Prod_{K_\tc,1}} &= \sum_{j=1}^{D_{K_\tc}} \mu'_{K_\tc,j}   \langle \Gs_{K_\tc} \ket{\Prod'_j}  \le  \sqrt{\sum_{j=1}^{D_{K_\tc}} \mu_{K_\tc,j}^{'2}} \sqrt{ \sum_{j=1}^{D_{K_\tc}}|   \langle \Gs_{K_\tc} \ket{\Prod'_j}  |^2 }   
=\| K_\tc \ket{\Prod_{K_\tc,1}}\| \sqrt{ \sum_{j=1}^{D_{K_\tc}}| \langle \Gs_{K_\tc} \ket{\Prod'_j} |^2 }, \label{erefer_inequality_boot121}
\end{align}  
where the first inequality is given by the Cauchy-Schwartz inequality.
We now have
\begin{align}
&\bra{\Gs_{K_\tc}} K_\tc \ket{\Prod_{K_\tc,1}} =\langle \Gs_{K_\tc} \ket{\Prod_{K_\tc,1}} = \mu_{K_\tc,1} , \notag \\
&\sum_{j=1}^{D_{K_\tc}}| \langle \Gs_{K_\tc} \ket{\Prod'_j} |^2  \le \sum_{j=1}^{D_{K_\tc}} \mu_{K_\tc,1} ^2 = D_{K_\tc} \mu_{K_\tc,1}^2, \notag \\ 
& \| K_\tc \ket{\Prod_{K_\tc,1}}\| \le \sqrt{\mu_{K_\tc,1}^2 + \epsilon_{K_\tc}^2} ,
\end{align}  
where the first equation is given by the definition $K_\tc \ket{\Gs_{K_\tc}} =\ket{\Gs_{K_\tc}}$, the second inequality is derived from Ineq.~\eqref{Schmidt_decomp_arbtirary_overlap_gs}, and the third inequality  is derived from Ineq.~\eqref{K_p_2Prod_p_1} with $\bra{\psi_{K_\tc,\bot}}K_\tc^2\ket{\psi_{K_\tc,\bot}} \le \epsilon_{K_\tc}^2 $.
Thus, the inequality~\eqref{erefer_inequality_boot121} reduces to
\begin{align}
\mu_{K_\tc,1} \le  \sqrt{\mu_{K_\tc,1}^2 + \epsilon_{K_\tc}^2}  \sqrt{D_{K_\tc} \mu_{K_\tc,1}^2}  , 
\end{align}
which gives the inequality
\begin{align}
\mu_{K_\tc,1}^2 \ge  \frac{1}{D_{K_\tc}}-\epsilon^2_{K_\tc} = \frac{1-D_{K_\tc} \epsilon^2_{K_\tc}}{D_{K_\tc}} \ge  \frac{1}{2D_{K_\tc}}  , \label{tilde_mu_1_DK}
\end{align}
where we utilized $D_{K_\tc} \epsilon^2_{K_\tc}  \le 1/2$.
This completes the proof. $\square$

\subsection{Upper bound of the entanglement entropy by the AGSP operators}

We here relate the AGSP operator to the ground-state entropy $S(\ket{\Gs})$.
For this purpose, we make a sequence of  the AGSP operators $\{K_p\}_{p=1}^\infty$ for $\ket{\Gs}$ each of which has a  state $\ket{\Gs_{K_p}}$ such that  $K_p\ket{\Gs_{K_p}}=\ket{\Gs_{K_p}}$ with Eq.~\eqref{Def_AGSP_error_chap2}.
For simplicity, we denote $\{\delta_{K_p}, \epsilon_{K_p}, D_{K_p} \}$ by $\{\delta_p, \epsilon_p, D_p \}$.
We choose $K_p$ so that $K_\infty$ may satisfy $\epsilon_{\infty}=0$, $\delta_{\infty}=0$; in other words, $K_\infty$ is the exact ground-state projector.
We denote the exact ground state $\ket{\Gs}=\ket{\Gs_{\infty}}$ by
 \begin{align}
\ket{\Gs} = \sum_{m=1}^{D_\infty} \mu_{m}  \ket{\Prod_{m}}.
\label{g_s_schmidt_decomp}
\end{align}
Note that because of $K_\infty = \ket{\Gs} \bra{\Gs}$ the Schmidt rank of $\ket{\Gs}$ is equal to $D_\infty$.
%We then expand the state $\ket{\Gs_{K_p}}$ by the use of the Schmidt decomposition:
% \begin{align}
%\ket{\Gs_{K_p}} = \sum_{m=1} \mu_{s,m}  \ket{\Prod_{s,m}}, \label_{K_\tc,
%\end{align}
%where $\{\ket{\Prod_{s,m}}\}$ are product states with respect to the partition $\Lambda=L\sqcup R$.

We now obtain the following proposition:
\begin{prop} \label{prop:entropy_and_AGSP}
Let $\ket{\psi_D}$ be an arbitrary quantum state with 
 \begin{align}
\| \ket{\psi_D} -\ket{\Gs}\|= \nu_0 \quad {\rm and} \quad  {\rm SR}(\ket{\psi_D})=D. 
\label{def:nu_0_psiD}
\end{align}
Also, we define $\{K_p\}_{p=1}^{\infty}$ as ($\{\delta_p, \epsilon_p, D_p\}_{p=1}^\infty$)-AGSP operators, respectively, where errors $\epsilon_{p}$ and $\delta_{p}$ decrease with
the index $p$, namely $\epsilon_1\ge \epsilon_2\ge\cdots$ and $\delta_1\ge \delta_2\ge\cdots$. 
Then, we prove for each of $\{K_p\}_{p=1}^{\infty}$
 \begin{align}
 \left \| \frac{ K_p e^{-i\theta_p}\ket{\psi_D} }{\| K_p \ket{\psi_D}\|} - \ket{\Gs} \right\|  \le \gamma_p 
 \label{norm_distance_AGSP_pth}
\end{align}
with $\theta_p \in \mathbb{R}$ given in Eq.~\eqref{choice_of_theta_p}, where $\{\gamma_p\}_{p=1}^\infty$ are defined as 
 \begin{align}
\gamma_p := \frac{\epsilon_{p}}{1- \nu_0- \delta_p} + \delta_{p}. \label{Def:gamma_p_bar}
\end{align}
Moreover, under the condition $\gamma_p\le 1$ for all $p$, the entanglement entropy $S(E_0)$ is bounded from above by
 \begin{align}
S(\ket{\Gs}) &\le \log (D) -  \sum_{p=0}^\infty \gamma_p^2  \log  \frac{\gamma_p^2 }{3D_{p+1}} , \label{Basic_inequality_for_entropy_bound_fin}
\end{align}
where we set $\gamma_0:=1$.
\end{prop}

\textit{Proof of Proposition~\ref{prop:entropy_and_AGSP}.}
For the proof, we construct an approximate ground state by means of $ K_p e^{-i\theta_p} \ket{\psi_D}$ with $e^{-i\theta_p}$ an appropriate phase factor such that 
 \begin{align}
\bra{\Gs_{K_p}} e^{-i\theta_p} \ket{\psi_D} =| \bra{\Gs_{K_p}}\psi_D\rangle | .\label{choice_of_theta_p}
\end{align}  
We now want to know how close it is to the exact ground state $\ket{\Gs}$.
We first define the Schmidt rank of $K_p e^{-i\theta_p}\ket{\psi_D}$ by $D'_p$ which is smaller than $D D_p$:
 \begin{align}
D'_p:= {\rm SR}( K_pe^{-i\theta_p}\ket{\psi_D})={\rm SR}(K_p\ket{\psi_D}) \le  {\rm SR}( \ket{\psi_D}) {\rm SR}(K_p)  = D D_p. \label{Schmidt_rank_tilde_D__p}
\end{align} 
Second, we apply the Eckart-Young theorem by letting
$\ket{\psi}=\ket{\Gs}$, $\ket{\phi}=\frac{  K_p e^{-i\theta_p} \ket{\psi_D} }{\| K_p \ket{\psi_D}\|}$ and $D_\phi= D'_p$ in \eqref{thm:The Eckart-Young theorem}: 
 \begin{align}
 \sum_{m > D'_p } \mu_{m}^2 \le  \left \| \frac{ K_p e^{-i\theta_p}\ket{\psi_D} }{\| K_p \ket{\psi_D}\|} - \ket{\Gs} \right\|  ^2 =: \Gamma_p^2 ,
 \label{Young_eckart_gamma_p}
\end{align}
where the Schmidt decomposition for the ground state has been given in Eq.~\eqref{g_s_schmidt_decomp}.
Therefore, in order to derive the inequality~\eqref{norm_distance_AGSP_pth}, we need to prove $\Gamma_p \le \gamma_p$ with $\gamma_p$ defined in Eq.~\eqref{Def:gamma_p_bar}.

In order to upper-bound $\Gamma_p$  by using the AGSP parameters ($\{\delta_p, \epsilon_p, D_p\}_{p=1}^\infty$), we start from the triangle inequality as follows:
 \begin{align}
\Gamma_p := \biggl \| \ket{\Gs} - \frac{ K_pe^{-i\theta_p} \ket{\psi_D}}{\| K_p \ket{\psi_D}\|} \biggr \|
 &\le \left \| \ket{\Gs_{K_p}} - \frac{ K_p e^{-i\theta_p}\ket{\psi_D}}{\| K_p \ket{\psi_D}\|} \right\| + \| \ket{\Gs_{K_p}} - \ket{\Gs}\|\notag \\ 
 & \le  \left \| \ket{\Gs_{K_p}} - \frac{ K_pe^{-i\theta_p} \ket{\psi_D}}{\| K_p \ket{\psi_D}\|}\right\|  + \delta_{p}, \label{upper_bound_gamma_p_chap2_ineq}
\end{align}
where the last inequality is derived from the definition of the AGSP parameter as in the inequality~\eqref{Def_AGSP_error_chap2}.
We, in the following, derive the upper bound of the first term in \eqref{upper_bound_gamma_p_chap2_ineq}. 
By using Eq.~\eqref{choice_of_theta_p}, we decompose the quantum state $e^{-i\theta_p}\ket{\psi_D}$ by
\begin{align}
e^{-i\theta_p} \ket{\psi_D} = \nu_{p} \ket{\Gs_{K_p}} + \sqrt{1-|\nu_{p}|^2} \ket{\psi_{p,\bot}} \with \nu_p:=| \bra{\Gs_{K_p}}\psi_D\rangle |,
\end{align}  
where $\ket{\psi_{p,\bot}}$ is a state orthogonal to $\ket{\Gs_{K_p}}$.
We then derive the upper bound of $\nu_p$, which is given by
 \begin{align}
 \nu_p = | (\bra{\Gs}+\bra{\Gs_{K_p}}-\bra{\Gs}) \ket{\psi_D})| \ge |\bra{\Gs}\psi_D\rangle| - \| \ket{\Gs_{K_p}}-\ket{\Gs}\| \ge1- \nu_0- \delta_p,
 \label{upper_bound_abs_nu_p}
\end{align} 
where $\nu_0$ has been defined in  Eq.~\eqref{def:nu_0_psiD} and we use $ |\bra{\Gs}\psi_D\rangle| \ge 1 - \|\ket{\psi_D} - \ket{\Gs}\| =1-\nu_0$ in the last inequality.
We then follow the same steps as the derivations of Ineq.~\eqref{K_p_2Prod_p_1}, \eqref{K_p_2Prod_p_1_2} and \eqref{K_p_2Prod_p_1_2_3}; 
in these inequalities, we replace as 
 \begin{align}
\ket{\Prod_{K_\tc,1}} \to \ket{\psi_D},\quad \ket{\Gs_{K_\tc}} \to \ket{\Gs_{K_p}} ,\quad \mu_{K_\tc,1} \to \nu_p.
\end{align} 
Thus, we obtain 
 \begin{align}
\left \| \ket{\Gs_{K_p}} - \frac{K_p e^{-i\theta_p} \ket{\psi_D}}{\| K_p \ket{\psi_D}\|}\right\| \le \frac{\epsilon_{p}}{\nu_{p}} \le \frac{\epsilon_{p}}{1- \nu_0- \delta_p},\label{K_p_2Prod_p_1_2_3_4}
\end{align}
where we use \eqref{upper_bound_abs_nu_p} in the second inequality.
By combining the inequalities~\eqref{upper_bound_gamma_p_chap2_ineq} and \eqref{K_p_2Prod_p_1_2_3_4}, we obtain $\Gamma_p \le \gamma_p$.
%Thus, the inequality~\eqref{Basic_inequality_for_entropy_bound} reduces to \eqref{Basic_inequality_for_entropy_bound_fin}.

The remaining task is to upper-bound the entanglement entropy to derive the inequality~\eqref{Basic_inequality_for_entropy_bound_fin}.
We first define 
 \begin{align}
\Gamma_{p,p+1}^2 := \sum_{D'_{p}<m \le D'_{p+1} } \mu_{m}^2,
% \le \Gamma_p^2\le \gamma_p^2 ,\label{inequality_Gamma_gamma}
\end{align}
where we define $D'_{0}=0$. 
Note that from Eq.~\eqref{g_s_schmidt_decomp} we have 
 \begin{align}
\sum_{p=0}^\infty \Gamma_{p,p+1}^2 = \sum_{0<m \le D'_{\infty} } \mu_{m}^2= \sum_{m=1}^{D_{\infty}} \mu_{m}^2 = 1. \label{summation_Gamma_p2_eq_1}
\end{align}
From the inequality~\eqref{Young_eckart_gamma_p}, we have
 \begin{align}
\Gamma_{p,p+1}^2  \le  \Gamma_p^2\le \gamma_p^2 \le 1 ,\label{inequality_Gamma_gamma}
\end{align}
where the last inequality is given by the condition in the proposition.
From the above definition, we have
 \begin{align}
- \sum_{D'_{p}<m \le D'_{p+1}}  \mu_{m}^2 \log (\mu_{m}^2 )
&\le  - \sum_{D'_{p}<m \le D'_{p+1} } \frac{\Gamma_{p,p+1}^2}{D'_{p+1}-D'_{p}} \log  \frac{\Gamma_{p,p+1}^2}{D'_{p+1}-D'_{p}}\notag \\
&= - \Gamma_{p,p+1}^2\log  \frac{\Gamma_{p,p+1}^2}{D'_{p+1}-D'_{p}} \le - \Gamma_{p,p+1}^2\log  \frac{\Gamma_{p,p+1}^2}{D_{p+1}} + \Gamma_{p,p+1}^2 \log D,
\label{upp_bound_entropy_Gamma_p}
\end{align}
where in the last inequality we use $D'_p\le DD_p$ in \eqref{Schmidt_rank_tilde_D__p}.

By using $\{\Gamma_{p,p+1}\}_{p=1}^\infty$ and the ineuqlaity~\eqref{upp_bound_entropy_Gamma_p}, the entanglement entropy of $\ket{\Gs}$ is bounded from above by
 \begin{align}
S(\ket{\Gs}) = - \sum_{m\ge 1}  \mu_{m}^2 \log (\mu_{m}^2 )= -  \sum_{p=0}^\infty\sum_{D'_{p}<m \le D'_{p+1}}  \mu_{m}^2 \log (\mu_{m}^2 )
&\le  \log D - \sum_{p=0}^\infty \Gamma_{p,p+1}^2  \log  \frac{\Gamma_{p,p+1}^2 }{D_{p+1}} ,
\label{upp_bound_entropy_Gamma_p_full}
\end{align}
where we use Eq.~\eqref{summation_Gamma_p2_eq_1} in the second equation. 
We have $-x \log (x/3) \le -y \log (y/3)$ for $0< x \le y \le 1$, and hence 
 \begin{align}
- \Gamma_{p,p+1}^2  \log \Gamma_{p,p+1}^2  \le - \Gamma_{p,p+1}^2  \log ( \Gamma_{p,p+1}^2 /3) \le   - \gamma_p^2  \log (\gamma_p^2/3),
\end{align}
which reduces the inequality~\eqref{upp_bound_entropy_Gamma_p_full} to the main inequality~\eqref{Basic_inequality_for_entropy_bound_fin}.
This completes the proof of Proposition~\ref{prop:entropy_and_AGSP}. $\square$

\subsection{Schmidt rank of the polynomials of the truncated Hamiltonian}

We first show the following lemma:
\begin{lemma} \label{thm:Schmit_rank_Ham_power_0}
The Schmidt rank of the power of the truncated Hamiltonian ${\rm SR} (H_\tc^m)$ is bounded from above by
\begin{align}
{\rm SR} (H_\tc^m) \le   [2+(2dl)^{k}]^{m} .\label{ineq:Schmit_rank_Ham_power_0}
\end{align}
\end{lemma}

{~}\\
\textit{Proof of Lemma~\ref{thm:Schmit_rank_Ham_power_0}.} 
We first decompose $H_\tc$  into 
\begin{align}
 H_\tc = H_{q/2} + H_{<q/2} + H_{>q/2}, 
\end{align}
where $H_{<q/2}:= \sum_{s<q/2} H_s$ and  $H_{>q/2}:= \sum_{s>q/2} H_s$. 
Note that $H_{<q/2}$ and $H_{>q/2}$ are supported on the subsystems $L$ and $R$, respectively.
From Lemmas~\ref{thm:Schmit_rank_basic} and \ref{thm:Schmit_rank_Ham}, we have 
\begin{align}
&{\rm SR}(H_\tc^m) \le  [{\rm SR}(H_\tc)]^m  , \notag \\
&{\rm SR}(H_{q/2})\le  (2dl)^{k}, \quad {\rm SR}(H_{<q/2}) = {\rm SR}(H_{>q/2})=1 ,
\end{align}
which yield the inequality~\eqref{ineq:Schmit_rank_Ham_power_0}. This completes the proof.

Roughly speaking, the inequality~\eqref{ineq:Schmit_rank_Ham_power_0} gives the Schmidt rank of order of $\exp[\orderof{m} \log(dl)]$. 
In fact, when $q$ is large, we obtain much better bound for ${\rm SR} (H_\tc^m)$ as shown in the following proposition:
\begin{prop} \label{thm:Schmit_rank_Ham_power}
The Schmidt rank of the power of the truncated Hamiltonian ${\rm SR} (H_\tc^m)$ is bounded from above by
\begin{align}
{\rm SR} (H_\tc^m) \le d^{ql}(q+m+1)^{q+1}  [ e(q+1)^2 (2dl)^{k}]^{m/(q+1)} \le d^{2ql}   [ e(q+1)^2 (2dl)^{k}]^{m/(q+1)} ,
\end{align}
where for the simplicity we assume $(q+m+1)^{q+1} \le d^{ql}$ which yields the second inequality.
\end{prop}
\noindent
The above estimation gives the Schmidt rank of order of $\exp [\orderof{ql} \log(d) +\orderof{m/q} \log(dl)]$.

\textit{Proof of Proposition~\ref{thm:Schmit_rank_Ham_power}.}
We can prove the proposition by extending the original argument in Ref.~\cite{arad2013area} to the present long-range interacting case. 
In order to estimate the Schmidt rank of $H_\tc^m$, we first describe it as 
  \begin{align}
H_\tc^m = \sum_{s_0+s_1+\cdots +s_{q}=m}O_{s_0,s_1,\ldots,s_{q}}  ,
\end{align}
where each of $\{ O_{s_0,s_1,\ldots,s_{q}}\}$ is given by summation of the operator products in which $H_i$ appears $s_i$ times for $0\le i\le q$.
We here define $O_{i,s}$ as the summation of $\{ O_{s_0,s_1,\ldots,s_{q}}\}$ such that $s_i=s$, $\min(s_0,s_1,\ldots,s_{i-1})>s$ and $\min(s_{i+1},s_{i+2},\ldots,s_{q})\ge s $; that is, the number of appearance of $H_i$ is minimum. 
Notice that the definition of $O_{i,s}$ implies $s\le \lfloor m/(q+1) \rfloor$.
Explicitly, $O_{i,s}$ is given by
  \begin{align}
O_{i,s}= \sum_{\substack{ s_0+s_1+\cdots +s_{q}=m \\ s_i=s,\  \min(s_0,s_1,\ldots,s_{i-1})>s \ \min(s_{i+1},s_{i+2},\ldots,s_{q})\ge s }} O_{s_0,s_1,\ldots,s_{q}}.
\end{align}

By using the notation of $O_{i,s}$, we obtain
\begin{align}
H_\tc^m = \sum_{i=0}^{q} \sum_{s=0}^{\lfloor m/(q+1) \rfloor}O_{i,s}.
\end{align}
From the basic property of the Schmidt rank, we obtain
\begin{align}
{\rm SR} (H_\tc^m) \le \sum_{i=0}^{q} \sum_{s=0}^{\lfloor m/(q+1) \rfloor}{\rm SR}(O_{i,s}). \label{SR__H_t^m}
\end{align}

Our task is to estimate the Schmidt rank ${\rm SR}(O_{i,s})$.
For the purpose, instead of considering $O_{i,s}$ in itself, we consider the following alternative operator $P_{i,s}$ which depends on parameters $\vec{z}=\{z_i\}_{i=0}^{q}\in \mathbb{C}^{\otimes q+1}$: 
  \begin{align}
P_{i,s}(\vec{z})= \sum_{\substack{ s_0+s_1+\cdots +s_{q}=m \\ s_i=s}} z_0^{s_0} z_1^{s_1} \cdots z_{q}^{s_{q}} O_{s_0,s_1,\ldots,s_{q}}  . 
\label{P_i_s_(vec_z)_introduction}
\end{align}
We notice that $P_{i,s}(\vec{1})$ is not generally equal to $O_{i,s}$ since the conditions $\min(s_0,\ldots,s_{i-1})>s$ and $\min(s_{i+1},\ldots,s_{q})\ge s$
are not imposed for $P_{i,s}(\vec{1})$.

In the following, we aim to express $O_{i,s}$ by using $P_{i,s}(\vec{z})$ for specific choices of $\{\vec{z}_\alpha\}_{\alpha=1}^{\mathcal{N}_s}$:
  \begin{align}
O_{i,s}= \sum_{\alpha=1}^{\mathcal{N}_s} \lambda_{\alpha} P_{i,s}(\vec{z}_\alpha) ,  \label{Eq:O_im_0_decomp}
\end{align}
where $\lambda_{\alpha} \in \mathbb{C}$ for $\alpha=1,2,\ldots,\mathcal{N}_s$.
The equation~\eqref{Eq:O_im_0_decomp} gives the upper bound of ${\rm SR} (O_{i,s})$ as 
  \begin{align}
{\rm SR} (O_{i,s})&\le  \mathcal{N}_{s}  \sup_{\vec{z}\in \mathbb{C}^{\otimes q+1}  } {\rm SR} [ P_{i,s}(\vec{z}) ] .
\label{upp_bound_SR_O_i_s_formally}
\end{align}
As shown in the following lemma, the Schmidt rank of $P_{i,s}(\vec{z})$ can be efficiently estimated:
\begin{lemma} \label{lemma:Schmidt_rank__P_i_j}
The Schmidt rank of $ P_{i,s}(\vec{z})$ is bounded from above as 
 \begin{align}
{\rm SR} [ P_{i,s}(\vec{z}) ]\le d^{ql} (m+1)[ em^2 (2dl)^{k}/s^2]^s  \label{ineq_lemma:Schmidt_rank__P_i_j}
\end{align}
for $\forall \vec{z}\in \mathbb{C}^{\otimes q+1} $.
\end{lemma}
Second, we can prove the following lemma:
\begin{lemma} \label{lemma:O_im_0_decomp_necceary_number}
There exists a set of $\{\vec{z}_\alpha\}_{\alpha=1}^{\mathcal{N}_s}$ which gives Eq.~\eqref{Eq:O_im_0_decomp} as long as
  \begin{align}
\mathcal{N}_{s}= \binom{q+m-s-1}{q-1} \le (q+m)^{q-1}. \label{Eq:O_im_0_decomp_necceary_number}
\end{align}
\end{lemma}

From the lemmas~\ref{lemma:Schmidt_rank__P_i_j} and \ref{lemma:O_im_0_decomp_necceary_number}, the inequality~\eqref{upp_bound_SR_O_i_s_formally} reduces to
  \begin{align}
{\rm SR} (O_{i,s})&\le (q+m)^{q-1} d^{ql} (m+1)[ em^2 (2dl)^{k}/s^2]^s ,
\end{align}
which monotonically increases with $s$ for $s\le m/(q+1)$. 
From Eq.~\eqref{SR__H_t^m} and $\lfloor m/(q+1) \rfloor\le m/(q+1)$, we have
\begin{align}
{\rm SR} (H_\tc^m) &\le    \sum_{i=0}^{q} \sum_{s=0}^{\lfloor m/(q+1) \rfloor}   (q+m)^{q-1} d^{ql} (m+1)[ e(q+1)^2 (2dl)^{k}]^{m/(q+1)}  \notag \\
&\le  d^{ql} (q+m)^{q-1}(q+m+1)(m+1) [ e(q+1)^2 (2dl)^{k}]^{m/(q+1)} \notag \\
&\le d^{ql} (q+m+1)^{q+1}[ e(q+1)^2 (2dl)^{k}]^{m/(q+1)} ,
\end{align}
where in the second inequality we use $(q+1) (\lfloor m/(q+1) \rfloor +1)\le q+m+1$.
This completes the proof of Proposition~\ref{thm:Schmit_rank_Ham_power}. $\square$

\subsubsection{Proof of Lemma~\ref{lemma:Schmidt_rank__P_i_j}}
We first define parametrized Hamiltonian $H(\vec{z})$ as follows:
  \begin{align}
H(\vec{z})= \sum_{j=0}^{q} z_j H_j =: z_i H_i  + H_{\neq i}(\vec{z}),
\end{align}
where we define $ H_{\neq i}(\vec{z}):=\sum_{j\neq i} z_j H_j$.
Then, $P_{i,s}(\vec{z})$ is given by
  \begin{align}
P_{i,s}(\vec{z}) = \sum_{t_1+t_2+\cdots +t_{s+1}=m-s} [H_{\neq i}(\vec{z})]^{t_1}  ( z_i H_i)   [H_{\neq i}(\vec{z})]^{t_2}  ( z_i H_i)  \cdots   [H_{\neq i}(\vec{z})]^{t_s}  ( z_i H_i) [H_{\neq i}(\vec{z})]^{t_{s+1}}.
\end{align}

We now estimate the Schmidt rank of $[H_{\neq i}(\vec{z})]^{t}$ and $z_i H_i$.
The latter one has been already given by Lemma~\ref{thm:Schmit_rank_Ham} as ${\rm SR} ( z_i H_i, L_i) \le (2dl)^{k}$. 
Recall that the subset $L_i \in \Lambda$ has been defined in Lemma~\ref{thm:Schmit_rank_Ham} as $L_i := \bigcup_{j=0}^i B_j $. 
In order to estimate ${\rm SR} ( [H_{\neq i}(\vec{z})]^{t}, L_i)$, we define $H(\vec{z})_{<i}$ and $H(\vec{z})_{>i}$ as 
\begin{align}
H_{<i} (\vec{z})= \sum_{j<i} z_j H_j  ,\quad H_{>i} (\vec{z})= \sum_{j>i} z_j H_j ,
\end{align}
where $H_{<i} (\vec{z})$ and $H_{>i} (\vec{z})$ are supported on the subsets $L_i$ and $L_i^\co$, respectively.
Note that $ H_{\neq i}(\vec{z})= H_{<i} (\vec{z})+ H_{>i} (\vec{z})$. 
We have 
\begin{align}
{\rm SR} ([H_{<i} (\vec{z})]^j \otimes [H_{>i} (\vec{z})]^{t-j} ,L_i)=1
\end{align}
for $\forall j$, and hence the Schmidt rank of
\begin{align}
[H_{\neq i}(\vec{z})]^{t} = \sum_{j=0}^t  \binom{t}{j} [H_{<i} (\vec{z})]^j [H_{>i} (\vec{z})]^{t-j} 
\end{align}
is bounded from by
\begin{align}
{\rm SR}([H_{\neq i}(\vec{z})]^{t},L_i) \le t+1.
\end{align}

We thus obtain
  \begin{align}
{\rm SR}[P_{i,s}(\vec{z}) ,L_i] &\le  \sum_{t_1+t_2+\cdots +t_{s+1}=m-s} (2dl)^{ks} \prod_{j=1}^{s+1} (t_j+1)  \notag \\
&\le  \sum_{t_1+t_2+\cdots +t_{s+1}=m-s} (2dl)^{ks} (m+1) [m/(s+1)]^{s+1} \notag \\
&= \multiset{s+1}{m-s}(2dl)^{ks}(m+1) [m/(s+1)]^{s+1}  \le (m+1)[ em^2 (2dl)^{k}/s^2]^s , \label{Schmidt_L_i_P_is}
\end{align}
where the summation with respect to $\{t_1, t_2, \cdots , t_{s+1}\}$ such that $t_1+t_2+\cdots +t_{s+1}=m-s$ is equal to the $(m-s)$-multicombination from a set of $s+1$ elements, and in the inequality, we use
  \begin{align}
&\prod_{j=1}^{s+1} (t_j+1)  \le  \left(\frac{m+1}{s+1}\right)^{s+1} \le (m+1) (m/s)^s ,\notag \\
&\multiset{s+1}{m-s} = \binom{m}{s} \le (em/s)^s
\end{align}
for $t_1+t_2+\cdots +t_{s+1}=m-s$.
Finally, by applying the inequality~\eqref{Schmidt_rank_X_Y_Zupp} to \eqref{Schmidt_L_i_P_is},  we obtain the inequality~\eqref{ineq_lemma:Schmidt_rank__P_i_j}.
Note that $\max(|L_i \setminus L|, |L\setminus L_i|) \le ql/2$ for $0\le i \le q$.
This completes the proof. 
 $\square$

\subsubsection{Proof of Lemma~\ref{lemma:O_im_0_decomp_necceary_number}}

For the proof, we first choose each of $\{z_i\}_{i=0}^{q}$ as $ z_i= x^{(m+1)^i}$ with $x$ a parameter which is fixed afterward. 
It reduces Eq.~\eqref{P_i_s_(vec_z)_introduction} to
  \begin{align}
P_{i,s}(x)= \sum_{\substack{ s_0+s_1+\cdots +s_{q}=m \\ s_i=s}} x^{d(\vec{s})} O_{\vec{s}}   \label{P_i_s_(vec_z)_proof}
\end{align}
with $d(\vec{s}):= \sum_{i=0}^{q}s_i (m+1)^i $, where $\vec{s}=\{s_j \}_{j=0}^{q}$ and $O_{\vec{s}} :=O_{s_0,s_1,\ldots,s_{q}}$. 
We notice that we have $d(\vec{s}) \neq d(\vec{s}')$ if $\vec{s} \neq \vec{s}'$ because of $s_j \le m$ for $\forall j$.  

We label different $\vec{s}=\{s_j \}_{j=0}^{q}$ such that $s_0+s_1+\cdots +s_{q}=m$ with $s_i=s$ by $\{\vec{s}_{u}\}_{u=1}^\mathcal{N}$, 
where the total number $\mathcal{N}$ is equal to the $(m-s)$-multicombination from a set of $q$ elements: 
  \begin{align}
\mathcal{N}= \multiset{q}{m-s}= \binom{q+m-s-1}{m-s} = \mathcal{N}_{s}.
\end{align}
We also order $\{\vec{s}_{u}\}_{u=1}^\mathcal{N}$ so that $d(\vec{s}_1) > d(\vec{s}_2) > \cdots >d(\vec{s}_{\mathcal{N}_{s}}) >0$.
In this notation, Eq.~\eqref{P_i_s_(vec_z)_proof} reduces to 
  \begin{align}
P_{i,s}(x)= \sum_{u=1}^{\mathcal{N}_{s}} x^{d(u)} O_{u} ,
\end{align}
where $d(u):= d(\vec{s}_u)$ and $O_{u}:= O_{\vec{s}_u}$.
Therefore, if there exists a set of $\{x_v\}_{v=1}^{\mathcal{N}_s}$ such that the matrix 
  \begin{align}
M=\begin{pmatrix} 
x_1^{d(1)} & x_1^{d(2)} & \cdots &x_1^{d(\mathcal{N}_{s})}  \\
x_2^{d(1)} & x_2^{d(2)} & \cdots &x_2^{d(\mathcal{N}_{s})}  \\
\vdots & \vdots & \cdots & \vdots \\
x_{\mathcal{N}_{s}}^{d(1)} & x_{\mathcal{N}_{s}}^{d(2)} & \cdots &x_{\mathcal{N}_{s}}^{d(\mathcal{N}_{s})}
\end{pmatrix} \label{matrix_x_1_x_2_schur}
\end{align}
has full rank, an arbitrary $O_{u}$ is described by 
  \begin{align}
O_{u} = \sum_{v=1}^{\mathcal{N}_{s}} \lambda_{u,v} P_{i,s}(x_v).
\end{align}
Now, $\lambda_{u,v}$ is given by $\lambda_{u,v}= (M^{-1})_{u,v}$ with $M^{-1}$ the inverse matrix of $M$.

In order to show the full rank of $M$, we prove $\det (M) \neq 0$ for a particular choice of $\{x_v\}_{v=1}^{\mathcal{N}_s}$. 
Because of $d(1) > d(2) > \cdots > d(\mathcal{N}_{s}) $, $\det (M)$ is equal to the product of the Vandermonde's determinant and the Schur polynomial. 
The former one is given by $\prod_{v < v'} (x_v -x_{v'})$ and is non-zero as long as $x_v \neq x_{v'}$ for $v \neq v'$. 
Moreover, the latter one is positive if $x_v >0$ for $\forall v$ 
since the Schur polynomial is composed of monomials with positive coefficients~\cite{BIEDENHARN1989396,Biedenharn1990}. 
Hence, if we choose $\{x_v\}_{v=1}^\mathcal{N}$ such that $x_v >0$ for $\forall v$ and $x_v \neq x_{v'}$ for $v \neq v'$, we have $\det (M) \neq 0$.

This completes the proof of Lemma~\ref{lemma:O_im_0_decomp_necceary_number}. $\square$

\subsection{Construction of the AGSP}
We here discuss how we can find the AGSP operator satisfying \eqref{cond:Boot_strapping_lemma}.
In order to construct the AGSP, we utilize a polynomial of the Hamiltonian like ${\rm Poly}(H_\tc)$.
For example, one of the candidates for AGSP $\{K_p\}_{p=1}^\infty$ is given by
$
K_p = \left(1- \frac{H_\tc}{\|H_\tc\|} \right)^p.
$
In this case, $\epsilon_{K_p}$ in Ineq.~\eqref{Def_AGSP_error_chap2} is upper-bounded by $e^{-p \Delta/\|H_\tc \|}$, and we thereby obtain the exact ground-state projection in the limit of $p\to \infty$.
However, this AGSP cannot satisfy the condition $D_p \epsilon_p^2 \le 1/2$ of the bootstrapping lemma. 
For the proof of the area law, we need to construct an AGSP operator with a higher accuracy and a lower Schmidt rank.

   \begin{figure}
  \begin{center}
    \includegraphics[scale=0.6]{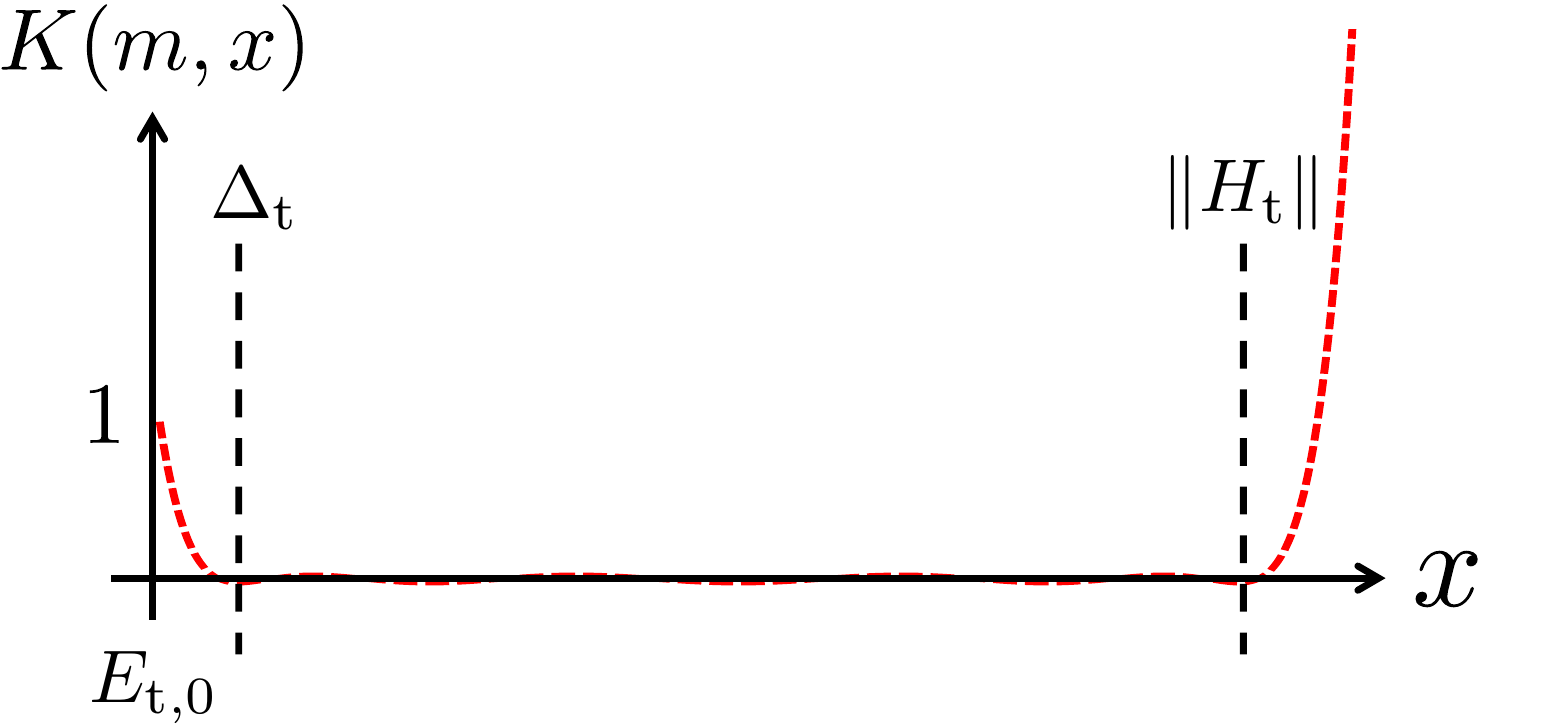}
  \end{center}
  \caption{By the use of the Chebyshev polynomials, we can construct a function $K(m,x)$ which approximately satisfies $K(m,0)=1$ and $K(m,x) \simeq 0$ for $\Delta_\tc \le x\le \|H_\tc\|$. The error of the approximation is bounded from above as in \eqref{error_k_m_poly_AGSP}.
  }\label{fig:Chebyshev} 
\end{figure}

For this purpose, we first define an $m$th-order polynomial $K(m,x)$ such that $K(m,0) =1$ and 
 \begin{align}
|K(m,x)| \le \epsilon_m \label{k_m_poly_AGSP}
\end{align}
for $\Delta_\tc \le x\le \|H_\tc\|$ with $\epsilon_m$ a positive number  (see Supplementary Figure~\ref{fig:Chebyshev} for the schematic picture).
From the definition $E_{\tc,0}=0$, this polynomial gives 
 \begin{align}
\| K(m,H_\tc) -\ket{\Gs_\tc}\bra{\Gs_\tc} \| \le \epsilon_m .
\end{align}

In the construction of the polynomial $K(m,x)$, we employ the Chebyshev polynomial~\cite{PhysRevB.85.195145,arad2013area,Kuwahara_2017}:
 \begin{align}
T_m(x):=\frac{\left(x+\sqrt{x^2-1}\right)^m + \left(x-\sqrt{x^2-1}\right)^m}{2}. \label{AGSP_Chebyshev_polynomial}
\end{align}
The first few polynomials are given by 
 \begin{align}
&T_1(x)= x, \quad T_2(x)=2x^2-1, \quad T_3(x)=4x^3-3x, \notag \\
&T_4(x)=8x^4-8x^2+1 , \quad T_5(x)=16x^5-20x^3+5x .
\end{align}
As shown in the following lemma, the Chebyshev polynomial $T_m(x)$ approximately behaves as a boxcar function in the range $[-1, 1]$. 
\begin{lemma} [Lemma~B.2 in Kuwahara, Arad, Amico and Vedral~\cite{Kuwahara_2017}] \label{ChebyShev_box}
The Chebyshev polynomial $T_m(x)$ satisfies 
 \begin{align}
&|T_m(x)| \le 1 \quad  {\rm for} \quad |x|\le1, \\
&\frac{1}{2} \exp \left(2m\sqrt{\frac{|x|-1}{|x|+1}} \right) \le |T_m(x)|  \le \frac{(2x)^m  }{2}  \quad  {\rm for} \quad |x|\ge 1. \label{Chebyshev_basic_prop_up_low}
\end{align}
\end{lemma}

We now choose $K(m,x) $ as follows:
 \begin{align}
&K(m,x) = \frac{T_m \Bigl [ \frac{2x- (\|H_\tc\|+\Delta_\tc) }{\|H_\tc\|-\Delta_\tc}\Bigr] }{T_m \Bigl [-\frac{\|H_\tc\|+\Delta_\tc}{\|H_\tc\|-\Delta_\tc}\Bigr] },   \label{Chebyshev_based_AGSP_func}\\
& \frac{2x- (\|H_\tc\|+\Delta_\tc) }{\|H_\tc\|-\Delta_\tc} \begin{cases}
=-1 &\for x=\Delta_\tc , \notag \\
\in (-1,1) &\for \Delta_\tc <x <\|H_\tc\| ,\notag \\
=1&\for x=\|H_\tc\|,
\end{cases}       
\end{align}
where $K(m,0)=1$ and the lemma~\ref{ChebyShev_box} implies
 \begin{align}
&T_m \Bigl [-\frac{\|H_\tc\|+\Delta_\tc}{\|H_\tc\|-\Delta_\tc}\Bigr] \ge  \frac{1}{2} \exp \left(2m\sqrt{\frac{\Delta_\tc}{\|H_\tc\|}} \right).\label{Chebyshev_basic_prop_up_low_chap2}  
\end{align}
Thus, $\epsilon_m$ in the inequality~\eqref{k_m_poly_AGSP} is upper-bounded by
 \begin{align}
\epsilon_m = \sup_{\Delta_\tc \le x \le \|H_\tc\|} |K(m,x)|  \le \frac{1}{T_m \Bigl [-\frac{\|H_\tc\|+\Delta_\tc}{\|H_\tc\|-\Delta_\tc}\Bigr]}\le 2e^{-2m \sqrt{\Delta_\tc/\|H_\tc\|}} . \label{error_k_m_poly_AGSP}
\end{align}

We therefore conclude that the error of the AGSP $K(m,H_\tc)$ decreases as $e^{-2m \sqrt{\Delta_\tc/\orderof{n}}}$ because of $\|H_\tc\|=\orderof{n}$. 
However, in this case, we have to take $m$ as large as $\orderof{\sqrt{n}}$ for a good approximation, which may result in a high Schmidt rank of the AGSP operator.
We thus need to achieve a good approximation with smaller $m$. 
We thereby consider an effective Hamiltonian instead of the original Hamiltonian.

\subsection{Effective Hamiltonian with a small norm} \label{Effective Hamiltonian with a small norm}

In order to construct the AGSP operator that satisfies the condition~\eqref{cond:Boot_strapping_lemma} for the bootstrapping lemma, it is convenient to utilize an effective Hamiltonian $\tilde{H}_\tc$ instead of the original Hamiltonian $H_\tc$.
Here, the effective Hamiltonian $\tilde{H}_\tc$ has almost the same ground state as the original one.
The points are the followings:
\begin{enumerate}
\item{} The effective Hamiltonian has the norm much smaller than that of the original Hamiltonian, namely $\|\tilde{H}_\tc\| \ll \|H_\tc\|$.
\item{} The Schmidt rank of $\tilde{H}_\tc^m$ should be as small as that of $H_\tc^m$.
This condition implies that the effective Hamiltonian~$\tilde{H}_\tc$ should still have the similar locality to the original one $H_\tc$.
\end{enumerate}
Note that because of the inequality~\eqref{error_k_m_poly_AGSP} the norm of the Hamiltonian critically determines the error of the AGSP operators.
By applying the Chebyshev-based AGSP construction~\eqref{Chebyshev_based_AGSP_func} to the effective Hamiltonian, we can construct the AGSP $K(m,\tilde{H}_\tc)$ which satisfies the condition~\eqref{cond:Boot_strapping_lemma}.

We, in the following, analyze the fundamental property of the effective Hamiltonian given in Eq.~\eqref{truncation_effective_Hamiltonian_H_tc} (see also Supplementary Figure~\ref{fig:effective_Ham}):
 \begin{align}
\tilde{H}_\tc =\sum_{s=0}^{q+1} \tilde{h}_{s} + \sum_{s=0}^q h_{s,s+1}  \label{truncation_effective_Hamiltonian_H_tc_restate}
\end{align}
with 
 \begin{align}
&\tilde{h}_{s}  =\sum_{E_{s,j} < \tau_s} E_{s,j} \ket{E_{s,j}}\bra{E_{s,j}} + \sum_{E_{s,j} \ge \tau_s} \tau_s\ket{E_{s,j}}\bra{E_{s,j}}  , \notag \\
& \tau_s = E_{s,0} + \tau, \label{truncation_effective_Hamiltonian_h_s_again}
\end{align}
where $\{ E_{s,j}, \ket{E_{s,j}}\}_{j\ge 0}$ are the eigenvalues and the eigenstates of $h_{s}$, respectively.

We first notice that Lemma~\ref{thm:Schmit_rank_Ham_power_0} and Proposition~\ref{thm:Schmit_rank_Ham_power} on the Schmidt rank 
are applicable to the effective Hamiltonian $\tilde{H}_\tc$.

From the definition~\eqref{truncation_effective_Hamiltonian_h_s_again}, we immediately obtain $\|\tilde{h}_{s}\| \le \tau$. 
Also, the inequality~\eqref{truncated_Hamiltonian_block_interaction} gives $\| h_{s,s+1} \| \le g_0$,
and hence the norm of the effective Hamiltonian$\|\tilde{H}_\tc\|$ is upper-bounded by 
 \begin{align}
\| \tilde{H}_\tc \| \le \sum_{s=0}^{q+1} \| \tilde{h}_s \| + \sum_{s=0}^{q} \|h_{s,s+1}\| 
\le   \sum_{s=0}^{q+1} (\tau + |E_{s,0}| ) + (q+1)g_0 .
\end{align}
The ground-state energy $|E_{s,0}|$ can have a value of $\orderof{l}$, which gives the upper bound of $\| \tilde{H}_\tc \| \le \orderof{ql} + \orderof{q\tau}$.
However, by appropriately shifting each of the energy origins of $\{h_s\}_{s=0}^{q+1}$ as $\{h_s+ \mathcal{E}_s\}_{s=0}^{q+1}$, 
we can achieve $|E_{s,0} + \mathcal{E}_s|=\orderof{1}$ for $s=0,1,2,\ldots,q+1$.

One can obtain the following lemma (see~\ref{sec_proof_ground_energy_of_block} for the proof). 
\begin{lemma} \label{ground_energy_of_block}
There exists an energy shift from $\{h_s\}_{s=0}^{q+1}$ to $\{h_s+ \mathcal{E}_s\}_{s=0}^{q+1}$ such that 
  \begin{align}
\sum_{s=0}^{q+1}\mathcal{E}_s=0 ,
\label{shifting_energy_mathcal_E_s}
\end{align}
and the absolute values of the ground-state energy $\{E_{s,0}\}_{s=0}^{q+1}$ are bounded from above by
  \begin{align}
|E_{s,0}+\mathcal{E}_s |  \le \frac{q+1}{q+2} g_0 \le g_0 , \label{lower_bound_of_E_s_0}
\end{align}
where $g_0$ has been defined in \eqref{cond_area_LR_sup}.
Note that the condition~\eqref{shifting_energy_mathcal_E_s} guarantees that $H_\tc$ remains the same as before shifting the energies.
\end{lemma}
By following Lemma~\ref{ground_energy_of_block}, we shift the energy origin so that the inequality~\eqref{lower_bound_of_E_s_0} is satisfied.
We then obtain the upper bound of $\| \tilde{H}_\tc \| $ as follows:
  \begin{align}
\| \tilde{H}_\tc \| \le \tau (q+2)  +2g_0 (q+1) \le (q+2)(\tau+2g_0). \label{upp_bound_tilde_H_tc}
\end{align}
which is roughly as large as $\orderof{q \tau}$.

If the cut-off energy $\tau$ becomes sufficiently large, we expect that the low-energy behavior of both Hamiltonians $H_\tc$ and $\tilde{H}_\tc$ are approximately identical.
We now want to know the $\tau$-dependence of the accuracy of the low-energy spectrum of $\tilde{H}$ compared to that of the original Hamiltonian.
The accuracy has been investigated by Arad, Kuwahara and Landau~\cite{Arad_2016} when the energy cut-off is considered only for a single block Hamiltonian.
Unfortunately, the accuracy of the multi-energy cut-off has not been considered so far, and the generalization to the multi-energy cut-off necessitates highly intricate analyses (see~\ref{sec:main_theorem_effective_Hamiltonian_multi_truncation}).

%We intuitively expect that in the case of $\tau \gg \|V_{X,Y} \| $ we can obtain the good approximation.
We prove the following theorem, which ensures the exponentially accurate approximation with respect to the value of $\tau \gg \log (q)$:
\begin{theorem} \label{Effective Hamiltonian_multi_truncation} %{thm:Arad_kuwahara_Landau}
Let us choose $\tau$ such that
\begin{align}
\tau \ge \max \left[ 8g_0 + \frac{1}{\lambda'} \log \left(\frac{88g_0(q+1)(q+2)}{\Delta_\tc}\right) , 4g_0 + \frac{1}{\lambda} \log
\left( \frac{432  (q+2)}{\lambda \Delta_\tc}  \right)    \right],
\label{assumption_of_choice_tau_eff}
\end{align}
where $g_0$ has been defined in \eqref{cond_area_LR_sup} and $\{\lambda, \lambda'\}$ are defined as follows:
\begin{align}
& \lambda:=\frac{1}{12 k^2 +4g_0} ,\quad \lambda' :=\min \left(\frac{1}{112g_0} , \frac{1}{12k^2} \right).
\label{definition_lambda_lambda'}
\end{align}
Then, the spectral gap $\tilde{\Delta}_\tc$ of the effective Hamiltonian is preserved as 
\begin{align}
\tilde{\Delta}_\tc \ge \frac{1}{2} \Delta_\tc. \label{gap_preservation_eff}
\end{align}
Moreover, the norm distance between the original ground state $\ket{\Gs_\tc}$ and the effective one $\ket{\tilde{\Gs}_\tc}$ is exponentially small with respect to the cut-off energy $\tau$:
\begin{align}
\| \ket{\tilde{\Gs}_\tc}-\ket{\Gs_\tc} \| \le  \frac{54(q+2)}{\lambda \Delta_\tc}e^{-\lambda (\tau - 4g_0)} . 
\label{effective_ground_state_overlap_main_thm}
\end{align}
\end{theorem}
We show the proof of this theorem in~\ref{sec:main_theorem_effective_Hamiltonian_multi_truncation}.

\subsubsection{Proof of Lemma~\ref{ground_energy_of_block}} \label{sec_proof_ground_energy_of_block}

For the proof, we aim to find energy shifts $\{\mathcal{E}_s\}_{s=0}^{q+1}$ with $\sum_{s=0}^{q+1}\mathcal{E}_s=0$ such that the inequality~\eqref{lower_bound_of_E_s_0} is satisfied.
For the purpose, we first consider a quantum state $\ket{\psi}:=\bigotimes_{s=0}^{q+1} \ket{E_{0,s}}$.
We then obtain 
  \begin{align}
\bra{\psi} H_\tc \ket{\psi} \ge E_{\tc,0}=0
\end{align}
and
 \begin{align}
\bra{\psi}  \left( \sum_{s=0}^{q+1} h_s  + \sum_{s=0}^{q} h_{s,s+1}\right)  \ket{\psi}
= \sum_{s=0}^{q+1} E_{s,0} + \sum_{s=0}^{q}  \bra{\psi}  h_{s,s+1} \ket{\psi} \le \sum_{s=0}^{q+1} E_{s,0} + (q+1) g_0,
\end{align}
where we use $\|h_{s,s+1}\|\le g_0$ as shown in Ineq.~\eqref{truncated_Hamiltonian_block_interaction}.
Also, we have
  \begin{align}
0= E_{\tc,0} \ge \sum_{s=0}^{q+1} E_{s,0}  - \sum_{s=0}^{q} \| h_{s,s+1}\| \ge \sum_{s=0}^{q+1} E_{s,0} -(q+1)g_0.
\end{align}

By combining the above three inequalities, we have
 \begin{align}
-(q+1) g_0 \le \sum_{s=0}^{q+1} E_{s,0} \le (q+1) g_0. \label{upper_bound_sum_E_s0}
\end{align}
We here shift the energy origins such that $E_{s,0}+\mathcal{E}_s=E_{s',0}+\mathcal{E}_{s'}$ for $\forall s,s'$, which implies  
 \begin{align}
&(q+2) (E_{s,0}+\mathcal{E}_s)= \sum_{s=0}^{q+1}(E_{s,0}+\mathcal{E}_s) = \sum_{s=0}^{q+1} E_{s,0}, \notag \\
&\quad {\rm or} \quad \mathcal{E}_s= - E_{s,0} + \frac{1}{q+2} \sum_{s=0}^{q+1} E_{s,0}
\end{align}
for arbitrary $\{\mathcal{E}_s\}_{s=0}^{q+1}$, where we use the condition $\sum_{s=0}^{q+1}\mathcal{E}_s=0$.
Conversely, the above choice satisfies $\sum_{s=0}^{q+1} \mathcal{E}_s= 0$.
From the inequality~\eqref{upper_bound_sum_E_s0}, the above choice of $\mathcal{E}_s$ leads to 
 \begin{align}
-\frac{q+1}{q+2} g_0 \le E_{s,0} +\mathcal{E}_s \le \frac{q+1}{q+2} g_0
\end{align}
for $s=0,1,2,\ldots,q+1$. 
This completes the proof.  $\square$

\section{Proof of Main Theorem~\ref{main_theorem_area_law}}

We now have all the ingredients to prove the main theorem.
Proof of Theorem~\ref{main_theorem_area_law} consists of the following two Propositions which we will prove in the subsequent subsections.
In the first proposition, we prove the existence of a quantum state which has an $\orderof{1}$ overlap with the exact ground state and has a small Schmidt rank.
\begin{prop} \label{prop1:truncate_gs_overlap}
There exists a quantum state $\ket{\phi}$ such that 
\begin{align}
\|\ket{\Gs} -\ket{\phi} \| \le \frac{1}{2}
\end{align}
with
\begin{align}
\log \left[{\rm SR}(\ket{\phi}) \right]\le  c^\ast  \log^2 (d) \left(\frac{\log (d)}{\Delta}\right)^{1 +2/\bar{\alpha}}\log^{3+3/\bar{\alpha}}\left(\frac{\log (d)}{\Delta} \right),
\end{align}
where $ c^\ast$ is a constant which depends only on $k$, $g_0$, $\bar{\alpha}$, which is finite in the limit of $\bar{\alpha}\to \infty$.
\end{prop}

In the second proposition, by using the quantum state given in Proposition~\ref{prop1:truncate_gs_overlap}, 
we construct an approximate ground state with a desired accuracy and estimate the Schmidt rank of the state. 
Based on this approximation, we also give the upper bound of the entanglement entropy.
\begin{prop} \label{prop0:overlap_AGSP_entropy_bound}
Let $\ket{\phi}$ be an arbitrary quantum state such that 
\begin{align}
\|\ket{\Gs} -\ket{\phi} \| \le \frac{1}{2} \label{def_phi_Prop_theorem_main}
\end{align}
with $D_\phi := {\rm SR}(\ket{\phi})$. 
Then, there exists a quantum state which approximates the ground state $\ket{\Gs}$ by
\begin{align}
\|\ket{\Gs} -\ket{\psi} \| \le \delta
\end{align}
with the state $\ket{\psi}$ satisfying
\begin{align}
\log [{\rm SR}(\ket{\psi})] \| \le \log (D_\phi)+ c_1 \bar{\alpha}^{-1} \frac{\log^{5/2} [2/(\delta \Delta)]}{\sqrt{\Delta}} +c_2 \frac{\log^{3/2} [2/(\delta \Delta)] \log (d) }{\sqrt{\Delta}}  .
\label{schmidt_rank_error_delta_SR}
\end{align}
Also, the entanglement entropy $S(\ket{\Gs})$ is bounded from above by
\begin{align}
S(\ket{\Gs}) &\le \log ( D_\phi)  + c_3 \bar{\alpha}^{-1}  \frac{\log^{5/2} (3/\Delta)}{\sqrt{\Delta}}  +c_4 \frac{\log^{3/2} (3/\Delta) \log(d)}{\sqrt{\Delta}}.
\label{area_law_phi_Gs_d}
\end{align}
Here, $c_1,c_2,c_3,c_4$ are constants of $\orderof{1}$ which depend only on $k$, $g_0$.
\end{prop}

By applying Proposition~\ref{prop1:truncate_gs_overlap} to Proposition~\ref{prop0:overlap_AGSP_entropy_bound}, we immediately prove Theorem~\ref{main_theorem_area_law}.
$\square$

\subsection{Proof of Proposition~\ref{prop1:truncate_gs_overlap}}
In the following, we choose the number of the blocks $q$ such that 
 \begin{align}
\|\delta H_\tc\| \le \frac{\Delta}{8} \Or \Delta_\tc \ge \frac{3}{4}\Delta  ,\label{condition_delta_H_t}
\end{align}
where the second inequality is derived from \eqref{ineq_Delta_t_gap_delta_H_t}, namely $\Delta_\tc\ge \Delta - 2\|\delta H_\tc\|$.
The inequality~\eqref{condition_delta_H_t} implies from Ineq.~\eqref{lemma:truncate__ineq}
 \begin{align}
&\|\delta H_\tc\| \le g_0 q l^{-\bar{\alpha}}   \le \frac{\Delta}{8}, \notag \\
&{\rm or} \quad q = \left \lfloor \frac{\Delta}{8g_0} l^{\bar{\alpha}} \right \rfloor =\orderof{ l^{\bar{\alpha}}\Delta }. \label{choice_of_q}
\end{align}
From Ineq.~\eqref{overlap_Gs_phi_norm} in Lemma~\ref{overlap_Lemma_truncate_original}, the above choice of $q$ gives the following inequality: 
\begin{align}
\| \ket{\Gs}-\ket{\phi} \|\le  \| \ket{\Gs_\tc}-\ket{\phi} \| +\frac{1}{4}  \label{Gs_phi_distance_ineq_4/15}
\end{align}
for an arbitrary quantum state $\ket{\phi}$.
From Proposition~\ref{prop:Overlap between the ground state and  low-entangled state}, if we find an AGSP operator $K_\tc$ such that 
 \begin{align}
&\epsilon_{K_\tc}^2 D_{K_\tc} \le  \frac{1}{2}, \label{boot_strapping_cond_prop5} 
\end{align}
there exists a quantum state $\ket{\psi}$ which satisfies
 \begin{align}
&\| \ket{\Gs_\tc}-\ket{\psi} \|  \le \epsilon_{K_\tc} \sqrt{2 D_{K_\tc}} + \delta_{K_\tc} \quad {\rm with} \quad {\rm SR}(\ket{\psi}) \le D_{K_\tc}.
\end{align}
Hence, if we can prove the existence of the AGSP operator which satisfies
 \begin{align}
&\epsilon_{K_\tc} \sqrt{2 D_{K_\tc}} + \delta_{K_\tc} \le  \frac{1}{4} , \label{trace_distance_cond_prop5}\\
& \log(D_{K_\tc}) \le c^\ast  \log^2 (d) \left(\frac{\log (d)}{\Delta}\right)^{1 +2/\bar{\alpha}}\log^{3+3/\bar{\alpha}}\left(\frac{\log (d)}{\Delta} \right)  \label{Schmidt_rank_cond_prop5}, 
\end{align}
we prove the Proposition~\ref{prop1:truncate_gs_overlap} by replacing $\ket{\phi}$ with $\ket{\psi}$  in Ineq.~\eqref{Gs_phi_distance_ineq_4/15}.

In the construction of the AGSP operator, we utilize the effective Hamiltonian from the truncated Hamiltonian $H_\tc$. 
We here consider the operator $K(m,\tilde{H}_\tc)$ as the AGSP operator $K_\tc$ for the ground state $\ket{\Gs_\tc}$, where 
the function $K(m,x)$ was defined in Eq.~\eqref{Chebyshev_based_AGSP_func}.
It has the parameters $(\delta_{K_\tc}, \epsilon_{K_\tc}, D_{K_\tc})$ as defined in \eqref{Def_AGSP_error_chap2}, which depend only on $m$, $q$ and $\tau$.
We need to appropriately determine these three parameters so that the conditions~\eqref{boot_strapping_cond_prop5}, \eqref{trace_distance_cond_prop5} and \eqref{Schmidt_rank_cond_prop5} are satisfied.

The parameters are bounded from above as follows.
First, from Ineq~\eqref{effective_ground_state_overlap_main_thm} in Theorem~\ref{Effective Hamiltonian_multi_truncation}, we obtain
\begin{align}
\delta_{K_\tc} \le \frac{54(q+2)}{\lambda \Delta_\tc}e^{-\lambda (\tau - 4g_0)} \le  \frac{216(q+2)}{3\lambda \Delta}e^{-\lambda (\tau - 4g_0)} \label{AGSP_delta_s_upper_bound}
\end{align}
under the condition of \eqref{assumption_of_choice_tau_eff}, where we use $\Delta_\tc \ge \frac{3}{4}\Delta $ as in \eqref{condition_delta_H_t}.
Second, from Ineqs.~\eqref{error_k_m_poly_AGSP} and \eqref{upp_bound_tilde_H_tc}, we obtain
\begin{align}
\epsilon_{K_\tc}  \le 2 e^{-2m \sqrt{\tilde{\Delta}_\tc/\|\tilde{H}_\tc \|}}
 \le 2\exp \left(-2m \sqrt{\frac{ \tilde{\Delta}_\tc}{(q+2)(\tau+2g_0)}} \right) \le 2\exp \left(-2m \sqrt{\frac{3\Delta}{8(q+2)(\tau+2g_0)}} \right) 
, \label{error_k_m_poly_AGSP_prop_area_1}
\end{align}
where we use $\tilde{\Delta}_\tc \ge \Delta_\tc /2 \ge \frac{3}{8}\Delta$ in the third inequality.
Third, from Proposition~\ref{thm:Schmit_rank_Ham_power}, we obtain
\begin{align}
D_{K_\tc}={\rm SR} [K(m,\tilde{H}_\tc)] \le   m d^{2ql}   [ e(q+1)^2 (2dl)^{k}]^{m/(q+1)}
\label{D_K_tc_Schmidt_rank}
\end{align}
under the assumption of $(q+m+1)^{q+1} \le d^{ql}$.

We choose $\tau$ such that $\delta_{K_\tc}  \le 1/8$, or equivalently from Ineq.~\eqref{AGSP_delta_s_upper_bound},
\begin{align}
\tau \ge 4g_0 + \frac{1}{\lambda} \log\left(\frac{1728}{3\lambda} \frac{q+2}{\Delta} \right) \quad \to \quad \tau = c_\tau\log (q/\Delta) , \label{delta_K_1/30_cond}
\end{align}
where $c_\tau$ is a constant depending only on $g_0$ and $k$ which is chosen to satisfy the condition of \eqref{assumption_of_choice_tau_eff}.
Then, for the inequality~\eqref{trace_distance_cond_prop5} to be satisfied, we need to choose $m$ and $q$ such that
\begin{align}
 \epsilon_{K_\tc} \sqrt{2D_{K_\tc}}\le \frac{1}{8} .   \label{proof_two_cond_prop5_reduce}
\end{align}
From the inequalities~\eqref{error_k_m_poly_AGSP_prop_area_1} and \eqref{D_K_tc_Schmidt_rank}, we can formally give  $\epsilon_{K_\tc}$ and $D_{K_\tc}$ as 
\begin{align}
\epsilon_{K_\tc} \le \exp \left(-b_1m \sqrt{\frac{\Delta}{q\log (q/\Delta) }}\right) , \quad 
D_{K_\tc} \le \exp\left( c_1\log(d)q  (q/\Delta)^{1/\bar{\alpha}} + c_2\frac{ \log (dq/\Delta) m}{q} \right) ,
\label{epsilon_K_D_K_formal_form}
\end{align}
where the constants $b_1$, $c_1$ depend only on $\{g_0,k\}$, and $c_2$ depends on $\{\bar{\alpha}, g_0,k\}$.
Also, the constant $c_2$ is proportional to $1/\bar{\alpha}$ and hence it is finite in the limit of $\bar{\alpha}\to \infty$.
Note that in Eq.~\eqref{epsilon_K_D_K_formal_form} we utilize $l \propto  (q/\Delta)^{1/\bar{\alpha}}$ from Eq.~\eqref{choice_of_q}.

In the following, we choose $m$ and $q$ as such $\epsilon_{K_\tc} D_{K_\tc} \le \epsilon_K^{1/2}$, which reduces the conditions~\eqref{boot_strapping_cond_prop5} and \eqref{proof_two_cond_prop5_reduce} to
\begin{align}
&\epsilon_{K_\tc}^2 D_{K_\tc}\le \epsilon_{K_\tc}^{3/2} \le  \frac{1}{2}, \quad \quad 
 \epsilon_{K_\tc} \sqrt{2D_{K_\tc} }  \le \sqrt{2} \epsilon_{K_\tc}^{3/4} \le \frac{1}{8}  .  \label{proof_two_cond_prop5}
\end{align}
From the inequalities in~\eqref{epsilon_K_D_K_formal_form}, the condition $\epsilon_{K_\tc} D_{K_\tc} \le \epsilon_{K_\tc}^{1/2}$ is satisfied for 
\begin{align}
&c_1\log(d)q  (q/\Delta)^{1/\bar{\alpha}} \le \frac{b_1}{4}m\sqrt{\frac{\Delta}{q\log (q/\Delta) }}    ,\notag \\
&c_2 \frac{\log (dq/\Delta) m}{q}  \le \frac{b_1}{4}m \sqrt{\frac{\Delta}{q\log (q/\Delta) }} . \label{condition_m_l_AGSP}
\end{align} 
The first inequality in \eqref{condition_m_l_AGSP} gives the lower bound of $m$ as follows:
\begin{align}
m \ge \frac{4c_1  \log (d)}{b_1} \Delta^{-1/2-1/\bar{\alpha}} q^{3/2 +1/\bar{\alpha} } \sqrt{\log (q/\Delta)}.
\label{condition_m_l_AGSP_first_one}
\end{align} 
The second one implies 
\begin{align}
\frac{1}{\log(q/\Delta)}\sqrt{\frac{q}{\log( q/\Delta)}} \ge \frac{4c_2}{b_1} \frac{\log (d)}{\sqrt{\Delta}},
\end{align} 
which is satisfied for 
\begin{align}
q \ge \frac{c_q c_2\log^2(d)}{\Delta} \log^3\left(\frac{\log (d)}{\Delta} \right),
\end{align} 
where $c_q$ is a constant depending only on  $k$ and $g_0$.
By choosing 
$
q =  \left \lceil \frac{c_qc_2\log^2(d)}{\Delta} \log^3\left(\frac{\log (d)}{\Delta} \right) \right\rceil,
$
the parameters $\epsilon_K$ exponentially decays with $m$. 
Hence, there exists a constant $c_m$ such that 
\begin{align}
&m= c_m c_2 \log (d) \Delta^{-1/2-1/\bar{\alpha}} \left(\frac{\log^2 (d)}{\Delta} \right)^{3/2 +1/\bar{\alpha} }   \log^{5+3/\bar{\alpha}}\left(\frac{\log (d)}{\Delta} \right)         
\end{align} 
satisfies~\eqref{proof_two_cond_prop5} and \eqref{condition_m_l_AGSP_first_one}, where $c_m$ is a constant depending only on  $k$ and $g_0$.

Finally, under the above choices of $m$ and $l$, the Schmidt rank $D_{K_\tc}$ is bounded from above as 
\begin{align}
\log (D_{K_\tc}) \le  c^\ast  \log^2 (d) \left(\frac{\log (d)}{\Delta}\right)^{1 +2/\bar{\alpha}}  \log^{3+3/\bar{\alpha}}\left(\frac{\log (d)}{\Delta} \right)  ,
\end{align}
 where $ c^\ast$ is a constant depending only on $k$, $g_0$, $\bar{\alpha}$ and $ c^\ast$ is finite in the limit of $\bar{\alpha}\to \infty$.
We thus obtain Eq.~\eqref{Schmidt_rank_cond_prop5}. This completes the proof.   $\square$

\subsection{Proof of Proposition~\ref{prop0:overlap_AGSP_entropy_bound}}
In the proof, we utilize Proposition~\ref{prop:entropy_and_AGSP}.
For the purpose, we set $q=2$ and construct the AGSP operator for $\ket{\Gs}$ by using $K(m,\tilde{H}_\tc)$.
From Ineq.~\eqref{lemma:truncate__ineq} with $q=2$, we obtain the upper bound of $\|\delta H_\tc\|$ as 
 \begin{align}
\|\delta H_\tc\|\le 2g_0 l^{-\bar{\alpha}} . \label{lemma:truncate__ineq_prop_area_law}
\end{align} 
From the inequality~\eqref{ineq_Delta_t_gap_delta_H_t}, namely $\Delta_\tc \ge \Delta- 2\|\delta H_\tc\| $, the condition
 \begin{align}
\|\delta H_\tc\| \le  2g_0 l^{-\bar{\alpha}} \le \frac{\Delta}{8} \label{cond_l_Ht_0}
\end{align}
implies $\Delta_\tc \ge 3\Delta/4 $.
In the following discussions, we choose $l$ as 
 \begin{align}
l\ge  \left( \frac{16g_0}{\Delta}  \right )^{1/\bar{\alpha}}\label{cond_l_Ht}
\end{align}
so that the condition~\eqref{cond_l_Ht_0} is satisfied. 
Also, from Lemma~\ref{overlap_Lemma_truncate_original},  for an arbitrary quantum state $\ket{\phi}$, we have
\begin{align}
\| \ket{\Gs}-\ket{\phi} \|&\le  \| \ket{\Gs_\tc}-\ket{\phi} \| + \frac{ \|\delta H_\tc\|}{\Delta -  4\|\delta H_\tc\|} \notag \\
&\le  \| \ket{\Gs_\tc}-\ket{\phi} \| + \frac{2\|\delta H_\tc\|}{\Delta} 
\le  \| \ket{\Gs_\tc}-\ket{\phi} \|  +\frac{4g_0  l^{-\bar{\alpha}}}{\Delta}   ,
\label{Gs_ket_phi_distance_last_prop}
\end{align}
where we use Ineq.~\eqref{cond_l_Ht_0} in the first inequality and Ineq.~\eqref{lemma:truncate__ineq_prop_area_law} in the second inequality.

For the construction of the AGSP operator, we utilize the effective Hamiltonian from the truncated Hamiltonian $H_\tc$ with $q=2$.
We here assume for $\tau$ the condition~\eqref{assumption_of_choice_tau_eff}, which is now given by
\begin{align} 
\tau \ge \max \left[ 8g_0 + \frac{1}{\lambda'} \log \left(\frac{1408g_0}{\Delta}\right) , 4g_0 + \frac{1}{\lambda} \log
\left( \frac{2304}{\lambda\Delta}  \right)    \right].
\label{assumption_of_choice_tau_eff_final_prop}
\end{align}
Then, Theorem~\ref{Effective Hamiltonian_multi_truncation} gives
\begin{align}
&\| \ket{\tilde{\Gs}_\tc}-\ket{\Gs_\tc} \| \le  \frac{216}{\lambda \Delta_\tc}e^{-\lambda (\tau - 4g_0)} \le   \frac{288}{\lambda \Delta}e^{-\lambda (\tau - 4g_0)}  , 
\label{tilde_Gs_Gs_distance_last_prop}  \\
&\tilde{\Delta}_\tc \ge \frac{\Delta_\tc}{2} \ge \frac{3\Delta}{8} ,
\label{tilde_Gs_Gs_distance_last_prop_gap_effective}  
\end{align}
where we use $\Delta_t \ge 3\Delta/4$ in the second inequality.

Here, the operator $K(m,\tilde{H}_\tc)$ depends only on the parameters $m$, $l$ and $\tau$, and hence we denote it by $K_{m,l,\tau}$ and define the AGSP parameters as $(\delta_{m,l,\tau},\epsilon_{m,l,\tau},D_{m,l,\tau})$.
We note that the parameters $(\delta_{m,l,\tau},\epsilon_{m,l,\tau},D_{m,l,\tau})$ do not depend on the choice of the energy origin for $H_\tc$.
Hence, when we consider the truncated Hamiltonians $H_\tc$ with different parameters $q$ and $l$, we will retake the energy origin so that $E_{\tc,0}=0$. 

The upper bounds of the parameters are given as follows.
First, $\delta_{m,l,\tau}$ is bounded from above by
\begin{align}
\delta_{m,l,\tau} =\| \ket{\Gs}-\ket{\tilde{\Gs}_\tc} \| &\le \frac{288}{\lambda \Delta}e^{-\lambda (\tau - 4g_0)}  +\frac{4g_0 l^{-\bar{\alpha}}}{\Delta} ,
%\le \frac{28}{ \lambda\Delta}e^{-\lambda \tilde{\tau} } +\frac{128  l^{-\bar{\alpha}}}{3\Delta}
\label{delta_k_m_poly_AGSP_prop_aRea_2}
\end{align}
where we use the inequality~\eqref{Gs_ket_phi_distance_last_prop} with $\ket{\phi}=\ket{\tilde{\Gs}_\tc}$ and apply Ineq.~\eqref{tilde_Gs_Gs_distance_last_prop} to $\| \ket{\Gs_\tc}-\ket{\tilde{\Gs}_\tc}\|$.
Second,  we obtain a similar inequality to \eqref{error_k_m_poly_AGSP_prop_area_1} as 
\begin{align}
\epsilon_{m,l,\tau}  \le 2\exp \left(-2m \sqrt{\frac{\tilde{\Delta}_\tc}{(q+2)(\tau+2g_0)}} \right) 
\le 2\exp \left(-m \sqrt{\frac{3\Delta}{8(\tau+2g_0)}} \right)  ,\label{error_k_m_poly_AGSP_prop_area_2}
\end{align}
where we use $q=2$ and $\tilde{\Delta}_\tc \ge 3\Delta/8$ in the inequality~\eqref{tilde_Gs_Gs_distance_last_prop_gap_effective}.
Third, from Lemma~\ref{thm:Schmit_rank_Ham_power_0}, we obtain
\begin{align}
D_{m,l,\tau} \le   [2+(2dl)^{k}]^{m} = e^{m \orderof{\log(ld)}}. 
\label{the Schmidt rank_prop_area_law}
\end{align}

In the following, we consider a sequence of the AGSP operator $\{K_p\}_{p=1}^\infty$.
We here characterize the AGSP parameters of $K_p$ by $\delta_p= \delta_{m_p,l_p,\tau_p}$, $\epsilon_p= \epsilon_{m_p,l_p,\tau_p}$ and $D_p=D_{m_p,l_p,\tau_p}$.
From Proposition~\ref{prop:entropy_and_AGSP}, the entanglement entropy $S(\ket{\Gs})$ is bounded from above by
\begin{align}
S(\ket{\Gs}) &\le \log (3D_\phi D_1) -  \sum_{p=1}^\infty \gamma_p^2  \log  \frac{\gamma_p^2 }{3D_{p+1}} \with \gamma_p := \frac{\epsilon_{p}}{1- \nu_0- \delta_p} + \delta_{p},
\label{entropy_s_gs_D}
\end{align}
where $\nu_0:=\|\ket{\Gs} -\ket{\phi} \|$ and $D_\phi = {\rm SR} (\ket{\phi})$. 
Note that the quantum state $\phi$ is given as in Eq.~\eqref{def_phi_Prop_theorem_main} and satisfies $\nu_0\le 1/2$.
Moreover, the quantity $\gamma_p$ characterizes the norm distance between $\ket{\Gs}$ and $K_p \ket{\phi}$ (see Ineq.~\eqref{norm_distance_AGSP_pth}).
We then choose $\{m_p,l_p,\tau_p\}$  such that
 \begin{align}
\gamma_p \le \frac{1}{p},
\end{align}
which is satisfied by choosing
\begin{align}
\delta_{p} \le \frac{1}{3p},\quad \frac{\epsilon_{p}}{1- \nu_0- \delta_p}\le \frac{2}{3p} \Or \epsilon_p \le \frac{1}{9p},  \label{choice_of_delta_epsilon_s}
\end{align}
where the second inequality is derived from
\begin{align}
\frac{\epsilon_{p}}{1- \nu_0- \delta_p} \le \frac{\epsilon_{p}}{1/2- \delta_p} \le \frac{\epsilon_{p}}{1/2- 1/3} =6\epsilon_p \le  \frac{2}{3p}.\label{choice_of_delta_epsilon_s_2}
\end{align}
In the first inequality, we use $\nu_0\le 1/2$.

From the inequality~\eqref{delta_k_m_poly_AGSP_prop_aRea_2}, the following choices of $\tau_p$ and $l_p$ ensure the condition $\delta_p\le 1/(3p)$:
 \begin{align}
\frac{288}{\lambda \Delta}e^{-\lambda (\tau_p - 4g_0)}    \le \frac{1}{6p}, \quad 
\frac{4g_0 l_p^{-\bar{\alpha}}}{\Delta} \le \frac{1}{6p},
\end{align}
 which implies 
 \begin{align}
\tau_p  \ge 4g_0+  \frac{\lambda_p}{\lambda} \log \left(\frac{1728p}{\lambda\Delta }\right) ,\quad
l_p \ge \left(\frac{24g_0 p}{\Delta}\right)^{1/\bar{\alpha}} ,
\label{tilde_p_l_p_agsp_p}
\end{align}
where $\lambda_p$ is a constant for the condition~\eqref{assumption_of_choice_tau_eff_final_prop} to be satisfied.
Note that $\lambda_p$ depends only on $k$ and $g_0$.
For $\epsilon_p\le 1/(9p)$, we choose $m_p$ such that
  \begin{align}
 2\exp \left(-m_p \sqrt{\frac{3\Delta}{8(\tau_p+2g_0)}} \right)  \le \frac{1}{9p},
\end{align}
which yields from Ineq.~\eqref{tilde_p_l_p_agsp_p}
  \begin{align}
 m_p \ge \frac{\log(18p)}{\sqrt{\Delta}}\sqrt{\frac{48g_0}{3} + \frac{8\lambda_p}{3\lambda} \log \left(\frac{1728p}{\lambda\Delta }\right)}.
\end{align}

Under the above choice, the Schmidt rank~\eqref{the Schmidt rank_prop_area_law} is formally given by
\begin{align}
\log (3D_p) \le c_1 \bar{\alpha}^{-1} \frac{\log^{5/2} (3p/\Delta)}{\sqrt{\Delta}} + c_2 \frac{\log^{3/2} (3p/\Delta) \log (d) }{\sqrt{\Delta}}  
\end{align}
with $c_1$ and $c_2$ constants which depend only on $k$, $g_0$.
We note that the dependence of $\bar{\alpha}$ results from \eqref{tilde_p_l_p_agsp_p}, and hence $\log (3D_p)$ are finitely bounded in the limit of $\bar{\alpha}\to \infty$.
Thus, by using Ineq.~\eqref{norm_distance_AGSP_pth} with Eq.~\eqref{Def:gamma_p_bar}, we obtain the inequality~\eqref{schmidt_rank_error_delta_SR} in Proposition~\ref{prop0:overlap_AGSP_entropy_bound}.

Finally, from the inequality~\eqref{entropy_s_gs_D}, we have
\begin{align}
S(\ket{\Gs}) &\le \log (3D_\phi  D_1) +  \sum_{p=1}^\infty \frac{1}{p^2} \left( \log (p^2) +  c_1 \bar{\alpha}^{-1} \frac{\log^{5/2} [3(p+1)/\Delta]}{\sqrt{\Delta}} + c_2 \frac{\log^{3/2} [3(p+1)/\Delta] \log (d) }{\sqrt{\Delta}}   \right) \notag \\
&\le  \log (D_\phi ) +c_3 \bar{\alpha}^{-1}  \frac{\log^{5/2} (3/\Delta)}{\sqrt{\Delta}}  +c_4 \frac{\log^{3/2} (3/\Delta) \log(d)}{\sqrt{\Delta}} ,
\end{align}
where $c_3$ and $c_4$ are constants which depend only on $k$, $g_0$.
We also use $\sum_{p=1}^\infty p^{-2} \log (p^2)= 1.875095 \cdots$.
We thus obtain the upper bound \eqref{area_law_phi_Gs_d} for the entanglement entropy.
This completes the proof of Proposition~\ref{prop0:overlap_AGSP_entropy_bound}. $\square$

\section{Proof of Theorem~\ref{Effective Hamiltonian_multi_truncation}: Accuracy of the effective Hamiltonian with multi-energy cut-off} \label{sec:main_theorem_effective_Hamiltonian_multi_truncation}

In this section, we show the proof of Theorem~\ref{Effective Hamiltonian_multi_truncation}.
Throughout the section, we explicitly take the parameter $g$ into account; in the setup, $g$ was defined in \eqref{g_exensiveness} and set to be $1$.  
For the convenience for the reader, we show the statement again  in the following: 

{\bf Theorem 5.}
\textit{
Let us choose $\tau$ such that
\begin{align}
\tau \ge \max \left[ 8g_0 + \frac{1}{\lambda'} \log \left(\frac{88g_0(q+1)(q+2)}{\Delta_\tc}\right) , 4g_0 + \frac{1}{\lambda} \log
\left( \frac{432  (q+2)}{\lambda \Delta_\tc}  \right)    \right],
\label{assumption_of_choice_tau_eff_sup}
\end{align}
where $g_0$ has been defined in \eqref{cond_area_LR_sup} and $\{\lambda, \lambda'\}$ are defined as follows:
\begin{align}
& \lambda:=\frac{1}{12 gk^2 +4g_0} ,\quad \lambda' :=\min \left(\frac{1}{112g_0} , \frac{1}{12gk^2} \right).
\end{align}
Then, the spectral gap $\tilde{\Delta}_\tc$ of the effective Hamiltonian is preserved as 
\begin{align}
\tilde{\Delta}_\tc \ge \frac{1}{2} \Delta_\tc. \label{gap_preservation_eff_sup}
\end{align}
Moreover, the norm distance between the original ground state $\ket{\Gs_\tc}$ and the effective one $\ket{\tilde{\Gs}_\tc}$ is exponentially small with respect to the cut-off energy $\tau$:
\begin{align}
\| \ket{\tilde{\Gs}_\tc}-\ket{\Gs_\tc} \| \le  \frac{54(q+2)}{\lambda \Delta_\tc}e^{-\lambda (\tau - 4g_0)} . 
\label{effective_ground_state_overlap_main_thm_sup}
\end{align}}

We notice that in~\ref{Effective Hamiltonian with a small norm} the above theorem was given by setting $g=1$ in the definitions of $\lambda$ and $\lambda'$.

\subsection{Preliminaries} \label{sec:preliminaries_eff_proof}

We first define the projection operator onto the eigenspace of $h_s$ as 
 \begin{align}
\Pi^{(s)}_{I}=\sum_{E_{s,j} \in I} \ket{E_{s,j}}\bra{E_{s,j}}  
\label{projection_to_spectrum}
\end{align}
for $I \subset \mathbb{R}$. Especially for $\Pi^{(s)}_{(-\infty,x)}$ and $\Pi^{(s)}_{(-\infty,x]}$, we denote them by $\Pi^{(s)}_{< x}$ and $\Pi^{(s)}_{\le x}$, respectively. 
In the same way, we define $\Pi^{(s)}_{> x}$ and $\Pi^{(s)}_{\ge x}$.
By using the above notations, we describe the effective Hamiltonian $\tilde{H}_\tc$ in Eq.~\eqref{truncation_effective_Hamiltonian_H_tc} as 
 \begin{align}
\tilde{H}_\tc =\sum_{s=0}^{q+1} \tilde{h}_s + \sum_{s=0}^q h_{s,s+1} ,\quad  \tilde{h}_s := h_s \Pi^{(s)}_{\le \tau_s}+ \tau_s \Pi^{(s)}_{> \tau_s} 
\label{explicit_tilde_H_eff}
\end{align}
for $s=0,1,2\ldots, q+1$, where we choose the cut-off energies $\{\tau_s\}_{s=0}^{q+1}$ as 
 \begin{align}
 \tau_s= E_{s,0} + \tau \for s=0,1,2\ldots, q+1
\end{align}
For the total Hamiltonian $H_\tc$ and the effective Hamiltonian $\tilde{H}_\tc$, we define $\Pi_{I}$ and $\tilde{\Pi}_I$ as 
 \begin{align}
&\Pi_{I}=\sum_{E_{\tc,j} \in I} \ket{E_{\tc,j}}\bra{E_{\tc,j}},   \notag \\
&\tilde{\Pi}_I=\sum_{\tilde{E}_{\tc,j} \in I} \ket{\tilde{E}_{\tc,j}}\bra{\tilde{E}_{\tc,j}}  . \label{projector_total_energy}
\end{align}
We here define $\{ E_{\tc,j}, \ket{E_{\tc,j}}\}_{j}$ are the eigenvalues and the eigenstates of $H_\tc$, respectively.
Also, we define $\{ \tilde{E}_{\tc,j}, \ket{\tilde{E}_{\tc,j}}\}_{j}$ are the eigenvalues and the eigenstates of $\tilde{H}_\tc$, respectively.

For $\Pi_{(-\infty,x)}$ (or $\Pi_{(-\infty,x]}$) and $\tilde{\Pi}_{(-\infty,x)}$ (or $\tilde{\Pi}_{(-\infty,x]}$), 
we denote them by $\Pi_{< x}$ (or $\Pi_{\le  x}$) and $\tilde{\Pi}_{< x}$ (or $\tilde{\Pi}_{\le x}$), respectively. 

\subsubsection{Upper bound of the spectral gap for $\tilde{H}_\tc$}

We first estimate the upper bound of the spectral gap $\tilde{\Delta}_\tc$ for the effective Hamiltonian $\tilde{H}_\tc$, which is necessary in utilizing the inequality~\eqref{Energy_dist_subsystem_Lc_effective_H_lc} in Proposition~\ref{The energy distribution of the subsystem L_c}.
We can prove the following lemma.
\begin{lemma} \label{upper_bound_energy_gap_eff}
The spectral gap $\tilde{\Delta}_\tc = \tilde{E}_{\tc,1}- \tilde{E}_{\tc,0}$ is bounded from above by
  \begin{align}
\tilde{\Delta}_\tc \le 4 g_0 , \label{lower_bound_of_tilde_Delta_tc}
\end{align}
where $g_0$ has been defined in \eqref{cond_area_LR_sup}.
\end{lemma}

\textit{Proof of Lemma~\ref{upper_bound_energy_gap_eff}.} 
For the proof, let us pick up the block $B_0$ and consider a quantum state 
\begin{align}
\ket{\psi} = (a_0 \ket{E_{0,0}}+a_1 \ket{E_{0,1}}) \otimes \ket{\tilde{E}_{\Lambda_0,0}},
\end{align}
where $\ket{\tilde{E}_{\Lambda_0,0}}$ is the minimum energy state for $\tilde{H}_\tc - \tilde{h}_{s=0} - h_{s=0,s=1}$ and is supported on $\Lambda_0:=\Lambda \setminus B_0$.
Note that $\ket{E_{0,0}}$ and $\ket{E_{0,1}}$ have been defined as the ground state and the first excited state of $h_{s=0}$, respectively.
 We define the energy $\bra{\tilde{E}_{\Lambda_0,0}}(\tilde{H}_\tc - \tilde{h}_{s=0} - h_{s=0,s=1})\ket{\tilde{E}_{\Lambda_0,0}}$ as $\tilde{E}_{\Lambda_0 , 0}$.
Here, we choose $a_0$ and $a_1$ such that $\bra{\psi} \tilde{\Gs}_\tc \rangle =0$.
We then obtain
\begin{align}
\bra{\psi}\tilde{H}_\tc \ket{\psi} &=  (|a_0|^2 E_{0,0} + |a_1|^2 E_{0,1}) +  \tilde{E}_{\Lambda_0,0} + \bra{\psi}h_{s=0,s=1}\ket{\psi}  \notag \\
&\le    E_{0,1} +  \tilde{E}_{\Lambda_0,0} + \|h_{s=0,s=1}\| \le E_{0,0} +  \tilde{E}_{\Lambda_0,0} + 2g +g_0, 
\end{align}
where we use $\|h_{s,s+1}\| \le g_0$ for $\forall s\in \{0,1,\ldots,q+1\}$ and $E_{0,1} - E_{0,0} \le 2g$ from the inequality~\eqref{gap_2g_bound}.

On the other hand, we have for the ground state $\ket{\tilde{\Gs}_\tc}$ 
\begin{align}
\bra{\tilde{\Gs}_\tc} \tilde{H}_\tc \ket{\tilde{\Gs}_\tc} \ge E_{0,0} +  \tilde{E}_{\Lambda_0,0} - \|h_{s=0,s=1}\| \ge E_{0,0} +  \tilde{E}_{\Lambda_0,0} -g_0.
\end{align}
We thus obtain the upper bound of $\tilde{\Delta}_\tc$ as follows 
\begin{align}
\tilde{\Delta}_\tc \le \bra{\psi}\tilde{H}_\tc \ket{\psi}-\bra{\tilde{\Gs}_\tc} \tilde{H}_\tc \ket{\tilde{\Gs}_\tc}\le 2g +2g_0 \le 4g_0,
\end{align}
where we use $g_0 \ge 1=g$ (see Assumption~\ref{assump:main_assumption}).
This completes the proof of Lemma~\ref{upper_bound_energy_gap_eff}. $\square$

\subsubsection{Lower bound of the spectral gap $\tilde{\Delta}_\tc$ for $\tilde{H}_\tc$}

In the following lemma, we obtain a lower bound of $\tilde{\Delta}_\tc$:

\begin{lemma} \label{energy_gap_upper_bound_eff}
Let us define the spectral decomposition of $\ket{\Gs_\tc}$ as follows:
\begin{align}
\ket{\Gs_\tc} = a_0 \ket{\tilde{\Gs}_{\tc}} +  a_1 \ket{\tilde{E}_{\tc, 1}} +  \sum_{j\ge 2} a_j  \ket{\tilde{E}_{\tc, j}}.\label{expand_phi_Gs_eff}
\end{align}
Then, the quantum state 
\begin{align}
\ket{\psi} =\frac{1}{\sqrt{|a_0|^2+ |a_1|^2}} \left( a_1^\ast \ket{\tilde{\Gs}_{\tc}} -  a_0^\ast \ket{\tilde{E}_{\tc, 1}} \right) 
\label{def:psi_ket_phi}
\end{align}
gives
\begin{align}
\ket{\psi} = \arg \inf_{\psi: \langle \psi \ket{\Gs_\tc}=0} ( \bra{\psi} \tilde{H}_\tc \ket{\psi} ). \label{psi_property_inf}
\end{align}
Also, the spectral gap of the Hamiltonian $\tilde{H}_\tc$ is bounded from below by
  \begin{align}
\tilde{\Delta}_\tc \ge \bra{\psi} \tilde{H}_\tc \ket{\psi}  - \bra{\Gs_\tc} \tilde{H}_\tc \ket{\Gs_\tc} .
\end{align}
\end{lemma}

\textit{Proof of Lemma~\ref{energy_gap_upper_bound_eff}.}
First, the quantum state as $\arg \inf_{\psi: \langle \psi \ket{\Gs_\tc}=0} ( \bra{\psi} \tilde{H}_\tc \ket{\psi} )$ is given in the form of 
\begin{align}
\ket{\psi} =b_0 \ket{\tilde{\Gs}_{\tc}} + b_1 \ket{\tilde{E}_{\tc, 1}} .
\end{align}
Then, due to the constraint of $\langle \psi \ket{\Gs_\tc}=0$, the coefficients $\{b_0,b_1\}$ are uniquely determined as 
\begin{align}
\{b_0,b_1\}= \frac{1}{\sqrt{|a_0|^2+ |a_1|^2}} \{a_1^\ast, -a_0^\ast\}
\end{align}
up to a phase factor. We thus prove Eq.~\eqref{psi_property_inf}.

Also, the definition \eqref{def:psi_ket_phi} implies
\begin{align}
 \bra{\psi} \tilde{H}_\tc \ket{\psi} = \frac{ |a_1|^2\tilde{E}_{\tc, 0} +  |a_0|^2\tilde{E}_{\tc, 1} }{|a_0|^2+ |a_1|^2} .\label{expec_psi_h_t_eff}
\end{align}
On the other hand, the decomposition~\eqref{expand_phi_Gs_eff} yields
\begin{align}
 \bra{\Gs_\tc} \tilde{H}_\tc \ket{\Gs_\tc} \ge  \frac{ |a_0|^2\tilde{E}_{\tc, 0} +  |a_1|^2\tilde{E}_{\tc, 1} }{|a_0|^2+ |a_1|^2} .\label{expec_phi_h_t_eff}
\end{align}
By combining the inequalities~\eqref{expec_psi_h_t_eff} and \eqref{expec_phi_h_t_eff}, we obtain
\begin{align}
 \bra{\psi} \tilde{H}_\tc \ket{\psi}  -  \bra{\Gs_\tc} \tilde{H}_\tc \ket{\Gs_\tc} \le  
 \frac{|a_0|^2-|a_1|^2}{|a_0|^2+ |a_1|^2}  (\tilde{E}_{\tc, 1}-\tilde{E}_{\tc, 0}) \le \tilde{\Delta}_{\tc}.
\end{align}
This completes the proof of Lemma~\ref{energy_gap_upper_bound_eff}. $\square$

\subsubsection{Lower bound of $\bra{\psi} \tilde{H}_\tc \ket{\psi}$}

In order to apply Lemma~\ref{energy_gap_upper_bound_eff}, we need to give a lower bound of $\bra{\psi} \tilde{H}_\tc \ket{\psi}$ with $\ket{\psi}$ defined in 
Eq.~\eqref{def:psi_ket_phi}.
It is given by the following lemma:
\begin{lemma} \label{energy_gap_upper_bound_projection}
Let $\ket{\psi}$ be in the form of Eq.~\eqref{def:psi_ket_phi}.
We then obtain
\begin{align}
\bra{\psi} \tilde{H}_\tc \ket{\psi}  \ge E_\bot  ,
\label{main_ineq_energy_gap_upper_bound_projection}
\end{align}
where we define 
\begin{align}
E_\bot&:= \Delta_\tc (1-  \kappa)^2 - 2 g_0   \kappa (1+\kappa) (q+1) , \label{Def_E_bot_lemma}\\  
\kappa&:=\sum_{s=0}^{q+1} \left\| \Pi^{(s)}_{> \tau_s} \tilde{\Pi}_{\le  \tilde{E}_{\tc,1}} \right\|.\label{Def_epsilon_0_lemma}
\end{align}
\end{lemma}

We also notice that because of Eq.~\eqref{psi_property_inf} we have 
\begin{align}
\bra{\phi} \tilde{H}_\tc \ket{\phi}  \ge \bra{\psi} \tilde{H}_\tc \ket{\psi}  \ge E_\bot   \label{main_ineq_energy_gap_upper_bound_projection_inf}
\end{align}
for an arbitrary quantum state such that $\bra{\phi} \Gs_\tc \rangle =0$.

%
%For the proof, we only need to combine Eq.~\eqref{psi_property_inf} with Ineq.~\eqref{main_ineq_energy_gap_upper_bound_projection} by choosing $\ket{\phi}=\ket{\Gs_\tc}$, where we use the fact that for arbitrary quantum state in the form of
%\begin{align}
%c_0  \ket{\tilde{E}_{\tc, 0}} +  c_1 \ket{\tilde{E}_{\tc, 1}} \ \ (c_0,c_1 \in \mathbb{C}) ,
%\end{align}
%we have
%\begin{align}
% \left\| \Pi^{(s)}_{> \tau_s} \left( c_0  \ket{\tilde{E}_{\tc, 0}} +  c_1 \ket{\tilde{E}_{\tc, 1}} \right)\right\| \le  \left \| \Pi^{(s)}_{> \tau_s} \tilde{\Pi}_{\le \tilde{E}_{\tc,1}} \right\| .
%\end{align}
%In the practical applications, we will utilize this corollary.
%

{~}

{~}

\textit{Proof of Lemma~\ref{energy_gap_upper_bound_projection}.}
Let $P_s$ and $Q_s$ be the projection operators such that
 \begin{align}
P_s := \Pi^{(s)}_{\le \tau_s} ,\quad Q_s := \Pi^{(s)}_{> \tau_s} 
\end{align}
for $s=0,1,\ldots, q+1$, where each of $\{P_s,Q_s\}_{s=0}^{q+1}$ is supported on the subset $B_s\subset \Lambda$.
We  define $P^{(m)}$ as follows:
 \begin{align}
P^{(m)}:= P_0 P_1 \cdots P_{m},\quad Q^{(m)}=1-P^{(m)}.
\end{align}
From Eq.~\eqref{explicit_tilde_H_eff}, we can express $\tilde{h}_s$ as $ \tilde{h}_s= h_s P_s + \tau_s Q_s$ (note that $\tilde{h}_s P_s= h_s P_s$), and hence
\begin{align}
\tilde{H}_\tc P^{(q+1)} = H_\tc P^{(q+1)} . \label{tilde_H_P_q+1}
\end{align}

We here prove the following inequality:
\begin{align}
\| P^{(q+1)}\ket{\psi} - \ket{\psi} \| \le \sum_{s=0}^{q+1} \| Q_s \ket{\psi}\| \le  \kappa   \label{P^q+1_ket_psi_psi}
\end{align}
with $\kappa$ defined in Eq.~\eqref{Def_epsilon_0_lemma}, where the second inequality is derived from
\begin{align}
\| Q_s \ket{\psi}\| = \|  \Pi^{(s)}_{> \tau_s} \ket{\psi}\| = \|  \Pi^{(s)}_{> \tau_s} \tilde{\Pi}_{\le \tilde{E}_{\tc,1}} \ket{\psi}\| \le \|  \Pi^{(s)}_{> \tau_s} \tilde{\Pi}_{\le \tilde{E}_{\tc,1}}\|
\end{align}
for $s\in \{0,1,2,\ldots, q+1\}$.
Note that the definition~\eqref{def:psi_ket_phi} implies $\ket{\psi}=\tilde{\Pi}_{\le \tilde{E}_{\tc,1}} \ket{\psi}$.
In order to prove the first inequality in \eqref{P^q+1_ket_psi_psi}, we consider 
\begin{align}
P^{(q+1)}\ket{\psi} - \ket{\psi} = \sum_{s=0}^{q+1} [P^{(s)} -P^{(s-1)} ] \ket{\psi},
\end{align}
where we set $P^{(-1)}=1$. 
Then, by combining the inequalities
\begin{align}
\| P^{(q+1)}\ket{\psi} - \ket{\psi}\| \le  \sum_{s=0}^{q+1} \| [P^{(s)} -P^{(s-1)} ] \ket{\psi}\|,
\end{align}
and 
\begin{align}
\| [P^{(s)} -P^{(s-1)} ] \ket{\psi}\| = \| P^{(s-1)} (P_s -1) \ket{\psi}\|  \le \| Q_s \ket{\psi}\|  ,
\end{align}
we obtain the first inequality in \eqref{P^q+1_ket_psi_psi}.

%From the inequality~\eqref{P^q+1_ket_psi_psi}, we immediately obtain
%\begin{align}
%| \bra{\Gs_\tc} P^{(q+1)}\ket{\psi} | \le \kappa,  \label{P^q+1_ket_psi_psi_overlap_gs}
%\end{align}
%where we use $\langle \psi\ket{\Gs_\tc}=0$.

By using the notations of $P^{(q+1)}$ and $Q^{(q+1)}$, we obtain
\begin{align}
\bra{\psi}\tilde{H}_\tc \ket{\psi}\ge &  \bra{\psi} P^{(q+1)} \tilde{H}_\tc P^{(q+1)}\ket{\psi} -  |\bra{\psi} P^{(q+1)} \tilde{H}_\tc Q^{(q+1)} \ket{\psi} |  \notag \\
&-  |\bra{\psi} Q^{(q+1)} \tilde{H}_\tc P^{(q+1)} \ket{\psi} | +  \bra{\psi} Q^{(q+1)} \tilde{H}_\tc Q^{(q+1)}\ket{\psi}. 
\label{psi_tilde_H_tc_psi}
\end{align}
We consider the lower bound of the first term in \eqref{psi_tilde_H_tc_psi}.
From Eq.~\eqref{tilde_H_P_q+1} and $E_{\tc,0}=0$, we obtain 
 \begin{align}
 \bra{\psi} P^{(q+1)} \tilde{H}_\tc P^{(q+1)}\ket{\psi} &=  \bra{\psi} P^{(q+1)}H_\tc P^{(q+1)}\ket{\psi} \notag \\
 &=  \bra{\psi} P^{(q+1)} \Pi_{\ge \Delta_\tc} H_\tc\Pi_{\ge \Delta_\tc}  P^{(q+1)}\ket{\psi} 
 \ge  \Delta_{\tc}\|  \Pi_{\ge \Delta_\tc}  P^{(q+1)}\ket{\psi} \|^2
 \end{align}
From the inequality~\eqref{P^q+1_ket_psi_psi}, we immediately obtain
\begin{align}
\| \Pi_{\ge \Delta_\tc} P^{(q+1)}\ket{\psi} \| &\ge 
\|\Pi_{\ge \Delta_\tc} \ket{\psi} \| - \left\| \Pi_{\ge \Delta_\tc} \left( P^{(q+1)}\ket{\psi} -  \ket{\psi} \right)\right\|    \notag \\
&\ge 1- \| P^{(q+1)}\ket{\psi} -  \ket{\psi} \|  \ge  1-\kappa ,   \label{Pi_ge_P_q+1_psi}
\end{align}
where we use $\Pi_{\ge \Delta_\tc} \ket{\psi}= \ket{\psi}$ due to $\bra{\Gs_\tc} \psi \rangle=0$.
By combining the above two inequalities, we obtain
 \begin{align}
 \bra{\psi} P^{(q+1)} \tilde{H}_\tc P^{(q+1)}\ket{\psi} \ge  \Delta_\tc (1-  \kappa)^2,
 \label{psi_tilde_H_tc_psi_first}
 \end{align}
 
We then consider the second and third terms in \eqref{psi_tilde_H_tc_psi}.
Because of $[P_s,\tilde{h}_s]=0$ and $P^{(q+1)}Q^{(q+1)}=0$, we have $P^{(q+1)} \tilde{h}_s Q^{(q+1)}=0$ for $s=0,1,2\ldots,q+1$. 
Hence, the second term is bounded from below as 
 \begin{align}
 - |\bra{\psi} P^{(q+1)} \tilde{H}_\tc Q^{(q+1)} \ket{\psi} | 
 &\ge - \sum_{s=0}^{q} |  \bra{\psi} P^{(q+1)} h_{s,s+1}Q^{(q+1)} \ket{\psi} |  \notag \\
 &\ge -g_0 (q+1) \| (1-P^{(q+1)})  \ket{\psi}\|  \ge - g_0 \kappa  (q+1),
  \label{psi_tilde_H_tc_psi_second}
 \end{align}
 where we use $\|h_{s,s+1}\| \le g_0$ and $P^{(q+1)}+Q^{(q+1)}=1$ in the second inequality, and use Ineq.~\eqref{P^q+1_ket_psi_psi} in the last inequality.
We obtain the same inequality for the third term. 
Finally, the fourth term is bounded from below by 
 \begin{align}
 \bra{\psi} Q^{(q+1)} \tilde{H}_\tc Q^{(q+1)}\ket{\psi} \ge  \| Q^{(q+1)}\ket{\psi} \|^2 \tilde{E}_{\tc,0} \ge  - 2g_0\kappa^2 (q+1),
  \label{psi_tilde_H_tc_psi_fourth}
 \end{align}
 where we use $\| Q^{(q+1)}\ket{\psi} \|=\|(1-P^{(q+1)})\ket{\psi}\| \le \kappa$ from Ineq.~\eqref{P^q+1_ket_psi_psi} and 
  \begin{align}
\tilde{E}_{\tc,0}  \ge  \sum_{s=0}^{q+1} E_{s,0}  -    \sum_{s=0}^{q} \|h_{s,s+1}\|  \ge  - 2(q+1) g_0.
 \end{align}
Note that we use Ineq.~\eqref{upper_bound_sum_E_s0} in order to bound $E_{s,0}$ from below.
By applying the inequalities~\eqref{psi_tilde_H_tc_psi_first}, \eqref{psi_tilde_H_tc_psi_second} and \eqref{psi_tilde_H_tc_psi_fourth} to \eqref{psi_tilde_H_tc_psi}, 
we prove the inequality~\eqref{main_ineq_energy_gap_upper_bound_projection}.
This completes the proof of Lemma~\ref{energy_gap_upper_bound_projection}. $\square$ 

\subsubsection{Upper bound of the norm difference $\| \ket{\Gs_\tc} - \ket{\tilde{\Gs}_\tc} \|$}

%
%By choosing $\ket{\phi}=\ket{\Gs_\tc}$ in Lemma~\ref{energy_gap_upper_bound_eff}, we have 
%\begin{align}
%\tilde{\Delta}_\tc \ge \bra{\psi} \tilde{H}_\tc \ket{\psi}  - \bra{\Gs} \tilde{H}_\tc \ket{\Gs} ,
%\end{align}
%with 
%\begin{align}
%\ket{\psi} =\frac{1}{\sqrt{|a_0|^2+ |a_1|^2}} \left( a_1^\ast \ket{\tilde{E}_{\tc, 0}} -  a_0^\ast \ket{\tilde{E}_{\tc, 1}} \right).
%\end{align}

Finally, we prove the following lemma which estimate $\| \ket{\Gs_\tc} - \ket{\tilde{\Gs}_\tc} \|$:
\begin{lemma} \label{overlap_Lemma_original_effective}
Under the assumption of
\begin{align}
 \bra{\Gs_\tc} \tilde{H}_\tc \ket{\Gs_\tc} < E_\bot ,\label{assumption_tilde_gs_E_bot}
\end{align}
the norm of $\ket{\Gs_\tc} - \ket{\tilde{\Gs}_\tc}$ is bounded from above as follows:
\begin{align}
\| \ket{\tilde{\Gs}_\tc}-\ket{\Gs_\tc} \| \le \frac{\|\tilde{H}_\tc \ket{\Gs_\tc } \|}{E_\bot -\bra{\Gs_\tc}\tilde{H}_\tc\ket{\Gs_\tc} }  .
\label{norm_distance_tilde_Gs_Gs}
\end{align}
\end{lemma}

%estimate the lower bound of the ground-state energy of $\tilde{H}_\tc$ (i.e., $\tilde{E}_{\tc,0}$). 

\textit{Proof of Lemma~\ref{overlap_Lemma_original_effective}.}
The proof is almost the same as that of Lemma~\ref{overlap_Lemma_truncate_original}.
For the convenience for readers, we show the full proof.
We first expand $\ket{\tilde{\Gs}_\tc}$ as follows:
\begin{align}
\ket{\tilde{\Gs}_\tc} =\zeta_1 \ket{\Gs_\tc } + \zeta_2 \ket{\phi_0} ,
\end{align}
where $\langle \Gs_\tc  \ket{\phi_0}=0$ and we choose the phase term of $\ket{\tilde{\Gs}_\tc}$ so that $\bra{ \Gs_\tc } \tilde{\Gs}_\tc  \rangle $ 
has a positive real value, namely $|\zeta_1|= | \bra{ \Gs_\tc } \tilde{\Gs}_\tc  \rangle |= \bra{ \Gs_\tc } \tilde{\Gs}_\tc  \rangle =\zeta_1 $. 
Then, the coefficients $\{\zeta_1,\zeta_2\}$ is determined by the eigen-problem of the following matrix:
\begin{align}
\begin{pmatrix}
\bra{\Gs_\tc}\tilde{H}_\tc\ket{\Gs_\tc} & \bra{\Gs_\tc }  \tilde{H}_\tc     \ket{\phi_0} \\
 \bra{\phi_0}   \tilde{H}_\tc     \ket{\Gs_\tc } &  \bra{\phi_0} \tilde{H}_\tc \ket{\phi_0} 
\end{pmatrix}
 =: 
 \begin{pmatrix}
f_0 & f \\
f^\ast &  f_\bot
\end{pmatrix}.
\end{align}
Then, the ground-state energy of $\tilde{H}_\tc$ is formally given by
\begin{align}
\tilde{E}_{\tc,0} = \frac{f_0+f_\bot - \sqrt{(f_0-f_\bot )^2 +4 |f|^2 } }{2}, \label{effective_ground_state_formal}
\end{align}
and the corresponding coefficients $\{\zeta_1,\zeta_2\}$ are 
\begin{align}
\{\zeta_1,\zeta_2\} \propto \left \{f_\bot -f_0 + \sqrt{(f_\bot -f_0)^2 +4 |f|^2 } , -2 f^\ast  \right \}.
\end{align}
By using Ineq.~\eqref{main_ineq_energy_gap_upper_bound_projection_inf}, we have $f_\bot \ge E_\bot$, and hence the assumption~\eqref{assumption_tilde_gs_E_bot} implies $f_\bot - f_0 \ge 0$.
We thus obtain 
\begin{align}
\frac{|\zeta_2|}{\zeta_1} = \frac{2|f|/(f_\bot -f_0)}{1 + \sqrt{1 +4 |f|^2/(f_\bot -f_0)^2}} \le \frac{|f|}{f_\bot -f_0}  
\le \frac{\|\tilde{H}_\tc \ket{\Gs_\tc } \|}{E_\bot -\bra{\Gs_\tc}\tilde{H}_\tc\ket{\Gs_\tc} } ,  \label{mu_2_mu_1_frac}
\end{align}
where we use $|f|=|\bra{\Gs_\tc }  \tilde{H}_\tc     \ket{\phi_0}|\le \|\tilde{H}_\tc \ket{\Gs_\tc } \|$.

From the equation $\zeta_1^2 +|\zeta_2|^2=1$, we obtain
\begin{align}
\zeta_1 = \frac{1}{\sqrt{1+|\zeta_2/\zeta_1|^2}} \ge 1- \frac{1}{2 }\left| \frac{\zeta_2}{\zeta_1} \right|^2. \label{lower_bound_mu_1_mu2/mu1}
\end{align}
On the other hand, we have
\begin{align}
\| \ket{\tilde{\Gs}_\tc}-\ket{\Gs_\tc} \|^2 = (\zeta_1-1)^2 + |\zeta_2|^2 = 2 - 2 \zeta_1, \label{norm_distance_Gs_Gs_tilde}
\end{align}
where we use the fact that $\zeta_1 \in \mathbb{R}$ and $\zeta_1 \ge 0$.
By combining the inequalities~\eqref{mu_2_mu_1_frac}, \eqref{lower_bound_mu_1_mu2/mu1} and \eqref{norm_distance_Gs_Gs_tilde}, we obtain
\begin{align}
\| \ket{\tilde{\Gs}_\tc}-\ket{\Gs_\tc} \|^2 \le \left| \frac{\zeta_2}{\zeta_1} \right|^2 \le \left( \frac{\|\tilde{H}_\tc \ket{\Gs_\tc } \|}{E_\bot -\bra{\Gs_\tc}\tilde{H}_\tc \ket{\Gs_\tc} }  \right)^2,
\end{align}
which reduces to the inequality~\eqref{norm_distance_tilde_Gs_Gs}.
This completes the proof. $\square$

\subsection{Outline of the proof}
We here prove Theorem~\ref{Effective Hamiltonian_multi_truncation}.
For the proof, we need to derive the following two proposition.
The first proposition is related to the energy distribution of the subsystem $B_s \subset \Lambda$ under the condition that the total energy is involved in an interval $(-\infty,E]$.
We can prove the exponential decay of the distribution of $h_s$:
\begin{prop} \label{The energy distribution of the subsystem L_c}
The overlap between the projections $\Pi^{(s)}_{> E'}$ and $\Pi_{\le E}$ is bounded from above by
\begin{align}
\left \|  \Pi^{(s)}_{> E'} \Pi_{\le E} \right\| \le \frac{4e^{3/2}}{e-1}  e^{-\lambda (\delta E'_s -\delta E - 4g_0)} \label{Energy_dist_subsystem_Lc}
\end{align}
with
\begin{align}
\lambda:= \frac{1}{12gk^2 +4g_0}, \label{def_Lambda_prop_sup}
\end{align}
where $\delta E'_s :=  E'- E_{s,0}$, $\delta E:=  E-E_{\tc ,0}$ with $E_{s,0}$ and $E_{\tc ,0}$ the ground-state energies of $h_s$ and $H_\tc$, respectively.
A similar inequality is satisfied for $\tilde{\Pi}_{\le E}$:
\begin{align}
\left \|  \Pi^{(s)}_{>E'} \tilde{\Pi}_{\le E} \right\|\le \frac{4e^{3/2}}{e-1} e^{-\lambda' (\min(E',\tau_s)   - E_{s,0} - \delta \tilde{E} - 4g_0  )}  \label{Energy_dist_subsystem_Lc_effective_H_lc}
\end{align}
with $\delta \tilde{E} :=  E-\tilde{E}_{\tc,0}$ and 
\begin{align}
\lambda' :=\min \left(\frac{1}{112g_0} , \frac{1}{12gk^2} \right).
\label{def_Lambda'_prop_sup}
\end{align}
\end{prop}
Second, we prove the norm difference between $H_\tc$ and $\tilde{H}_\tc$ under the constraint that the total energy is involved in an interval $(-\infty,E]$: 
\begin{prop} \label{The difference_original_ham_effective_ham}
Let us define $\tau_s$ such that $\tau_s -E_{s,0}=\tau$, where $\tau$ is a fixed constant. 
We then obtain
\begin{align}
\|(H_\tc-\tilde{H}_\tc) \Pi_{\le E}\| &\le \frac{27(q+2)}{\lambda}e^{-\lambda (\tau-\delta E - 4g_0)},
\end{align} 
where $\delta E$ has been defined as $\delta E:= E-E_{\tc,0}$.
\end{prop}

\textit{Remark.} We can derive the similar inequality for $\|(H_\tc-\tilde{H}_\tc) \tilde{\Pi}_{\le E}\| $. 
However, the inequality becomes rather weak as follow
\begin{align}
\|(H_\tc-\tilde{H}_\tc) \tilde{\Pi}_{\le E}\| &\le \|H_\tc\| (q+2) e^{-\lambda' (\tau-\delta E)},
\end{align} 
which does not work in the thermodynamic limit as $\|H_\tc\| \to \infty$.

Before going to the proof, we give the upper bound of $\kappa$ and $\|\tilde{H}_\tc \ket{\Gs_\tc}\|$.
From the definition~\eqref{Def_epsilon_0_lemma} of $\kappa$, we have
\begin{align}
\kappa := \sum_{s=0}^{q+1}  \left \| \Pi^{(s)}_{> \tau_s} \tilde{\Pi}_{\le \tilde{E}_{\tc,1}} \right\|.
\end{align}
We now obtain 
\begin{align}
\| \Pi^{(s)}_{>\tau_s}  \tilde{\Pi}_{\le \tilde{E}_{\tc, 1}} \|
&\le \frac{4e^{3/2}}{e-1} e^{-\lambda' (\tau_s   - E_{s,0} - \tilde{\Delta}_\tc - 4g_0  )}=  \frac{4e^{3/2}}{e-1} e^{-\lambda' (\tau - \tilde{\Delta}_\tc - 4g_0  )},
\end{align}
where we use the inequality~\eqref{Energy_dist_subsystem_Lc_effective_H_lc} with $E=\tilde{E}_{\tc,1}$ and $E'=\tau_s$. Note that $\tau_s=E_{s,0} + \tau$ and 
$\tilde{\Delta}_\tc =\tilde{E}_{\tc, 1}- \tilde{E}_{\tc, 0}$. 
Then, $\kappa$ is bounded from above by
\begin{align}
\kappa \le \frac{4e^{3/2}(q+2)}{e-1} e^{-\lambda' (\tau - \tilde{\Delta}_\tc - 4g_0  )} \le 11(q+2) e^{-\lambda' (\tau - 8g_0  )},
\end{align}
where we use $\tilde{\Delta}_\tc \le 4g_0$ in Lemma~\ref{upper_bound_energy_gap_eff}.
Also, by using Proposition~\ref{The difference_original_ham_effective_ham} with $E=0$, we obtain
\begin{align}
\|(H_\tc-\tilde{H}_\tc) \ket{\Gs_\tc} \| =\|\tilde{H}_\tc \ket{\Gs_\tc} \| \le \frac{27(q+2)}{\lambda}e^{-\lambda (\tau - 4g_0)},
\label{tilde_H_tc_Gs_t_norm_upper_bound}
\end{align}
where we use $H_\tc \ket{\Gs_\tc}=0$ in the first equation.

Under the assumption of \eqref{assumption_of_choice_tau_eff_sup}, we can obtain by straightforward calculations
\begin{align}
\kappa \le \frac{\Delta_\tc}{8g_0(q+1)} \le \frac{1}{12},  \label{Upp_bound_epsilon_0}
\end{align}
and 
\begin{align}
\|\tilde{H}_\tc \ket{\Gs_\tc} \| \le  \frac{\Delta_\tc}{16}, 
\label{Upp_bound_tilde_H_tc_Gs_t_under_assump}
\end{align}
where in the first inequality we use $g_0 \ge 1$ (Assumption~\ref{assump:main_assumption}), $q\ge 2$ and $\Delta_\tc \le 2$.
In particular, from the inequality~\eqref{Upp_bound_epsilon_0}, we have
\begin{align}
E_\bot = \Delta_\tc (1-  \kappa)^2 - 2 g_0   \kappa (1+\kappa) (q+1) \ge  \frac{41}{72}\Delta_\tc,
\label{Upp_bound_E_bot}
\end{align}
where $E_\bot$ has been defined in Eq.~\eqref{Def_E_bot_lemma}.

We now have all the ingredients to prove the main inequalities in Theorem~\ref{Effective Hamiltonian_multi_truncation}.
First, from Lemmas~\ref{energy_gap_upper_bound_eff} and \ref{energy_gap_upper_bound_projection}, we obtain
\begin{align}
\tilde{\Delta}_\tc \ge \bra{\psi} \tilde{H}_\tc \ket{\psi}  - \bra{\Gs_\tc} \tilde{H}_\tc \ket{\Gs_\tc} \ge E_\bot -\|\tilde{H}_\tc \ket{\Gs_\tc} \|.
\end{align}
By applying the upper bounds of \eqref{Upp_bound_tilde_H_tc_Gs_t_under_assump} and \eqref{Upp_bound_E_bot} to the above inequality, we prove the inequality~\eqref{gap_preservation_eff_sup}.
In addition, from Lemma~\ref{overlap_Lemma_original_effective}, we have
\begin{align}
\| \ket{\tilde{\Gs}_\tc}-\ket{\Gs_\tc} \| \le \frac{\|\tilde{H}_\tc \ket{\Gs_\tc } \|}{E_\bot -\bra{\Gs_\tc}\tilde{H}_\tc\ket{\Gs_\tc} }  \le  \frac{\|\tilde{H}_\tc \ket{\Gs_\tc } \|}{E_\bot -\| \tilde{H}_\tc\ket{\Gs_\tc} \|}.
\end{align}
By applying the inequalities~\eqref{tilde_H_tc_Gs_t_norm_upper_bound}, \eqref{Upp_bound_tilde_H_tc_Gs_t_under_assump} and \eqref{Upp_bound_E_bot} to the above inequality, we have 
\begin{align}
\| \ket{\tilde{\Gs}_\tc}-\ket{\Gs_\tc} \| \le \frac{1}{\frac{41}{72}\Delta_\tc - \frac{1}{16} \Delta_\tc} \frac{27(q+2)}{\lambda}e^{-\lambda (\tau - 4g_0)} 
\le  \frac{54(q+2)}{\lambda \Delta_\tc}e^{-\lambda (\tau - 4g_0)} .
\end{align}
This gives the proof of Ineq.~\eqref{effective_ground_state_overlap_main_thm}.

This completes the proof of Theorem~\ref{Effective Hamiltonian_multi_truncation}. $\square$

\subsection{Proof of Proposition~\ref{The difference_original_ham_effective_ham} by utilizing Proposition~\ref{The energy distribution of the subsystem L_c}}

We first prove the Proposition~\ref{The difference_original_ham_effective_ham} by assuming Proposition~\ref{The energy distribution of the subsystem L_c}.
We will prove Proposition~\ref{The energy distribution of the subsystem L_c} afterward.
From the definition of the effective Hamiltonian~\eqref{explicit_tilde_H_eff}, we calculate $\|(H_\tc-\tilde{H}_\tc)\Pi_{\le E}\|$ as 
\begin{align}
\|(H_\tc-\tilde{H}_\tc) \Pi_{\le E}\| &\le \sum_{s=0}^{q+1} \left \|  (h_s-\tau_s) \Pi^{(s)}_{>{\tau_s}}  \Pi_{\le E} \right\|.
\label{H_tilde_H_pi_le_E_s_q+1}
\end{align}
Hence, we need to estimate the upper bound of $\|  (h_s-\tau_s) \Pi^{(s)}_{> {\tau_s}}  \Pi_{\le E} \|$.
For a given $y\in \mathbb{R}^+$ which we set afterward, we have
\begin{align}
\left \|  (h_s-\tau_s) \Pi^{(s)}_{> \tau_s}  \Pi_{\le E} \right\|
&= \left  \| \sum_{j=0}^{\infty}  (h_s-\tau_s) \Pi^{(s)}_{(\tau_s+j{ y }, \tau_s+(j+1){ y }]} \Pi_{\le E} \right\| 
\notag \\
&\le  \sum_{j=0}^{\infty} \left \|  (h_s-\tau_s)  \Pi^{(s)}_{(\tau_s+j{ y }, \tau_s+(j+1){ y }]} \right\| \cdot  \left \|  \Pi_{(\tau_s+j{ y }, \tau_s+(j+1){ y }]}^{(s)} 
\Pi_{\le E} \right\|  \notag \\
&\le \sum_{j=0}^{\infty} (j+1) y \cdot  \left \|  \Pi^{(s)}_{> \tau_s+j{ y }} \Pi_{\le E} \right\| , 
\label{upp_bound_h_s_minus_tau_Pi_tau_Pi_E}
\end{align}
where in the last inequality we use
 \begin{align}
\left \|  \Pi_{(\tau_s+j{ y }, \tau_s+(j+1)y]}^{(s)}  \Pi_{\le E} \right\| \le  \left \|  \Pi^{(s)}_{(\tau_s+j{ y }, \infty)} \Pi_{\le E} \right\| =  \left \|  \Pi^{(s)}_{>\tau_s+j{ y }} \Pi_{\le E} \right\|  .
\end{align}
Thus, we can calculate the upper bound of $\left \|  (h_s-\tau_s) \Pi^{(s)}_{> \tau_s}  \Pi_{\le E} \right\|$ by estimating the norm of
\begin{align}
\left \|  \Pi^{(s)}_{> E'} \Pi_{\le E} \right\|  . 
\end{align}

%For $\|(H_\tc-\tilde{H}_\tc) \tilde{\Pi}_{\le E}\|$, the same inequalities as Ineqs~\eqref{H_tilde_H_pi_le_E_s_q+1} and \eqref{upp_bound_h_s_minus_tau_Pi_tau_Pi_E} hold
%\begin{align}
%&\|(H_\tc-\tilde{H}_\tc)  \tilde{\Pi}_{\le E}\| \le \sum_{s=0}^{q+1} \left \|  (h_s-\tau_s) \Pi^{(s)}_{> \tau_s}  \tilde{\Pi}_{\le E} \right\|, \quad \left \|  (h_s-\tau_s) \Pi^{(s)}_{> \tau_s}   \tilde{\Pi}_{\le E} \right\| \le \sum_{j=0}^{\infty} (j+1){ y } \cdot  \left \|  \Pi^{(s)}_{> \tau_s+j{ y }} \tilde{\Pi}_{\le E} \right\| . 
%\end{align}

By using the Proposition~\ref{The energy distribution of the subsystem L_c}, we obtain
\begin{align}
\bigl \|  \Pi_{>\tau_s +j{ y }}^{(s)} \Pi_{\le E} \bigr\|  \le \frac{4e^{3/2}}{e-1}e^{-\lambda (\tau_s+j{ y } - E_{s,0} -\delta E - 4g_0 )} 
= \frac{4e^{3/2}}{e-1} e^{-\lambda (\tau -\delta E - 4g_0)}e^{-\lambda  j{ y } }  , \label{chap3_com_ene_part1}
\end{align}
where we use the definition $\tau_s := \tau + E_{s,0}$ for each of $s\in \{0,1,2,\ldots, q+1\}$.
It reduces the inequality~\eqref{upp_bound_h_s_minus_tau_Pi_tau_Pi_E} to
\begin{align}
\left \|  (h_s-\tau_s) \Pi^{(s)}_{> \tau_s}  \Pi_{\le E} \right\| 
&\le \frac{4e^{3/2}}{e-1} e^{-\lambda (\tau-\delta E - 4g_0)} \sum_{j=0}^{\infty} y(j+1)e^{-\lambda  j{ y } } \notag \\
&=\frac{4e^{3/2}}{e-1}e^{-\lambda (\tau-\delta E - 4g_0)} \frac{y \cdot e^{2\lambda y}}{(e^{\lambda y}-1)^2}.
\end{align}
By choosing $y= 1/\lambda$, we obtain 
\begin{align}
\left \|  (h_s-\tau_s) \Pi^{(s)}_{> \tau_s}  \Pi_{\le E} \right\|  &\le 
\frac{4e^{3/2}}{e-1}\cdot  \frac{e^2}{\lambda(e-1)^2}e^{-\lambda (\tau-\delta E - 4g_0)} 
\le \frac{27}{\lambda}e^{-\lambda (\tau-\delta E - 4g_0)}.
\end{align}
Finally, by applying the above inequality to \eqref{H_tilde_H_pi_le_E_s_q+1}, we obatain
\begin{align}
\|(H_\tc-\tilde{H}_\tc) \Pi_{\le E}\| &\le \frac{27(q+2)}{\lambda}e^{-\lambda (\tau-\delta E - 4g_0)}.
\end{align}
This completes the proof of Proposition~\ref{The difference_original_ham_effective_ham}. $\square$

\subsection{Proof of Proposition~\ref{The energy distribution of the subsystem L_c}: the first part~\eqref{Energy_dist_subsystem_Lc}}

For the proof, we first consider a normalized quantum state $\ket{\psi}$ and construct the following quantum state $\ket{\phi}$:
\begin{align}
\ket{\phi} :=\Pi_{> E'}^{(s)} \Pi_{\le E} \ket{\psi}.\label{Def:of_phi}
\end{align}
Note that this state $\ket{\phi}$ may not be normalized.
The norm $\| \Pi_{> E'}^{(s)} \Pi_{\le E}\|$ is now given by
\begin{align}
\| \Pi_{> E'}^{(s)} \Pi_{\le E} \| = \sup_{\ket{\psi}} \|\phi\|,\label{Basic_theorem_for_effe_Hami}
\end{align}
where $\|\phi\|$ denotes the norm of the state $\ket{\phi}$.
We then utilize the following inequality which we will prove below:
\begin{align}
 \| \phi \| \le \frac{4e^{3/2}}{e-1} e^{-\lambda (\langle H_\tc \rangle_{\phi}- E)}    \label{Norm_of_Psi_tilde}
\end{align}
with
\begin{align}
\langle H_\tc \rangle_{\phi} = \frac{\bra{\phi}H_\tc \ket{\phi} }{\| \phi \|^2}, \label{Definition_H_phi_ave}
\end{align}
where $\lambda:= 1/(12gk^2 +4g_0)$ as in Eq.~\eqref{def_Lambda_prop_sup}.

In order to obtain an upper bound of $\| \phi \|^2$ from \eqref{Norm_of_Psi_tilde}, we have to calculate a lower bound of $\langle H_\tc \rangle_{\phi}$:
\begin{align}
\langle H_\tc \rangle_{\phi} =\langle h_s \rangle_{\phi}   + \langle (h_{s,s+1} +h_{s-1,s})  \rangle_{\phi}  +\langle \delta H_s \rangle_{\phi},
\label{explicit_H_tc_phi_ave}
\end{align}
where we define $\delta H_s:= H_\tc - h_s -h_{s,s+1} -h_{s-1,s}$, which acts on the sites $\Lambda_s:=\Lambda\setminus B_s$.
We denote the ground state and the ground-state energy of $\delta H_s$ by $\ket{E_{\Lambda_s,0}}$ and $E_{\Lambda_s,0}$, respectively. 
From the definition of $\ket{\phi}$, we obtain 
\begin{align}
&\langle h_s \rangle_{\phi}  = \frac{1}{\|\phi\|^2}\bra{\psi} \Pi_{\le E}  \Pi_{> E'}^{(s)}h_s\Pi_{> E'}^{(s)} \Pi_{\le E} \ket{\psi}  \ge E',     \notag \\
&\langle (h_{s,s+1} +h_{s-1,s})  \rangle_{\phi} \ge   - (\| h_{s,s+1}\| + \| h_{s-1,s}|\|)  \ge -2g_0,  \notag  \\
&\langle \delta H_s \rangle_{\phi}  \ge E_{\Lambda_s,0} \ge E_{\tc,0}- E_{s,0} -2g_0 , 
\label{delta_H_s_ave_phi}
\end{align}
where the last inequality is derived from
\begin{align}
E_{\tc,0} \le& \bigl(\bra{E_{s,0}}\otimes \bra{E_{\Lambda_s,0}}\bigr) H_\tc\bigl( \ket{E_{s,0}}\otimes \ket{E_{\Lambda_s,0}}\bigr) \notag \\
\le & E_{s,0}+ E_{\Lambda_s,0} +\| h_{s,s+1}\| +\|h_{s-1,s}\| \le E_{s,0}+ E_{\Lambda_s,0} + 2g_0  .
\end{align}
The inequalities in~\eqref{delta_H_s_ave_phi} imply the lower bound of $\langle H_\tc \rangle_{\phi}$ in~\eqref{explicit_H_tc_phi_ave} as follows:
\begin{align}
\langle H_\tc \rangle_{\phi} \ge  E_{\tc,0}+E'   - E_{s,0}- 4g_0 . \label{lower_bound_of_ave_of_H}
\end{align}
By applying the inequality~\eqref{Norm_of_Psi_tilde} with \eqref{lower_bound_of_ave_of_H} to Eq.~\eqref{Basic_theorem_for_effe_Hami}, 
we prove the inequality~\eqref{Energy_dist_subsystem_Lc}.

%%%%%%%%%%%%%%%%%%%%%%%%%%%%%%%%%%%%%%%%%%%%%%%%%%%%%%%%%%%%%%%%%%%%%%%%%%%%%%%%%%%%%%%%%%%%%%%%%%
{~}

{~}

{~}\\
%%%%%%%%%%%%%%%%%%%%%%%%%%%%%%%%%%%%%%%%%%%%%%%%%%%%%%%%%%%%%%%%%%%%%%%%%%%%%%%%%%%%%%%%%%%%%%%%%%
%%%%%%%%%%%%%%%%%%%%%%%%%%%%%%%%%%%%%%%%%%%%%%%%%%%%%%%%%%%%%%%%%%%%%%%%%%%%%%%%%%%%%%%%%%%%%%%%%%
{\bf[Proof of the inequality~\eqref{Norm_of_Psi_tilde}]}

We, in the following, prove the inequality~\eqref{Norm_of_Psi_tilde}.
We start from the following equality:
\begin{align}
\bra{\phi}H_\tc \ket{\phi}=\bra{\phi}  \Pi_{\le x }H_\tc  \Pi_{\le x} \ket{\phi} + \sum_{j=1}^\infty \bra{\phi}  \Pi_{[x +(j-1)y,x + j y) } H_\tc  \Pi_{[{ x } +(j-1){ y }, { x } + j{ y }) }   \ket{\phi} ,
\end{align}
where ${ x }$ and $y$ are parameters which we set afterward.
We here obtain
\begin{align}
\bra{\phi}H_\tc \ket{\phi}&\le x  \| \Pi_{< x}  \ket{\phi}\|^2 + \sum_{j=1}^\infty  (x + j{ y })
\| \Pi_{[{ x } +(j-1){ y }, { x } + j{ y }) }   \ket{\phi} \|^2 \notag  \\
&= { x } \biggl ( \| \Pi_{<x}\ket{\phi}\|^2 + \sum_{j=1}^\infty 
\| \Pi_{[{ x } +(j-1){ y }, { x } + j{ y }) }   \ket{\phi} \|^2 \biggr)+ y \sum_{j=1}^\infty   j
\| \Pi_{[{ x } +(j-1){ y }, { x } + j{ y }) }   \ket{\phi} \|^2 \notag \\
&={ x } \| \phi \|^2+y\sum_{j=1}^\infty   j
\| \Pi_{[{ x } +(j-1){ y }, { x } + j{ y }) }   \ket{\phi} \|^2.                                                               \label{Cal_Ave_H_tilde_psi}
\end{align}

The definition of $\ket{\phi}$ in Eq.~\eqref{Def:of_phi} implies 
\begin{align}
\| \Pi_{[{ x } +(j-1){ y }, { x } + j{ y }) }   \ket{\phi} \|^2&= \| \Pi_{[{ x } +(j-1){ y }, { x } + j{ y }) }\Pi_{> E'}^{(s)}   \Pi_{\le E} \ket{\psi}\|^2 \notag \\
&\le \| \Pi_{[{ x } +(j-1){ y }, { x } + j{ y }) }\Pi_{> E'}^{(s)}   \Pi_{\le E} \|^2 .
\label{phi_2_Pi_jm1_j_norm}
\end{align}
In order to obtain the upper bound of $\| \Pi_{[{ x } +(j-1){ y }, { x } + j{ y }) }\Pi_{> E'}^{(s)}   \Pi_{\le E} \|^2$, we utilize the following lemma (see~\ref{Proof of Lemma_ref_thm:expE} for the proof):

\begin{lemma}\label{thm:expE}
Let $O_s$ be an arbitrary operator such that $[O_s,h_s]=0$.
Then, we have
\begin{align}
 \label{eq:general-expE}
  \| \Pi_{\ge E'}O_s  \Pi_{\le E} \| \le 4 \| O_s\| \cdot e^{-\lambda(E'-E)} 
  \end{align}
 with $\lambda:= 1/(12gk^2 +4g_0)$.
\end{lemma}
By choosing $O_s=\Pi_{\ge E'}^{(s)}$ in Lemma~\ref{thm:expE}, we have $[\Pi_{\ge E'}^{(s)}, h_s]=0$ from the definition~\eqref{projection_to_spectrum}. 
Hence,  the inequality~\eqref{eq:general-expE} implies
\begin{align}
 \| \Pi_{[{ x } +(j-1){ y }, { x } + j{ y }) }\Pi_{> E'}^{(s)}   \Pi_{\le E} \|^2  \le 16 e^{-2\lambda \bigl[{ x } -E+ (j-1){ y } \bigr]}, \label{inequality_91_97_proof_for}
\end{align}
where we use the fact that $\|\Pi_{\ge E'}^{(s)}\|=1$.
By the use of this inequality, we have from \eqref{phi_2_Pi_jm1_j_norm}
\begin{align}
\sum_{j=1}^{\infty}   j \| \Pi_{[{ x } +(j-1){ y }, { x } + j{ y }) }   \ket{\phi} \|^2 &\le 16 e^{-2\lambda ({ x } -E )}  \sum_{j=1}^{\infty}    j e^{-2\lambda (j-1){ y } } \notag \\
&=16e^{-2\lambda ({ x } -E)} \frac{e^{4y \lambda}}{(e^{2y \lambda}-1)^2}.
\end{align}
This inequality reduces the inequality~\eqref{Cal_Ave_H_tilde_psi} to
\begin{align}
\bra{\phi}H_\tc \ket{\phi}&\le { x } \|\phi \|^2+16y \cdot e^{-2\lambda ( { x } -E)} \frac{e^{4y \lambda}}{(e^{2y \lambda}-1)^2} .
\end{align}
By choosing $y=1/(2\lambda)$, we obtain
\begin{align}
\bra{\phi}H_\tc \ket{\phi}&\le { x } \| \phi \|^2+\frac{8e^2}{\lambda(e-1)^2} e^{-2\lambda ({ x } -E )}   .
\end{align}
From the definition~\eqref{Definition_H_phi_ave}, we have $\bra{\phi}H_\tc \ket{\phi}= \|\phi\|^2 \cdot \langle H_\tc \rangle_{\phi}$,  
which reduces the above inequality to
\begin{align}
 \|  \phi \|^2  \le\frac{8e^2}{\lambda(e-1)^2(\langle H_\tc \rangle_{\phi}-{ x })}e^{-2\lambda ({ x } -E)}   .
\end{align}
By choosing ${ x }$ so that $\langle H_\tc \rangle_{\phi} -{ x } = 1 /(2\lambda)$, we finally obtain 
\begin{align}
 \| \phi \|  \le \frac{4e^{3/2}}{e-1} e^{-\lambda (\langle H_\tc \rangle_{\phi}- E)}    .
\end{align}
This completes the proof. $\square$

\subsubsection{Proof of Lemma~\ref{thm:expE}} \label{Proof of Lemma_ref_thm:expE}
In order to derive the lemma, we utilize the following additional lemma~\cite{Kuwahara_2016_njp}:
\begin{lemma}[Theorem 2.1 in Ref.~\cite{Kuwahara_2016_njp}]\label{thm:njp_kuwahara}
Let $H$ be an arbitrary Hamiltonian in the form of 
\begin{align}
H=\sum_{Z: |Z| \le k} h_Z, \quad \sum_{Z:Z\ni i} \|h_Z\| \le g. \label{k_local_g_extensive}
\end{align}
Then, for an arbitrary $q$-local operator $\Gamma^{(q)}$ (i.e., at most $q$-body interactions are included), we have
\begin{align}
\| [H,\Gamma^{(q)}]\| \le 6gk q \|\Gamma^{(q)}\|.
  \end{align}
\end{lemma}

{~}

In order to prove Lemma~\ref{thm:expE}, we follow Ref.~\cite{Arad_2016}.
We start from the inequality
\begin{align}
\|\Pi_{\ge E'}  O_s  \Pi_{\le E}\| &= 
\left \|\Pi_{\ge E'} e^{-\nu H_\tc} e^{\nu H_\tc} O_s e^{-\nu H_\tc} e^{\nu H_\tc}\Pi _{\le E} \right\|
\le  e^{-\nu (E' - E)}\left \|e^{\nu H_\tc}  O_s   e^{-\nu H_\tc}\right\|.   \label{upper_bound_norm_Pi_E'_O_s_Pi_E}
\end{align}
Then, we need to upper-bound
\begin{align}
\left \|e^{\nu H_\tc}  O_s  e^{-\nu H_\tc}\right\|\le  \sum_{m=0}^\infty \frac{\nu^m}{m!} \| \ad_{H_\tc}^m (O_s)\|,
\label{bch_expansion_H_tc_Os}
\end{align}
where $\ad$ is the commutation operator, namely $\ad_{H_\tc} (\cdot) := [H_\tc,\cdot]$.
On the norm of $\|\ad_{H_\tc}^m (O_s)\|$, we prove the following inequality:
\begin{align}
\|\ad_{H_\tc}^m (O_s)\| \le 2(6gk^2 + 2g_0)^m m!   \|O_s\|.
\label{assump_kappa_form}
\end{align}

For the proof of \eqref{assump_kappa_form}, we use the mathematical induction.
For $m=1$, we prove the inequality~\eqref{assump_kappa_form} as follows:
\begin{align}
\|\ad_{H_\tc} (O_s)\| = \|[h_{s-1,s} +h_{s,s+1}, O_s]\|  \le 2 \|h_{s-1,s} +h_{s,s+1}\| \cdot \|O_s\| \le 4g_0 \|O_s\| \le 2(6gk^2 + 2g_0) \|O_s\|.
\end{align}
We then assume the inequality~\eqref{assump_kappa_form} for $m\le m_0$ and consider the case of general $m_0+1$:
\begin{align}
\|\ad_{H_\tc}^{m_0+1} (O_s)\| &= \|\ad_{H_\tc}^{m_0} \left( [h_{s-1,s} +h_{s,s+1}, O_s] \right)\|   \notag \\
&\le 2 \sum_{m_1+m_2=m_0}  \|\ad_{H_\tc}^{m_1}(O_s)\| \cdot \|\ad_{H_\tc}^{m_2} \left( h_{s-1,s} +h_{s,s+1} \right)\|  .
%+ 2 \|h_{s-1,s} +h_{s,s+1}\| \cdot\|\ad_{H_\tc}^m(O_s)\| \le 4g_0 \|O_s\|
\label{induction_commutator_os}
\end{align}
By applying Lemma~\ref{thm:njp_kuwahara} to $\ad_{H_\tc}^{m} ( h_{s-1,s} +h_{s,s+1})$, we have 
\begin{align}
 \|\ad_{H_\tc}^m ( h_{s-1,s} +h_{s,s+1})\| \le (6gk^2)^m m! \|h_{s-1,s} +h_{s,s+1}\| \le  2g_0 (6gk^2)^m m!   ,
 \label{commutator_h_tc_h_s_s}
\end{align}
where we use the fact that $\ad_{H_\tc}^j ( h_{s-1,s} +h_{s,s+1})$ ($j\in \mathbb{N}$) has at most $(jk)$-body interactions.
By applying the inequalities~\eqref{assump_kappa_form} and \eqref{commutator_h_tc_h_s_s} to \eqref{induction_commutator_os}, we obtain
\begin{align}
\|\ad_{H_\tc}^{m_0+1} (O_s)\| &\le  
4g_0 \|O_s\|  \sum_{m_1+m_2=m_0}(6gk^2+2g_0)^{m_1} (6gk^2)^{m_2} m_1! m_2!  \notag \\
&\le 2(6gk^2+2g_0)^{m_0+1} \|O_s\|  m_0! \sum_{j=0}^{m_0} \frac{1}{\binom{m_0}{j}} \le 2(6gk^2+2g_0)^{m_0+1} \|O_s\|   (m_0+1)! .
\end{align}
This completes the proof of \eqref{assump_kappa_form}.

By applying the inequality~\eqref{bch_expansion_H_tc_Os} to \eqref{assump_kappa_form}, we obtain 
\begin{align}
\left \|e^{\nu H_\tc}  O_s  e^{-\nu H_\tc}\right\|\le  2 \|O_s\| \sum_{m=0}^\infty [\nu (6gk^2 + 2g_0)]^m  = \frac{2\|O_s\|}{1-\nu (6gk^2 + 2g_0)},
\end{align}
which yields $\left \|e^{\nu H_\tc}  O_s  e^{-\nu H_\tc}\right\| \le 4\|O_s\|$ for $\nu=1/(12gk^2 + 4g_0)=:\lambda$.
Therefore, from \eqref{upper_bound_norm_Pi_E'_O_s_Pi_E}, we prove Lemma~\ref{thm:expE}. $\square$

%%%%%%%%%%%%%%%%%%%%%%%%%%%%%%%%%%%%%%%%%%%%%%%%%%%%%%%%%%%%%%%%%%%%%%%%%%%%%%%%%%%%%%%%%%%%%%%%%%
%%%%%%%%%%%%%%%%%%%%%%%%%%%%%%%%%%%%%%%%%%%%%%%%%%%%%%%%%%%%%%%%%%%%%%%%%%%%%%%%%%%%%%%%%%%%%%%%%%
%%%%%%%%%%%%%%%%%%%%%%%%%%%%%%%%%%%%%%%%%%%%%%%%%%%%%%%%%%%%%%%%%%%%%%%%%%%%%%%%%%%%%%%%%%%%%%%%%%
%%%%%%%%%%%%%%%%%%%%%%%%%%%%%%%%%%%%%%%%%%%%%%%%%%%%%%%%%%%%%%%%%%%%%%%%%%%%%%%%%%%%%%%%%%%%%%%%%%
%%%%%%%%%%%%%%%%%%%%%%%%%%%%%%%%%%%%%%%%%%%%%%%%%%%%%%%%%%%%%%%%%%%%%%%%%%%%%%%%%%%%%%%%%%%%%%%%%%
%%%%%%%%%%%%%%%%%%%%%%%%%%%%%%%%%%%%%%%%%%%%%%%%%%%%%%%%%%%%%%%%%%%%%%%%%%%%%%%%%%%%%%%%%%%%%%%%%%
% Effective terms : we should be most careful to this part
%%%%%%%%%%%%%%%%%%%%%%%%%%%%%%%%%%%%%%%%%%%%%%%%%%%%%%%%%%%%%%%%%%%%%%%%%%%%%%%%%%%%%%%%%%%%%%%%%%
%%%%%%%%%%%%%%%%%%%%%%%%%%%%%%%%%%%%%%%%%%%%%%%%%%%%%%%%%%%%%%%%%%%%%%%%%%%%%%%%%%%%%%%%%%%%%%%%%%
%%%%%%%%%%%%%%%%%%%%%%%%%%%%%%%%%%%%%%%%%%%%%%%%%%%%%%%%%%%%%%%%%%%%%%%%%%%%%%%%%%%%%%%%%%%%%%%%%%
%%%%%%%%%%%%%%%%%%%%%%%%%%%%%%%%%%%%%%%%%%%%%%%%%%%%%%%%%%%%%%%%%%%%%%%%%%%%%%%%%%%%%%%%%%%%%%%%%%
%%%%%%%%%%%%%%%%%%%%%%%%%%%%%%%%%%%%%%%%%%%%%%%%%%%%%%%%%%%%%%%%%%%%%%%%%%%%%%%%%%%%%%%%%%%%%%%%%%
%%%%%%%%%%%%%%%%%%%%%%%%%%%%%%%%%%%%%%%%%%%%%%%%%%%%%%%%%%%%%%%%%%%%%%%%%%%%%%%%%%%%%%%%%%%%%%%%%%

\subsection{Proof of Proposition~\ref{The energy distribution of the subsystem L_c}: the second part~\eqref{Energy_dist_subsystem_Lc_effective_H_lc}}

We can prove the inequality~\eqref{Energy_dist_subsystem_Lc_effective_H_lc} in the same way.
We define $\ket{\tilde{\phi}}$ as follows:
\begin{align}
\ket{\tilde{\phi}} :=  \Pi_{> E'}^{(s)} \tilde{\Pi}_{\le E} \ket{\psi},\label{Def:of_tilde_phi}
\end{align}
which implies
\begin{align}
\| \Pi_{> E'}^{(s)} \tilde{\Pi}_{\le E} \| = \max_{\ket{\psi}} \|\tilde{\phi}\|,
\end{align}
where $\|\tilde{\phi}\|$ denotes the norm of the state $\ket{\tilde{\phi}}$.
We then prove the following inequality similar to Ineq.~\eqref{Norm_of_Psi_tilde}:
\begin{align}
 \|  \tilde{\phi} \| \le \frac{4e^{3/2}}{e-1} e^{-\lambda' (\langle \tilde{H}_\tc \rangle_{\tilde{\phi} }- E )} , \label{Norm_of_Psi_tilde21}
\end{align}
with
\begin{align}
\langle \tilde{H}_\tc \rangle_{\tilde{\phi} }= \frac{\bra{\tilde{\phi}} \tilde{H}_\tc \ket{\tilde{\phi}} }{\| \tilde{\phi} \|^2}, \label{Definition_H_phi_ave_eff}
\end{align}
where $\lambda'$ was defined in Eq.~\eqref{def_Lambda'_prop_sup}.
We show the proof of the inequality~\eqref{Norm_of_Psi_tilde21} below.

We estimate a lower bound of $\langle \tilde{H}_\tc \rangle_{\tilde{\phi}}$. We first obtain
\begin{align}
\langle \tilde{H}_\tc \rangle_{\tilde{\phi}} := \frac{\bra{\tilde{\phi}}\tilde{H}_\tc\ket{\tilde{\phi}} }{\|\tilde{\phi}\|^2}=
\langle \tilde{h}_s \rangle_{\tilde{\phi}}   + \langle (h_{s,s+1} +h_{s-1,s})  \rangle_{\tilde{\phi}}  +\langle \delta \tilde{H}_s \rangle_{\tilde{\phi}},
\label{explicit_H_tc_phi_ave_effective}
\end{align}
where we define $\delta \tilde{H}_s:= \tilde{H}_\tc -\tilde{h}_s -h_{s,s+1} -h_{s-1,s}$.
We here define the ground state and the ground-state energy of $\delta \tilde{H}_s$ as $\ket{\tilde{E}_{\Lambda_s,0}}$ and $\tilde{E}_{\Lambda_s,0}$, respectively. 
Note that $\delta \tilde{H}_s$ acts on the sites $\Lambda_s:=\Lambda\setminus B_s$.
We can obtain the similar inequalities to \eqref{delta_H_s_ave_phi} as follows:
\begin{align}
&\langle \tilde{h}_s \rangle_{\tilde{\phi}}   \ge \frac{1}{\|\tilde{\phi}\|^2}\bra{\psi} \tilde{\Pi}_{\le E}  \Pi_{> E'}^{(s)} \tilde{h}_s\Pi_{> E'}^{(s)} \tilde{\Pi}_{\le E} \ket{\psi}  \ge \min (E', \tau_s ),     \notag \\
& \langle (h_{s,s+1} +h_{s-1,s})  \rangle_{\tilde{\phi}} \ge   - (\| h_{s,s+1}\| + \| h_{s-1,s}|\|)  \ge -2g_0 , \notag  \\
&\langle \delta \tilde{H}_s \rangle_{\tilde{\phi}}  \ge \tilde{E}_{\Lambda_s,0} \ge \tilde{E}_{\tc,0}- E_{s,0} -2g_0 , 
\label{delta_H_s_ave_phi_eff}
\end{align}
where the first inequality and the third inequality are derived from
\begin{align}
\tilde{h}_s\Pi_{> E'}^{(s)} = \begin{cases}
h_s\Pi_{(E',\tau_s)}^{(s)} + \tau_s \Pi_{[\tau_s,\infty)}^{(s)} &\for E'< \tau_s \\
\tau_s \Pi_{(E',\infty)}^{(s)} &\for E'\ge \tau_s .
\end{cases}
\end{align}
and
\begin{align}
\tilde{E}_{\tc,0} \le \bigl(\bra{E_{s,0}}\otimes \bra{\tilde{E}_{\Lambda_s,0}}\bigr) \tilde{H}_\tc \bigl( \ket{E_{s,0}}\otimes \ket{\tilde{E}_{\Lambda_s,0}}\bigr)
\le E_{s,0}+ \tilde{E}_{\Lambda_s,0} + 2g_0  .
\end{align}

We therefore obtain
\begin{align}
\langle \tilde{H}_\tc \rangle_{\tilde{\phi}} \ge  \tilde{E}_{\tc,0}+ \min(E',\tau_s)   - E_{s,0}- 4g_0 . \label{lower_bound_of_ave_of_H_eff}
\end{align}
By combining the inequalities~\eqref{Norm_of_Psi_tilde21} and \eqref{lower_bound_of_ave_of_H_eff}, we prove the inequality~\eqref{Energy_dist_subsystem_Lc_effective_H_lc}. $\square$

%%%%%%%%%%%%%%%%%%%%%%%%%%%%%%%%%%%%%%%%%%%%%%%%%%%%%%%%%%%%%%%%%%%%%%%%%%%%%%%%%%%%%%%%%%%%%%%%%%
{~}

{~}

{~}\\
%%%%%%%%%%%%%%%%%%%%%%%%%%%%%%%%%%%%%%%%%%%%%%%%%%%%%%%%%%%%%%%%%%%%%%%%%%%%%%%%%%%%%%%%%%%%%%%%%%
%%%%%%%%%%%%%%%%%%%%%%%%%%%%%%%%%%%%%%%%%%%%%%%%%%%%%%%%%%%%%%%%%%%%%%%%%%%%%%%%%%%%%%%%%%%%%%%%%%
{\bf[Proof of the inequality~\eqref{Norm_of_Psi_tilde21}]}

We follow the same step as the proof of the inequality~\eqref{Norm_of_Psi_tilde}.
We start from the following equality:
\begin{align}
\bra{\tilde{\phi}}\tilde{H}_\tc \ket{\tilde{\phi}}=
\bra{\tilde{\phi}} \tilde{\Pi}_{\le x } \tilde{H}_\tc \tilde{\Pi}_{\le x } \ket{\tilde{\phi}} 
+ \sum_{j=1}^\infty \bra{\tilde{\phi}} \tilde{\Pi}_{[x +(j-1)y,x + j y) } \tilde{H}_\tc \tilde{\Pi}_{[{ x } +(j-1){ y }, { x } + j{ y }) }   \ket{\tilde{\phi}} ,
\end{align}
which yields 
\begin{align}
\bra{\tilde{\phi}}\tilde{H}_\tc \ket{\tilde{\phi}} &\le x  \| \tilde{\Pi}_{\le x } \ket{\tilde{\phi}} \|^2 + \sum_{j=1}^\infty  (x + j{ y })
\| \tilde{\Pi}_{[{ x } +(j-1){ y }, { x } + j{ y }) }   \ket{\tilde{\phi}} \|^2 \notag  \\
&= { x } \biggl ( \| \tilde{\Pi}_{<x}\ket{\phi}\|^2 + \sum_{j=1}^\infty 
\| \tilde{\Pi}_{[{ x } +(j-1){ y }, { x } + j{ y }) }   \ket{\tilde{\phi}} \|^2 \biggr)+ y \sum_{j=1}^\infty   j
\| \tilde{\Pi}_{[{ x } +(j-1){ y }, { x } + j{ y }) }   \ket{\tilde{\phi}} \|^2 \notag \\
&={ x } \| \tilde{\phi}\|^2+y\sum_{j=1}^\infty   j
\| \tilde{\Pi}_{[{ x } +(j-1){ y }, { x } + j{ y }) }   \ket{\tilde{\phi}} \|^2.                                                               \label{Cal_Ave_H_tilde_psi_eff}
\end{align}
From the definition of $\ket{\tilde{\phi}}$ as in Eq.~\eqref{Def:of_tilde_phi}, we have 
\begin{align}
\| \tilde{\Pi}_{[{ x } +(j-1){ y }, { x } + j{ y }) }   \ket{\tilde{\phi}} \|^2&= \| \tilde{\Pi}_{[{ x } +(j-1){ y }, { x } + j{ y }) } \Pi_{> E'}^{(s)} \tilde{\Pi}_{\le E} \ket{\psi}\|^2 \notag \\
&\le \| \tilde{\Pi}_{[{ x } +(j-1){ y }, { x } + j{ y }) } \Pi_{> E'}^{(s)} \tilde{\Pi}_{\le E} \|^2 .
\label{phi_2_Pi_jm1_j_norm_eff}
\end{align}
Now, the problem is to bound the norm $\| \tilde{\Pi}_{[{ x } +(j-1){ y }, { x } + j{ y }) } \Pi_{> E'}^{(s)} \tilde{\Pi}_{\le E} \|$ from above.
Because the effective Hamiltonian $\tilde{H}_\tc$ is no longer given in the form of $k$-local Hamiltonian as \eqref{k_local_g_extensive}, we cannot use Lemma~\ref{thm:expE}.
Instead, we can prove the following proposition which is similar to Lemma~6.2 in Ref.~\cite{Arad_2016}:

\begin{prop}\label{thm:expE_effective}
Let $O_s$ be an arbitrary operator supported on the subset $B_s \subset \Lambda$ such that $[O_s,h_s] = 0$.
Then, we have
\begin{align}
 \label{eq:general-expE_effective}
  \|\tilde{\Pi}_{\ge E'} O_s  \tilde{\Pi}_{\le E}\| \le  4 \| O_s\| \cdot e^{-\lambda'(E'-E)} ,
  \end{align}
 where $\lambda':= \min \left(\frac{1}{112g_0} , \frac{1}{12gk^2} \right)$.
\end{prop}

By choosing $O_s=\Pi_{\ge E'}^{(s)}$ in Proposition~\ref{thm:expE_effective},  we obtain 
\begin{align}
 \| \tilde{\Pi}_{[{ x } +(j-1){ y }, { x } + j{ y }) }\Pi_{> E'}^{(s)}  \tilde{\Pi} _{\le E} \|^2  \le 
 16 e^{-2\lambda' [{ x } -E+ (j-1){ y }  ]}, \label{inequality_91_97_proof_for_eff}
\end{align}
where we use $\| \Pi_{\ge E'}^{(s)}\| =1$.
By the use of this inequality, we have from \eqref{phi_2_Pi_jm1_j_norm_eff}
\begin{align}
\sum_{j=1}^{\infty}   j \| \tilde{\Pi}_{[{ x } +(j-1){ y }, { x } + j{ y }) }   \ket{\tilde{\phi}} \|^2
 &\le 16e^{-2\lambda' ({ x } -E )}  \sum_{j=1}^{\infty}    j e^{-2\lambda' (j-1){ y } } =16e^{-2\lambda'({ x } -E)} \frac{e^{4y\lambda'}}{(e^{2y\lambda'}-1)^2}.
\end{align}
This inequality reduces the inequality~\eqref{Cal_Ave_H_tilde_psi_eff} to
\begin{align}
\bra{\tilde{\phi}}\tilde{H}_\tc \ket{\tilde{\phi}} &\le { x }  \| \tilde{\phi}\|^2+16y  \cdot e^{-2\lambda' ( { x } -E)} \frac{e^{4y \lambda'}}{(e^{2y\lambda'}-1)^2} .
\end{align}
By choosing $y=1/(2\lambda')$, we obtain
\begin{align}
\bra{\tilde{\phi}}\tilde{H}_\tc \ket{\tilde{\phi}}&\le { x } \| \tilde{\phi} \|^2+\frac{8e^2}{\lambda'(e-1)^2} e^{-2\lambda' ({ x } -E )}   .
\end{align}
From the definition~\eqref{Definition_H_phi_ave_eff}, we have $\bra{\tilde{\phi}}\tilde{H}_\tc \ket{\tilde{\phi}}= \|\tilde{\phi}\|^2 \cdot \langle \tilde{H}_\tc\rangle_{\phi}$,  
which reduces the above inequality to
\begin{align}
 \|  \tilde{\phi} \|^2  \le\frac{8e^2}{\lambda'(e-1)^2(\langle \tilde{H}_\tc \rangle_{\tilde{\phi}}-{ x })}e^{-2\lambda' ({ x } -E )}   .
\end{align}
By choosing ${ x }$ so that $\langle \tilde{H}_\tc \rangle_{\tilde{\phi}} -{ x } = 1 /(2\lambda')$, we finally obtain 
\begin{align}
 \| \tilde{\phi} \|  \le \frac{4e^{3/2}}{e-1} e^{-\lambda' (\langle \tilde{H}_\tc \rangle_{\tilde{\phi}}- E)}    .
\end{align}
This completes the proof. $\square$

\subsection{Proof of Proposition~\ref{thm:expE_effective}}

We start from the following inequality:
\begin{align}
\|\tilde{\Pi}_{\ge E'}  O_s  \tilde{\Pi}_{\le E}\| &= 
\left \|\tilde{\Pi}_{\ge E'} e^{-\nu \tilde{H}_\tc} e^{\nu \tilde{H}_\tc}  O_s    e^{-\nu \tilde{H}_\tc} e^{\nu \tilde{H}_\tc}\tilde{\Pi}_{\le E} \right\|  \notag \\
&\le  e^{-\nu (E' - E)}\left \|e^{\nu \tilde{H}_\tc}  O_s    e^{-\nu \tilde{H}_\tc}\right\|,   \label{upper_bound_norm_tilde_Pi_E'_O_s_tilde_Pi_E}
\end{align}
where $\nu$ is a parameter satisfying $\nu\le 1/(12gk^2)$ which is determined afterward.
We then decompose $\tilde{H}_\tc$ as $\tilde{H}_\tc=G+F$: 
\begin{align}
G= \sum_{s=0}^{q+1} \tilde{h}_s ,\quad  F= \sum_{s=0}^{q} h_{s,s+1}. 
\end{align}
By using the above decomposition, we have  
\begin{align}
e^{\nu \tilde{H}_\tc}=\mathcal{T}_{\rightarrow} \left[e^{\int_0^\nu F(x)dx} \right] e^{\nu G} ,\quad 
e^{-\nu \tilde{H}_\tc}=e^{-\nu G} \mathcal{T}_{\leftarrow} \left[e^{-\int_0^\nu F(x)dx} \right] ,
\end{align}
where $F(x):= e^{x G} F e^{-xG}$ and $\mathcal{T}_{\rightarrow}$, $\mathcal{T}_{\leftarrow}$ are the ordering operator:
\begin{align}
&\mathcal{T}_{\rightarrow} \left[ F(x_1) F(x_2) \cdots F(x_m) \right]= F(x_{p_1}) F(x_{p_2}) \cdots F(x_{p_m}) ,\notag \\
&\mathcal{T}_{\leftarrow} \left[ F(x_1) F(x_2) \cdots F(x_m) \right]= F(x_{p_m}) F(x_{p_{m-1}}) \cdots F(x_{p_1}),
\end{align}
with $p$ a permutation such that $x_{p_1} \le x_{p_2} \le \cdots \le x_{p_m}$.
From the above expressions, we obtain
\begin{align}
e^{\nu \tilde{H}_\tc}  O_s    e^{-\nu \tilde{H}_\tc}= \mathcal{T}_{\rightarrow} \left[e^{\int_0^\nu F(x)dx} \right]  e^{\nu G} O_s    e^{-\nu G} \mathcal{T}_{\leftarrow} \left[e^{\int_0^\nu F(x)dx} \right] . \label{dyson_exp_A_tilde_H}
\end{align} 

Because the operator $O_s  $ is supported on the subset $B_s\subset \Lambda$ and satisfies $[O_s  ,\tilde{h}_s] = 0$, we have $[O_s  ,G]=0$ and hence 
\begin{align}
e^{\nu G} O_s    e^{-\nu G} =O_s  .
\end{align} 
Also, from $[\tilde{h}_s,\tilde{h}_{s'}]=0$ for $\forall s, s'$, we obtain 
\begin{align}
&F(x)= \sum_{s=0}^{q} h_{s,s+1}(x) ,  \notag \\
&h_{s,s+1}(x)= e^{x G} h_{s,s+1} e^{-x G} =  e^{x (\tilde{h}_s + \tilde{h}_{s+1})} h_{s,s+1}e^{-x (\tilde{h}_s + \tilde{h}_{s+1})} ,
\end{align}
where we use the fact that $[h_{s,s+1},\tilde{h}_{s'}]=0$ as long as $s'\neq s,s+1$.
Therefore, $h_{s,s+1}(x)$ is still supported on the subset $B_s \sqcup B_{s+1} \subset \Lambda$. We now define
\begin{align}
\mathfrak{g} := \sup_{ \substack{0<x<\nu \\ s \in \{ 0,1,\ldots,q \}} }\| h_{s,s+1}(x)\| .
\label{def:tilde:g}
\end{align}

In order to calculate $\mathfrak{g}$ in Eq.~\eqref{def:tilde:g}, we need to consider the norm of
\begin{align}
h_{s,s+1}(x)= e^{x \tilde{h}_s} e^{x \tilde{h}_{s+1}}  h_{s,s+1}  e^{-x \tilde{h}_{s+1}} e^{-x \tilde{h}_s}   . \label{h_Z_two_imaginary_time_evolution}
\end{align}
As we will prove in~\ref{Proof of the inequalityeqrefimaginary_time_effective_ham_ineq}, we prove the following inequality:
\begin{align}
\| e^{x \tilde{h}_s} e^{x \tilde{h}_{s+1}}  h_{s,s+1}  e^{-x \tilde{h}_{s+1}} e^{-x \tilde{h}_s} \|\le 28g_0 .
\label{tilde_g_28_g_0}
\end{align}
under the condition of 
\begin{align}
 x \le \nu \le \frac{1}{12gk^2}. \label{condition_for_x_nu_12gk2}
\end{align}
From the inequality~\eqref{tilde_g_28_g_0}, we obtain $\mathfrak{g}=28g_0$.

Then, we consider the Baker-Campbell-Hausdorff expansion as follows:
\begin{align}
e^{\nu \tilde{H}_\tc}  O_s    e^{-\nu \tilde{H}_\tc}= \sum_{m=0}^\infty \int_0^{\nu} dx_1\int_{x_1}^{\nu} dx_2 \cdots \int_{x_{m-1}}^{\nu} 
\ad_{F(x_1)} \ad_{F(x_2)} \cdots \ad_{F(x_m)} (O_s) dx_m , \label{complex_time_evo_O_s}
\end{align} 
where $\ad$ is the commutation operator, namely $\ad_{F(x)} (\cdot) := [F(x),\cdot]$.
By using Eq.~\eqref{def:tilde:g}, the norm of the commutators are bounded from above by
\begin{align}
&\| \ad_{F(x)} (O_s) \| \le  \| \ad_{h_{s,s+1}(x)} (O_s) \| + \| \ad_{h_{s-1,s}(x)} (O_s) \|  \le  2 (\|h_{s,s+1}(x) \| + \|h_{s-1,s}(x) \| )\|O_s\| \le 4\mathfrak{g} \|O_s\| , \notag \\
&\| \ad_{F(x_1)} \ad_{F(x_2)} (O_s) \| \le  2^2 \cdot 3\mathfrak{g} ( 2  \mathfrak{g}\|O_s\| )  , \quad
\| \ad_{F(x_1)}\ad_{F(x_2)} \ad_{F(x_3)} (O_s) \| \le 2^3 \cdot4\mathfrak{g} [  3\mathfrak{g} ( 2  \mathfrak{g}\|O_s\| )  ],
\end{align} 
which is generalized as 
\begin{align}
&\| \ad_{F(x_1)} \ad_{F(x_2)} \cdots \ad_{F(x_m)} (O_s) \| \le  (m+1)! (2\mathfrak{g})^m \|O_s\| . \label{multi_commu_norm_upper_O_s}
\end{align} 
By combining the inequality~\eqref{multi_commu_norm_upper_O_s} with Eq.~\eqref{complex_time_evo_O_s}, we obtain the upper bound of
\begin{align}
\| e^{\nu \tilde{H}_\tc}  O_s    e^{-\nu \tilde{H}_\tc}\| \le  \sum_{m=0}^\infty \frac{\nu^m}{m!}   (m+1)! (2\mathfrak{g})^m \|O_s\|  =\frac{ \|O_s\|}{(1-2\mathfrak{g} \nu)^2}.
\end{align} 
Also, by choosing $\nu$ such that $2\mathfrak{g} \nu \le 1/2$,  we have $\| e^{\nu \tilde{H}_\tc}  O_s    e^{-\nu \tilde{H}_\tc}\| \le 4\|O_s\|$ and reduce the inequality~\eqref{upper_bound_norm_tilde_Pi_E'_O_s_tilde_Pi_E} to 
\begin{align}
\|\tilde{\Pi}_{\ge E'}  O_s  \tilde{\Pi}_{\le E}\| \le 4 \|O_s\|e^{- \nu (E' - E)} .
 \label{upper_bound_norm_tilde_Pi_E'_O_s_tilde_Pi_E_2}
\end{align}
This is the inequality which we aim to prove. 
Because of $\mathfrak{g}=28g_0$, under the assumption of $\nu \le 1/(12gk^2)$, the condition $2\mathfrak{g} \nu \le 1/2$ is satisfied for 
\begin{align}
\nu  = \lambda':=  \min \left(\frac{1}{112g_0} , \frac{1}{12gk^2} \right).
\end{align} 
This completes the proof of Proposition~\ref{thm:expE_effective}. $\square$

%%%%%%%%%%%%%%%%%%%%%%%%%%%%%%%%%%%%%%%%%%%%%%%%%%%%%%%%%%%%%%%%%%%%%%%%%%%%%%%%%%%%%%%%%%%%%%%%%%

%%%%%%%%%%%%%%%%%%%%%%%%%%%%%%%%%%%%%%%%%%%%%%%%%%%%%%%%%%%%%%%%%%%%%%%%%%%%%%%%%%%%%%%%%%%%%%%%%%
%%%%%%%%%%%%%%%%%%%%%%%%%%%%%%%%%%%%%%%%%%%%%%%%%%%%%%%%%%%%%%%%%%%%%%%%%%%%%%%%%%%%%%%%%%%%%%%%%%
\subsubsection{Proof of the inequality~\eqref{tilde_g_28_g_0}}\label{Proof of the inequalityeqrefimaginary_time_effective_ham_ineq}

For the proof, we start from $\| e^{x \tilde{h}_s}  O  e^{-x \tilde{h}_s}\|$ for an arbitrary operator $O$.
From the definition of $\tilde{h}_s$,  the operator $e^{x \tilde{h}_s}$ is decomposed as follows:  
\begin{align}
e^{x \tilde{h}_s}= \Pi^{(s)}_{<\tau_s} e^{x h_s }  +\Pi^{(s)}_{\ge \tau_s} e^{x \tau_s},
\end{align}
where we utilized the equalities $\tilde{h}_s \Pi^{(s)}_{<\tau_s} = h_s \Pi^{(s)}_{<\tau_s} $ and  
$\tilde{h}_s \Pi^{(s)}_{\ge \tau_s} = \tau_s \Pi^{(s)}_{\ge \tau_s}$.
We then have
\begin{align}
e^{x \tilde{h}_s}  O  e^{-x \tilde{h}_s} =& \Pi^{(s)}_{\ge \tau_s}  O \Pi^{(s)}_{\ge \tau_s}  + \Pi^{(s)}_{< \tau_s} e^{x h_s } O  e^{-x h_s }\Pi^{(s)}_{< \tau_s} \notag \\
&+ \Pi^{(s)}_{< \tau_s} e^{x (h_s -\tau_s) } O \Pi^{(s)}_{\ge \tau_s} + \Pi^{(s)}_{\ge \tau_s} e^{x \tau_s}  O e^{-x h_s} \Pi^{(s)}_{< \tau_s} . \label{ineq_1_imaginary_time_OX}
\end{align}
The norms of the four terms are bounded from above by
\begin{align}
\| \Pi^{(s)}_{\ge \tau_s}  O \Pi^{(s)}_{\ge \tau_s} \| &\le \|O\| , \notag \\
\| \Pi^{(s)}_{< \tau_s} e^{x h_s } O  e^{-x h_s }\Pi^{(s)}_{< \tau_s} \| &\le \|e^{x h_s } O  e^{-x h_s }\| ,\notag \\
\| \Pi^{(s)}_{< \tau_s} e^{x (h_s -\tau_s) } O \Pi^{(s)}_{\ge \tau_s} \| &\le   \| \Pi^{(s)}_{< \tau_s} e^{x (h_s -\tau_s) }\|  \cdot  \|O\Pi^{(s)}_{\ge \tau_s} \|  \le \|O\| ,\notag \\ 
\| \Pi^{(s)}_{\ge \tau_s} e^{x \tau_s}  O_Xe^{-x h_s} \Pi^{(s)}_{< \tau_s} \| 
&=\| \Pi^{(s)}_{\ge \tau_s}  e^{x \tau_s} e^{-x h_s} e^{x h_s} Oe^{-xh_s} \Pi^{(s)}_{< \tau_s}\|  \notag \\
&\le \| \Pi^{(s)}_{\ge \tau_s}  e^{x (\tau_s - h_s )} \| \cdot \| e^{x h_s} Oe^{-xh_s}\|   \le \| e^{x h_s} Oe^{-xh_s}\| . \label{ineq_2_imaginary_time_OX}
\end{align}
By combining the above two inequalities~\eqref{ineq_1_imaginary_time_OX} and \eqref{ineq_2_imaginary_time_OX}, we obtain
\begin{align}
\| e^{x \tilde{h}_s}  O  e^{-x \tilde{h}_s}\|  \le  2 \|O\| + 2 \| e^{x h_s} Oe^{-xh_s}\| .\label{imaginary_time_OX_hs_x_upp0_effective}
\end{align}

From the inequality~\eqref{imaginary_time_OX_hs_x_upp0_effective}, we obtain
\begin{align}
\| e^{x \tilde{h}_{s+1}} e^{x \tilde{h}_s}  h_{s,s+1} e^{-x \tilde{h}_s} e^{-x \tilde{h}_{s+1}} \|  \le  2 \|e^{x \tilde{h}_s}  h_{s,s+1} e^{-x \tilde{h}_s} \| + 2 \| e^{x h_{s+1}} e^{x \tilde{h}_s}  h_{s,s+1} e^{-x \tilde{h}_s}  e^{-xh_{s+1}}\| ,\label{imaginary_time_h_s_sp1_hs_x_upp0_effective}
\end{align}
where we set $O=e^{x \tilde{h}_s}  h_{s,s+1} e^{-x \tilde{h}_s}$ in \eqref{imaginary_time_OX_hs_x_upp0_effective}.
Furthermore, the inequality~\eqref{imaginary_time_OX_hs_x_upp0_effective} gives
\begin{align}
\|e^{x \tilde{h}_s}  h_{s,s+1} e^{-x \tilde{h}_s} \| \le 2  \|h_{s,s+1}\| +2  \|e^{x h_s}  h_{s,s+1} e^{-x h_s} \|,
\label{imaginary_time_h_s_sp1_hs_x_upp0_effective_first}
\end{align}
and 
\begin{align}
\| e^{x h_{s+1}} e^{x \tilde{h}_s}  h_{s,s+1} e^{-x \tilde{h}_s}  e^{-xh_{s+1}}\|  
&=\| e^{x \tilde{h}_s}  e^{x h_{s+1}}  h_{s,s+1} e^{-xh_{s+1}} e^{-x \tilde{h}_s} \|  \notag \\
&\le 2 \| e^{x h_{s+1}}  h_{s,s+1} e^{-xh_{s+1}}\| + 2 \| e^{x h_s}  e^{x h_{s+1}}  h_{s,s+1} e^{-xh_{s+1}} e^{-x h_s} \| ,
\label{imaginary_time_h_s_sp1_hs_x_upp0_effective_second}
\end{align}
where we use $[\tilde{h}_s, h_{s+1}]=0$ which yields $e^{x h_{s+1}} e^{x \tilde{h}_s}=e^{x \tilde{h}_s}e^{x h_{s+1}} $.
By applying the inequalities~\eqref{imaginary_time_h_s_sp1_hs_x_upp0_effective_first} and \eqref{imaginary_time_h_s_sp1_hs_x_upp0_effective_second} to
\eqref{imaginary_time_h_s_sp1_hs_x_upp0_effective}, we have
\begin{align}
&\| e^{x \tilde{h}_{s+1}} e^{x \tilde{h}_s}  h_{s,s+1} e^{-x \tilde{h}_s} e^{-x \tilde{h}_{s+1}} \|   \notag \\
\le&  4 \|h_{s,s+1}\| +4  \|e^{x h_s}  h_{s,s+1} e^{-x h_s} \|+4 \| e^{x h_{s+1}}  h_{s,s+1} e^{-xh_{s+1}}\| + 4 \| e^{x h_s}  e^{x h_{s+1}}  h_{s,s+1} e^{-xh_{s+1}} e^{-x h_s} \| .
\label{imaginary_time_h_s_sp1_hs_x_upper_4term}
\end{align}

We here estimate an upper bound of each of the norms in \eqref{imaginary_time_h_s_sp1_hs_x_upper_4term}.
We first consider $\|e^{x h_s}  h_{s,s+1} e^{-x h_s} \|$ by using the Baker-Campbell-Hausdorff expansion:
\begin{align}
\|e^{x h_s}  h_{s,s+1} e^{-x h_s} \|\le \sum_{m=0}^\infty \frac{x^m}{m!} \| \ad_{h_s}^m ( h_{s,s+1} )\|.
\label{imaginary_time_h_s_sp1_xhs}
\end{align}
By applying Lemma~\ref{thm:njp_kuwahara} to $ \ad_{h_s}^m ( h_{s,s+1} )$, we have 
\begin{align}
 \|\ad_{h_s}^m ( h_{s,s+1} ) \| \le (6gk^2)^m m! \|h_{s,s+1}\| \le  g_0 (6gk^2)^m m!   ,
\end{align}
where we use $\| h_{s,s+1}\| \le g_0$ in \eqref{truncated_Hamiltonian_block_interaction}.
This inequality reduces \eqref{imaginary_time_h_s_sp1_xhs} to 
\begin{align}
\|e^{x h_s}  h_{s,s+1} e^{-x h_s} \|\le g_0 \sum_{m=0}^\infty (6gk^2x)^m= g_0 \frac{1}{1-6gk^2x} \le 2g_0,
\label{imaginary_time_hs_s_s+1_h_s_1}
\end{align}
where in the last inequality we use the condition of \eqref{condition_for_x_nu_12gk2}, namely $x\le \nu \le 1/(12gk^2)$.
We obtain the same inequality for $\| e^{x h_{s+1}}  h_{s,s+1} e^{-xh_{s+1}}\| $ and $\|e^{x h_s}  e^{x h_{s+1}}  h_{s,s+1} e^{-xh_{s+1}} e^{-x h_s}\|$.
Notice that we can apply Lemma~\ref{thm:njp_kuwahara} to $h_s+h_{s+1}$, which allows us to obtain the upper bound of $\|e^{x h_s}  e^{x h_{s+1}}  h_{s,s+1} e^{-xh_{s+1}} e^{-x h_s}\|= \|e^{x (h_s+ h_{s+1})}  h_{s,s+1} e^{-x(h_s +h_{s+1})}\|$ in the same way of \eqref{imaginary_time_hs_s_s+1_h_s_1}.
Therefore, we finally obtain 
\begin{align}
&\| e^{x \tilde{h}_{s+1}} e^{x \tilde{h}_s}  h_{s,s+1} e^{-x \tilde{h}_s} e^{-x \tilde{h}_{s+1}} \| \le  28g_0.
\end{align}
This completes the proof. $\square$

\clearpage

\section{List of notations and definitions} \label{Sec:List of notations and definitions}

We here give a list of definitions and notations which we use several times in the proof.

\begin{enumerate}
\item{} $\{\ket{\Gs},\Delta\}$ (\ref{Sec:Definition of the Hamiltonian}): the ground state and the spectral gap of the Hamiltonian $H$.

\item{} $g$ [Eq.~\eqref{g_exensiveness}]: upper bound of one-site energy, which is set to be equal to $1$ by choosing the energy unit appropriately. 

\item{} $V_{X,Y}(\Lambda_0)$ [Eq.~\eqref{Def:V_X_Y}]: interaction operator between two subsystems $X\subset \Lambda$ and $Y\subset \Lambda$. 

\item{} ${\rm SR} (O,X)$ and ${\rm SR} (\ket{\psi},X)$ (\ref{Subsec:Schmidt rank}): Schmidt rank of operator and quantum state. 

\item{} $K$ [Eq.~\eqref{AGSP:formal_def}]:  approximate ground state projection (AGSP) for the ground state $\ket{\Gs}$.

\item{} $\ket{\Gs_K}$ [Eq.~\eqref{def_of_Gs_K}]: quantum state which is invariant by AGSP $K$.

\item{} $\{\delta_K, \epsilon_K, D_K \}$ [Eq.~\eqref{Def_AGSP_error_chap2}]: three parameters which characterize the AGSP operator $K$. 

\item{} $\{B_s\}_{s=0}^{q+1}$ [Eq.~\eqref{Block_definition_LR}, Supplementary Figure~\ref{fig:Area_law_Ham_truncate}]: decomposed subsets which constitute the total system $\Lambda$. This decomposition has two parameters $q+2$ (the number of blocks) and $l$ (the length of the bulk blocks, i.e., $B_1,B_2,\ldots, B_q$). 

\item{} $h_s$ [Eq.~\eqref{def:truncated_Hamiltonian}]:  internal interactions in block $B_s$. 

\item{} $\{ E_{s,j}, \ket{E_{s,j}}\}_j$ [Eq.~\eqref{truncation_effective_Hamiltonian_h_s}]: the eigenvalues and the eigenstates of $h_{s}$, respectively.

\item{} $h_{s,s+1}$ [Eq.~\eqref{def:truncated_Hamiltonian}]: interactions between blocks $B_s$ and $B_{s+1}$. The norm is bounded from above by $g_0$ as in \eqref{truncated_Hamiltonian_block_interaction}.

\item{} $H_\tc$ [Eq.~\eqref{def:truncated_Hamiltonian}]: interaction-truncated Hamiltonian. 

\item{} $\delta H_\tc$ (Lemma~\ref{thm:locality_exp_effectiveHam}): difference between $H_\tc$ and $H$, namely $\delta H_\tc:=H-H_\tc$.

\item{} $\{ \ket{\Gs_\tc}, E_{\tc,0}, \Delta_\tc \}$ (Lemmas~\ref{thm:locality_exp_effectiveHam} and \ref{overlap_Lemma_truncate_original}): the ground state, the ground energy and the spectral gap of the Hamiltonian $H_\tc$. 

\item{} $K_\tc$ (Proposition~\ref{prop:Overlap between the ground state and  low-entangled state}): approximate ground state projection (AGSP) for the ground state $\ket{\Gs_\tc}$.

\item{} $\{\tilde{h}_{s},\tau_s,\tau\}$ [Eqs.~\eqref{truncation_effective_Hamiltonian_h_s} and \eqref{tau_s_tau}]: the block Hamiltonian in $B_s$ with the energy cut-off up to $\tau_s = \tau + E_{s,0}$, where $\tau$ is a control parameter. 

\item{} $\tilde{H}_\tc$ [Eq.~\eqref{truncation_effective_Hamiltonian_H_tc}, Supplementary Figure~\ref{fig:effective_Ham}]: effective Hamiltonian by using the multi-energy cut-off. The three parameters $\{q, l ,\tau\}$ characterize $\tilde{H}_\tc$.

\item{} $\{ \ket{\tilde{\Gs}_\tc}, \tilde{E}_{\tc,0}, \tilde{\Delta}_\tc \}$ (Theorem~\ref{Effective Hamiltonian_multi_truncation}): the ground state, the ground energy and the spectral gap of the effective Hamiltonian $\tilde{H}_\tc$.

\item{} $\{\lambda, \lambda'\}$ [Eq.~\eqref{definition_lambda_lambda'}]: constants given in Theorem~\ref{Effective Hamiltonian_multi_truncation}. 

 \item{} $K_m(x)$ [Eq.~\eqref{Chebyshev_based_AGSP_func}, Lemma~\ref{ChebyShev_box}]: approximate filter function which is utilized to construct the AGSP $K$. 

\item{} $\Pi^{(s)}_{I}$, $\Pi_{I}$ and $\tilde{\Pi}_I$ [Eqs.~\eqref{projection_to_spectrum} and \eqref{projector_total_energy}]: the projection operators onto the eigenspaces of $h_s$, $H$ and $\tilde{H}_\tc$, respectively. 

\item{} $\{E_\bot,\kappa\}$ [Eqs.~\eqref{Def_E_bot_lemma} and \eqref{Def_epsilon_0_lemma}]: constants given in Lemma~\ref{energy_gap_upper_bound_projection}. 

\end{enumerate}

%
%\def\bibsection{\section*{Supplementary References}} 
%%\section{Supplementary References}
%%\renewcommand\bibname{}
%
%
%
%\bibliography{Area_LR_main}
%%\bibliography{Area_LR_sup}
%
\end{widetext}

%\appendix

\end{document}